\pdfoutput=1
\PassOptionsToPackage{unicode}{hyperref}
\PassOptionsToPackage{hyphens}{url}
\PassOptionsToPackage{dvipsnames,svgnames,x11names}{xcolor}
\documentclass[
  11pt,
  letterpaper,
]{article}
\usepackage{xcolor}
\usepackage[margin=1in]{geometry}
\usepackage{amsmath,amssymb}
\usepackage{iftex}
\ifPDFTeX
  \usepackage[T1]{fontenc}
  \usepackage[utf8]{inputenc}
  \usepackage{textcomp} 
\else 
  \usepackage{unicode-math} 
  \defaultfontfeatures{Scale=MatchLowercase}
  \defaultfontfeatures[\rmfamily]{Ligatures=TeX,Scale=1}
\fi
\usepackage{lmodern}
\ifPDFTeX\else
\fi
\IfFileExists{upquote.sty}{\usepackage{upquote}}{}
\IfFileExists{microtype.sty}{
  \usepackage[]{microtype}
  \UseMicrotypeSet[protrusion]{basicmath} 
}{}
\makeatletter
\@ifundefined{KOMAClassName}{
  \IfFileExists{parskip.sty}{%
    \usepackage{parskip}
  }{
    \setlength{\parindent}{0pt}
    \setlength{\parskip}{6pt plus 2pt minus 1pt}}
}{
  \KOMAoptions{parskip=half}}
\makeatother
\makeatletter
\ifx\paragraph\undefined\else
  \let\oldparagraph\paragraph
  \renewcommand{\paragraph}{
    \@ifstar
      \xxxParagraphStar
      \xxxParagraphNoStar
  }
  \newcommand{\xxxParagraphStar}[1]{\oldparagraph*{#1}\mbox{}}
  \newcommand{\xxxParagraphNoStar}[1]{\oldparagraph{#1}\mbox{}}
\fi
\ifx\subparagraph\undefined\else
  \let\oldsubparagraph\subparagraph
  \renewcommand{\subparagraph}{
    \@ifstar
      \xxxSubParagraphStar
      \xxxSubParagraphNoStar
  }
  \newcommand{\xxxSubParagraphStar}[1]{\oldsubparagraph*{#1}\mbox{}}
  \newcommand{\xxxSubParagraphNoStar}[1]{\oldsubparagraph{#1}\mbox{}}
\fi
\makeatother

\usepackage{longtable,booktabs,array}
\usepackage{calc} 
\usepackage{etoolbox}
\makeatletter
\patchcmd\longtable{\par}{\if@noskipsec\mbox{}\fi\par}{}{}
\makeatother
\IfFileExists{footnotehyper.sty}{\usepackage{footnotehyper}}{\usepackage{footnote}}
\makesavenoteenv{longtable}
\usepackage{graphicx}
\makeatletter
\newsavebox\pandoc@box
\newcommand*\pandocbounded[1]{
  \sbox\pandoc@box{#1}%
  \Gscale@div\@tempa{\textheight}{\dimexpr\ht\pandoc@box+\dp\pandoc@box\relax}%
  \Gscale@div\@tempb{\linewidth}{\wd\pandoc@box}%
  \ifdim\@tempb\p@<\@tempa\p@\let\@tempa\@tempb\fi
  \ifdim\@tempa\p@<\p@\scalebox{\@tempa}{\usebox\pandoc@box}%
  \else\usebox{\pandoc@box}%
  \fi%
}
\def\fps@figure{htbp}
\makeatother

\usepackage[]{natbib}
\usepackage{amsmath,amssymb,amsthm}
\usepackage{mathtools}
\usepackage{dsfont}           
\usepackage{bbm}              
\usepackage{enumitem}         
\usepackage{booktabs}         
\usepackage{microtype}        
\usepackage{xcolor}

\providecommand{\cprime}{$'$}

\usepackage{fancyhdr}
\newcommand*{\doi}[1]{doi:~\href{https://doi.org/#1}{\nolinkurl{#1}}}
\makeatletter
\@ifpackageloaded{caption}{}{\usepackage{caption}}
\AtBeginDocument{%
\ifdefined\contentsname
  \renewcommand*\contentsname{Table of contents}
\else
  \newcommand\contentsname{Table of contents}
\fi
\ifdefined\listfigurename
  \renewcommand*\listfigurename{List of Figures}
\else
  \newcommand\listfigurename{List of Figures}
\fi
\ifdefined\listtablename
  \renewcommand*\listtablename{List of Tables}
\else
  \newcommand\listtablename{List of Tables}
\fi
\ifdefined\figurename
  \renewcommand*\figurename{Figure}
\else
  \newcommand\figurename{Figure}
\fi
\ifdefined\tablename
  \renewcommand*\tablename{Table}
\else
  \newcommand\tablename{Table}
\fi
}
\@ifpackageloaded{float}{}{\usepackage{float}}
\floatstyle{ruled}
\@ifundefined{c@chapter}{\newfloat{codelisting}{h}{lop}}{\newfloat{codelisting}{h}{lop}[chapter]}
\floatname{codelisting}{Listing}

\usepackage{amsthm}
\theoremstyle{plain}
\newtheorem{proposition}{Proposition}[section]
\theoremstyle{plain}
\newtheorem{lemma}{Lemma}[section]
\theoremstyle{plain}
\newtheorem{theorem}{Theorem}[section]
\theoremstyle{plain}
\newtheorem{corollary}{Corollary}[section]
\theoremstyle{definition}
\newtheorem{definition}{Definition}[section]
\theoremstyle{remark}
\AtBeginDocument{}
\newtheorem*{remark}{Remark}

\newtheorem{refremark}{Remark}[section]

\makeatother
\makeatletter
\@ifpackageloaded{caption}{}{\usepackage{caption}}
\@ifpackageloaded{subcaption}{}{\usepackage{subcaption}}
\makeatother
\usepackage{bookmark}
\IfFileExists{xurl.sty}{\usepackage{xurl}}{} 
\hypersetup{
  pdftitle={Quantum Robustness via Sandwiched Rényi Divergences},
  pdfauthor={Luc Rey-Bellet},
  pdfkeywords={quantum information, distributionally robust
optimization, sandwiched Rényi divergence, Pinsker inequality, quantum
risk measures, BMV conjecture},
  colorlinks=true,
  linkcolor={blue},
  filecolor={Maroon},
  citecolor={blue},
  urlcolor={blue},
  pdfcreator={LaTeX via pandoc}}

\title{Quantum Robustness via Sandwiched Rényi Divergences}
\author{Luc Rey-Bellet}
\date{November 9, 2026}
\begin{document}
\maketitle
\begin{abstract}
We develop the theory of distributional robustness for quantum states:
if a reference state is known only approximately, how much can the
expectation of an observable change? Uncertainty is measured by the
sandwiched Rényi divergence of order \(\alpha \in [\tfrac12, \infty]\),
and the distributionally robust value is the worst-case expectation over
the states whose divergence from the reference does not exceed a budget
\(q\). Convex duality expresses this value in closed form as a
one-dimensional optimization over a sandwiched Schatten norm, running
from a fidelity risk at \(\alpha = \tfrac12\), through the entropic
(Gibbs) risk at \(\alpha = 1\), to a quantum conditional value-at-risk
at \(\alpha = \infty\). Our main structural result is a classicalization
theorem: for every \(\alpha\) the quantum problem coincides exactly with
the classical problem for a probability measure \(\omega_\alpha\) on the
spectral interval of the observable, constructed from Heinävaara's
tracial joint spectral measure, given by Stahl's theorem (the
Bessis--Moussa--Villani conjecture) at \(\alpha = 1\), and reducing to
the Born measure only at \(\alpha = \tfrac12\). For small budgets the
value obeys a square-root law whose variance is the inverse of the
sandwiched Rényi Fisher metric and, by classicalization, the variance of
\(\omega_\alpha\). Classicalization also transfers the Hoeffding,
Bernstein and Markov--Krein bounds of classical probability to the
quantum problem; read on the Helstrom observable they become
Pinsker-type inequalities, recovering Pinsker and Fuchs--van de Graaf
and adding Bernstein-type refinements. The overall message is a quantum
deflation: for \(\alpha > \tfrac12\), noncommutativity of the observable
with the state strictly reduces the variance of \(\omega_\alpha\) below
the Born variance, and with it every variance-sensitive bound; at
\(\alpha = 1\) this gives a Bernstein--Pinsker inequality deflated by
the Kubo--Mori variance.
\end{abstract}

\section{Introduction}\label{sec-introduction}

How much can the expectation of an observable change when the state is
known only approximately? We take an information theoretic approach: fix
a reference state \(\sigma\), an observable \(L\), and a divergence
\(\mathcal{D}\); the \emph{distributionally robust} value is \[
V(q) \;=\; \sup\bigl\{\operatorname{tr}(L\rho) :\ \mathcal{D}(\rho\|\sigma) \leq q\bigr\},
\] the worst case over all states within a divergence ball of radius
\(q\) around \(\sigma\). Classically --- \(\sigma\) a probability
distribution, \(L\) a random variable --- this is the basic object of
distributionally robust optimization (DRO), which goes back to Scarf's
min--max inventory problem \citep{scarf1958}. The divergence-ball form
used here arose in robust control and in economics, with the
relative-entropy constraint of Petersen, James and Dupuis
\citep{petersen-james-dupuis2000} and the robustness program of Hansen
and Sargent \citep{hansen-sargent2008}, and became the
\(\varphi\)-divergence framework of Ben-Tal et al.
\citep{ben-tal-etal2013} and the variance-regularized learning of Duchi
and Namkoong \citep{duchi-namkoong2019, duchi-namkoong2021}; the other
families of ambiguity sets, fixed by moments
\citep{bertsimas-popescu2005, delage-ye2010, wiesemann-kuhn-sim2014} or
by optimal transport
\citep{pflug-wozabal2007, esfahani-kuhn2018, blanchet-murthy2019, gao-kleywegt2023},
are outside the scope of this paper, and the survey of Kuhn, Shafiee and
Wiesemann \citep{kuhn2025dro} gives an in-depth account of the subject.
The quantum theory developed here continues a line of work on
divergence-based robustness and uncertainty quantification via
variational formulas for Rényi and related divergences
\citep{dupuis-katsoulakis-pantazis-reybellet2020, birrell-renyi-variational2021, birrell-dro-rareevents2019, katsoulakis-reybellet-wang2017}.

In the quantum problem \(\sigma\) is a density matrix, \(L\) is a
Hermitian matrix, and \(\mathcal{D}\) a quantum divergence. In this work
\(\mathcal{D}\) is the sandwiched Rényi divergence of order \(\alpha
\in [\tfrac12, \infty]\) \citep{muller-lennert2013, wilde2014}, a
quantum extension of the classical Rényi divergence
\citep{vanerven-harremoes2014}. We use the Hellinger normalization
\(H_\alpha\) in Section~\ref{sec-main-theorem} (classically the
Cressie--Read power divergences \citep{cressie-read1984}): at
\(\alpha = \tfrac12\) the squared Bures distance (closely related to the
fidelity), at \(\alpha=1\) the relative entropy, at \(\alpha = \infty\)
the max-relative entropy \citep{datta2009}. It is one of several quantum
Rényi families that agree on commuting pairs and differ elsewhere
\citep{tomamichel2016}. Depending on the parameter regime, the
sandwiched Rényi divergence characterizes optimal operational
thresholds: for \(\alpha >1\), it determines the exact strong converse
exponents for binary quantum hypothesis testing
\citep{mosonyi-ogawa2015} and for \(\alpha \in (\frac{1}{2}, 1)\), it
yields the exact strong converse exponents for privacy amplification and
state smoothing when measured by fidelity \citep{li-yao2024}. At
\(\alpha = \tfrac12\) the ball is a fidelity ball, the form in which
state-estimation guarantees are stated. On the whole range the
divergence is monotone under quantum channels
\citep{frank-lieb2013, beigi2013} and in \(\alpha\)
\citep{muller-lennert2013}, so that the balls are nested and the robust
value decreases as the order increases
(Remark~\ref{rem-alpha-monotone}), its dual closes in a sandwiched
Schatten norm, and it classicalizes. It is also the smallest of the
quantum Rényi divergences with the natural properties --- additivity,
data processing and the classical value on commuting pairs --- on the
whole range \(\alpha \in [\tfrac12, \infty]\), by Tomamichel's pinching
argument \citep[§4.2.2]{tomamichel2016}; for the robust problem this
means that its ball is the largest and its value the most conservative:
every upper bound proved below for \(V_\alpha\) holds for the DRO value
of any such divergence at the same order and budget
(Remark~\ref{rem-minimality}). The DRO problem for other quantum Rényi
divergences, such as Petz's family \citep{petz1986, hiai-mosonyi2017}
(which governs the quantum Chernoff and Hoeffding bounds
\citep{nussbaum-szkola2009, audenaert-nussbaum2008, hayashi2007}) or the
measured Rényi divergence \citep{berta2017}, and for the
\(f\)-divergences built from hockey-stick divergences
\citep{hirche-tomamichel2024, liu-hirche-cheng2025}, is an open problem
(Section~\ref{sec-discussion}).

By convex duality \(V(q)\) can be written as a one-dimensional Legendre
transform of the free energy
\(G_\alpha(s) = \sup_\rho\{\operatorname{tr}(L\rho) -
s\,H_\alpha(\rho\|\sigma)\}\); at \(\alpha = 1\) this is the Gibbs
variational principle \citep{petz1988variational, ohya-petz2004}, and
for general \(\alpha\) using the sandwiched structure
\(\sigma^\beta(\,\cdot\,)
\sigma^\beta\), \(\beta = (\alpha-1)/2\alpha\), we prove,
Theorem~\ref{thm-main}, \[
V_\alpha(q) \;=\; \inf_{c \in \mathbb{R}}\Bigl\{c + \bigl(1 + (\alpha-1)q\bigr)^{1/\alpha}
\bigl\|\bigl(\sigma^\beta(L - cI)\sigma^\beta\bigr)_+\bigr\|_{\alpha'}\Bigr\},
\qquad \alpha > 1,
\] the quantum form of the optimized certainty equivalent of Ben-Tal and
Teboulle \citep{ben-tal-teboulle2007}, with the classical
\(L^{\alpha'}\)-norm of \((L - c)_+\) replaced by a sandwiched Schatten
norm, and, for \(\alpha < 1\), with the positive part replaced by a
resolvent and the norm by its dual exponent, \[
V_\alpha(q) \;=\; \inf_{c > \lambda_{\max}(L)}\Bigl\{c - \frac{\bigl(1 + (\alpha-1)q\bigr)^{1/\alpha}}
{\bigl\|\sigma^{\gamma}(cI - L)^{-1}\sigma^{\gamma}\bigr\|_p}\Bigr\},
\qquad \tfrac12 \le \alpha < 1,
\] with \(\gamma = -\beta\) and \(p = \alpha/(1-\alpha)\), for \(q\)
below an explicit threshold: a Stieltjes transform of order \(p\), which
at \(\alpha = \tfrac12\) (\(p = 1\), the norm a trace) is a fidelity DRO
(Corollary~\ref{cor-fidelity-dro}). The worst-case state is explicit in
terms of the dual minimizer (Corollary~\ref{cor-optimizer-large}); at
\(\alpha = 2\) the infimum closes in an exact square-root law, with a
sandwiched linear tilt of \(\sigma\) as worst case
(Corollary~\ref{cor-alpha-two-exact}); at \(\alpha = \infty\) the value
is a semidefinite program whose optimizer is a Holevo--Helstrom test, a
quantum conditional value-at-risk (Theorem~\ref{thm-quantum-cvar}).

The dual formula has a second reading, which is the main structural
result of the paper. Theorem~\ref{thm-spectral-rep}
(\emph{classicalization}) states that for every
\(\alpha \in [\tfrac12, \infty]\) there is a probability measure
\(\omega_\alpha\) on the spectral interval \([\lambda_{\min}(L),
\lambda_{\max}(L)]\), depending on \((\sigma, L, \alpha)\), such that \[
V_\alpha(q) \;=\; V_\alpha^{\mathrm{cl}}(q;\,\omega_\alpha)
\qquad \text{for all } q \geq 0 :
\] the quantum DRO \emph{is} the classical DRO of a scalar random
variable with law \(\omega_\alpha\). The measure is characterized by one
transform per regime --- the stop-loss transform of order \(\alpha'\)
for \(\alpha > 1\), the Laplace transform at \(\alpha = 1\), the
Stieltjes transform of order \(p\) for \(\alpha < 1\) --- each evaluated
exactly by the sandwiched trace of the dual. At \(\alpha = 1\) this is
Stahl's theorem \citep{stahl2013}, the former Bessis--Moussa--Villani
conjecture \citep{bessis1975}; for \(\alpha > 1\) it rests on
Heinävaara's tracial joint spectral measure
\citep{heinavaara2024tracial} for the pencil \((\sigma^\beta
L\sigma^\beta, \sigma^{2\beta})\), and for \(\alpha < 1\) on a Stieltjes
representation derived from it in the companion note
\citep{reybellet2026stieltjes}; at \(\alpha = \tfrac12\) the measure is
the Born measure of \(L\) in the state \(\sigma\), and at
\(\alpha = \infty\) the layer-cake measure recalled in the related work
below. The measures consist of atoms at the eigenvalues of \(L\) and an
absolutely continuous part that vanishes if and only if
\([\sigma, L] = 0\) or \(\alpha = \tfrac12\), with a density given by an
integral of Heinävaara's density
(Proposition~\ref{prp-explicit-density}).

At small budget the value linearizes, \(V_\alpha(q) =
\operatorname{tr}(L\sigma) + s_\alpha\sqrt{2q} + o(\sqrt q)\), and the
\emph{DRO variance} \(s_\alpha^2\) has two forms
(Theorem~\ref{thm-dro-variance}): intrinsically it is the inverse
Hessian \(\langle \hat L, \mathcal{H}_\alpha^{-1}\hat L\rangle\) of
\(H_\alpha(\cdot\|\sigma)\) at \(\sigma\), a generalized variance in the
sense of Petz dual to the sandwiched Rényi metric of Takahashi and
Fujiwara, and extrinsically it is \(\operatorname{Var}_{\omega_\alpha}
(\ell)/\alpha\), the variance of the classicalizing measure. At matched
curvature this variance decreases from the Born variance at \(\alpha =
\tfrac12\) to that of the conditional-value-at-risk measure at \(\alpha
= \infty\), strictly when \([\sigma, L] \neq 0\)
(Theorem~\ref{thm-variance-ordering}).

Finally, classicalization makes the concentration bounds of classical
probability available for \emph{non-asymptotic} statements, and the
comparison carries one message, a \emph{quantum deflation}:
noncommutativity reduces the worst case. Every classical bound evaluated
on the Born measure \(\nu_L\) of \(L\) in the state \(\sigma\) is a
valid quantum bound, using Golden--Thompson and Araki--Lieb--Thirring
inequalities (Proposition~\ref{prp-born-domination}). The spectral
measure \(\omega_\alpha\) shares mean and support with \(\nu_L\) but has
a strictly smaller variance for every \(\alpha > \tfrac12\) whenever
\([\sigma, L] \neq 0\), so the Bernstein bound built on
\(\omega_\alpha\), sharp over the moment class, lies strictly below the
same bound built on \(\nu_L\) whenever the measured certificate is below
saturation (Theorem~\ref{thm-spectral-bernstein},
Proposition~\ref{prp-deflation}, Figure~\ref{fig-deflation}), and
explicit bounds follow at every order from three anchors, the fidelity
at \(\alpha = \tfrac12\), Bernstein's inequality at \(\alpha = 1\) and
Cauchy--Schwarz at \(\alpha = 2\)
(Proposition~\ref{prp-explicit-bernstein}). Read on the Helstrom
observable, the same bounds are Pinsker-type inequalities: Pinsker, its
optimal form and Fuchs--van de Graaf are recovered with no quantum
penalty, the optimal trace-distance versus \(H_\alpha\) inequality being
the classical one (Remark~\ref{rem-no-penalty}); the deflated tier gives
a Bernstein--Pinsker inequality with the Kubo--Mori variance
(Corollary~\ref{cor-pinsker}); and at \(\alpha = \tfrac12\) the fidelity
robustness bound of Weber et al., read on the Helstrom observable, is
Fuchs--van de Graaf times the cosine of the mismatch between the
Helstrom weight and the extremal weight, with a spectrum-dependent
refinement (Theorem~\ref{thm-fuchs-vdg},
Proposition~\ref{prp-spectrum-fvdg}).

\textbf{Related work.} Guarantees of this form are what quantum state
estimation delivers: tomography and certification protocols return a
reconstructed state with a confidence region in fidelity or trace
distance
\citep{christandl-renner2012, faist-renner2016, flammia-liu2011, haah-etal2017, odonnell-wright2016},
classical shadows predict expectation values from few measurements with
error bars of the same kind \citep{huang-kueng-preskill2020}, and the
robust value converts such a guarantee on the state into a guarantee on
any observable. The same worst-case-over-a-ball structure underlies the
robustness certification of quantum classifiers
\citep{lu-duan-deng2020, weber-etal2021, guan-fang-ying2021}, the
smoothing of one-shot entropies \citep{tomamichel2016} and, at
\(\alpha = 1\), the sensitivity of Gibbs-state expectations to
misspecification of the Hamiltonian, the quantum form of the model-bias
question of \citep{gourgoulias-model-bias2020}. The problem itself has
been met at the endpoints of the family: at \(\alpha = 1\) in quantum
risk-sensitive control, where Yamamoto and Bouten
\citep{yamamoto-bouten2009} and Vladimirov, Petersen and James
\citep{vladimirov-petersen-james2018} bound the worst case over a
relative-entropy ball through the Golden--Thompson inequality, the
quantum counterpart of \citep{petersen-james-dupuis2000}; at
\(\alpha = \tfrac12\), where Weber et al. \citep[Thms. 1 and
2]{weber-etal2022} bound the expectation of an observable over a
fidelity ball by its mean and range, or its mean and variance, in the
reference state, the mean-only bound being implicit, for pure states, in
\citep{weber-etal2021}; and at \(\alpha = \infty\), where for
\(L \geq 0\) the value is, up to normalization, the quantum relative
Lorenz curve of Buscemi and Gour \citep{buscemi-gour2017}, represented
by the layer-cake measure of Gour \citep{gour2026} and of Liu, Hirche
and Cheng \citep{liu-hirche-cheng2025}, which Schmitt
\citep{schmitt2026} obtains independently and concurrently as a
push-forward of Heinävaara's measure. What the paper adds at each
endpoint is stated at the corresponding results
(Proposition~\ref{prp-born-domination},
Theorem~\ref{thm-spectral-hoeffding}, Theorem~\ref{thm-spectral-rep}).
Neighbouring lines of work are the thermodynamic uncertainty relations
for Petz \(f\)-divergences of \citep{salazar2026quantumf},
mean--variance inequalities for a different family, and the use of CVaR
\citep{barkoutsos2020cvar}, of the optimized certainty equivalent
\citep{zhu-simeone2026} and of the tilted loss \citep{qiu-etal2026} in
variational and learning settings, for the classical outcome of a
measurement.

The paper is organized as follows. Section~\ref{sec-framework} sets up
the DRO problem and its duality for an abstract divergence.
Section~\ref{sec-main-theorem} computes the dual for the sandwiched
Rényi family. Section~\ref{sec-spectral-measure} proves the
classicalization theorem and constructs the measures \(\omega_\alpha\).
Section~\ref{sec-linearization} treats the linearization and the quantum
Fisher metrics, Section~\ref{sec-concentration} the concentration,
comparison and state-discrimination bounds. Section~\ref{sec-discussion}
collects open problems, and \hyperref[sec-app-moment-tools]{Appendix A}
the classical proofs.

\section*{Notation}\label{sec-notation}
\addcontentsline{toc}{section}{Notation}

\(M_n^{\mathrm{sa}}\): Hermitian \(n \times n\) matrices.
\(\mathcal{S}_n \subset M_n^{\mathrm{sa}}\): density matrices
(trace-one, positive semidefinite). \(\sigma \in \mathcal{S}_n\),
\(\sigma > 0\): the fixed reference state, eigenvalues
\(p_1,\ldots,p_n\), \(\rho\): a generic state.
\(L = L^* \in M_n^{\mathrm{sa}}\): the observable, eigenvalues
\(l_1 \le \cdots \le l_n\). \(\mu := \operatorname{tr}(L\sigma)\),
\(\hat L := L - \mu I\), \(I\) is the identity;
\(\langle A, B\rangle = \operatorname{tr}(AB)\).

\(F(t), G(s), R(x), V(q)\): the abstract CGF, free energy, rate
function, and value function (Section~\ref{sec-framework}).
\(q_{\max}\): the saturation budget, \(V(q) = \lambda_{\max}(L)\) iff
\(q \ge q_{\max}\) (\ref{eq-qmax-def}).

\textbf{Rényi data} (\(\alpha \in [\tfrac12,\infty]\),
Section~\ref{sec-main-theorem}): \(\alpha' = \alpha/(\alpha-1)\);
\(\beta = (\alpha-1)/(2\alpha)\) for \(\alpha > 1\),
\(\gamma := -\beta\) for \(\alpha < 1\) (\ref{eq-exponent-conventions});
\(p := -\alpha' = \alpha/(1-\alpha) \ge 1\), the Stieltjes order on
\(\alpha < 1\). \(Q_\alpha\): sandwiched Rényi power
(\ref{eq-Q-alpha-def}). \(H_\alpha = (Q_\alpha - 1)/(\alpha-1)\): the
divergence dualized throughout
(Definition~\ref{def-sandwiched-hellinger});
\(D_\alpha = \log(Q_\alpha)/(\alpha-1)\): its logarithmic form
(\ref{eq-budget-dictionary}); \(q\): the budget,
\(\kappa_\alpha(q) := (1+(\alpha-1)q)^{1/\alpha}\).
\(\mathrm{Fid}(\rho,\sigma) := \operatorname{tr}|\sigma^{1/2}\rho^{1/2}|\):
root fidelity, \(H_{1/2} = 2(1-\mathrm{Fid})\).
\(T(\rho,\sigma) := \tfrac12\|\rho-\sigma\|_1\): trace distance.

\textbf{Spectral measures} (Section~\ref{sec-spectral-measure}):
\(\omega_\alpha\): the quantum spectral measure,
\(V_\alpha(q) = V_\alpha^{\rm cl}(q;\omega_\alpha)\). \(\omega_1\): the
Stahl/BMV measure.
\(\nu_L = \sum_i\langle i|\sigma|i\rangle\,\delta_{l_i} = \omega_{1/2}\):
the Born measure. \(\omega_\infty\): the CVaR endpoint measure.
\(V^{\rm cl}_\alpha(q;\nu)\): the classical DRO of a law \(\nu\)
(Lemma~\ref{lem-classical-dual});
\(V^{\rm meas}_\alpha := V^{\rm cl}_\alpha(\cdot;\nu_L)\): the measured
DRO.

\(\operatorname{Var}_\sigma(L) := \operatorname{tr}(\sigma\hat L^2)\):
the Born variance.

\textbf{Variance and metric} (Section~\ref{sec-linearization}):
\(s_\alpha^2(\hat L) = \langle \hat L,\mathcal{H}_\alpha^{-1}\hat L\rangle = \operatorname{Var}_{\omega_\alpha}(\ell)/\alpha\):
the DRO variance for \(\mathcal{D} = H_\alpha\)
(Corollary~\ref{cor-spectral-variance}), specializing the curvature
\(s_{\mathcal{D}}^2\) of Proposition~\ref{prp-square-root}.
\(\mathcal{H}_\alpha\): the Hessian of \(H_\alpha(\cdot\|\sigma)\) at
\(\sigma\), with kernel \(h_\alpha(p_i,p_j)\) (\ref{eq-fisher-kernel}).
Symbols used within a single section --- the tracial measure
\(\mu_{A,B}\) and its density (Section~\ref{sec-spectral-measure}), the
mean functions \(m_\alpha\) and representing functions
\(\mathfrak{f}_\alpha\) (Section~\ref{sec-linearization}), the data
\((a, b, \Delta, \theta)\), transforms, profiles and certificates of
Section~\ref{sec-concentration} --- are defined where they are used.

\section{The DRO framework}\label{sec-framework}

This section isolates the convex-duality skeleton common to all
divergences --- strong duality, the constraint--objective exchange, the
four functions \(F\), \(G\), \(R\), \(V\) and their interrelations ---
for any divergence \(\mathcal{D}\) satisfying (D1)--(D3) below. The
formalism applies equally to classical and quantum state spaces, with
the trace pairing replacing integration. Nothing in this section is new:
the results are standard for divergence balls of probability measures
\citep{ben-tal-etal2013, shapiro2017, rahimian-mehrotra2022, kuhn2025dro},
and references are given at the statements.

\subsection{The primal problem}\label{sec-primal}

Let \(M_n^{\text{sa}}\) be the Hermitian \(n \times n\) matrices, let
\(\mathcal{S}_n = \{\rho \in M_n^{\text{sa}} : \rho \ge 0,\,
\operatorname{tr}\rho = 1\}\) be the set of states (density matrices) on
\(\mathbb{C}^n\), let \(\sigma \in \mathcal{S}_n\) with \(\sigma > 0\)
be a faithful nominal state, and let
\(\mathcal{D}(\cdot\|\sigma) : \mathcal{S}_n \to [0, +\infty]\) be a
divergence satisfying:

(D1) \textbf{Non-negativity.} \(\mathcal{D}(\rho\|\sigma) \ge 0\) for
all \(\rho \in \mathcal{S}_n\).

(D2) \textbf{Faithfulness.} \(\mathcal{D}(\rho\|\sigma) = 0\) if and
only if \(\rho = \sigma\).

(D3) \textbf{Convexity and lower-semicontinuity.}
\(\rho \mapsto \mathcal{D}(\rho\|\sigma)\) is convex and lower
semicontinuous on \(\mathcal{S}_n\).

\begin{remark}
(D1) and (D3) suffice for the strong duality of
Proposition~\ref{prp-dro-duality}; (D2) is used only for the
normalization \(V(0) = \operatorname{tr}(L\sigma)\). All divergences in
this paper satisfy (D1)--(D3) and are finite on \(\mathcal{S}_n\)
because \(\sigma > 0\); sublevel sets
\(\{\mathcal{D}(\cdot\|\sigma) \le q\}\) are compact, so the extrema
over them below are attained.
\end{remark}

\begin{definition}[Distribution model
risk]\protect\hypertarget{def-distributionmodelrisk}{}\label{def-distributionmodelrisk}

For a Hermitian observable \(L \in M_n^{\text{sa}}\) and an uncertainty
level \(q \ge 0\), the \textbf{distribution model risk} is
\begin{equation}\phantomsection\label{eq-dro-primal}{
V(q;L) = \sup\bigl\{\operatorname{tr}(L\,\rho) :
  \rho \in \mathcal{S}_n,\;
  \mathcal{D}(\rho\|\sigma) \le q\bigr\}.
}\end{equation}

\end{definition}

This asks: what is the worst-case expected value of the observable \(L\)
over all states within a \(\mathcal{D}\)-ball of radius \(q\) around
\(\sigma\)? The name follows Breuer and Csiszár
\citep{breuer-csiszar2013, breuer-csiszar2016}, who define distribution
model risk classically as exactly this quantity; the same object is the
robust risk measurement of Glasserman and Xu \citep{glasserman-xu2014}
in finance and, for the relative entropy, the model-misspecification
problem of Petersen, James and Dupuis \citep{petersen-james-dupuis2000}
and of Hansen and Sargent \citep{hansen-sargent2008}.

\begin{refremark}
The infimum
\(U(q;L) = \inf\{\operatorname{tr}(L\rho) : \rho \in \mathcal{S}_n,\, \mathcal{D}(\rho\|\sigma) \le q\}\)
satisfies \(U(q;L) = -V(q;-L)\), so it suffices to study \(V\).
Translation equivariance \(V(q;L+cI) = V(q;L) + c\) allows us to work
with the \textbf{centered observable}
\(\hat{L} = L - \operatorname{tr}(L\sigma)\,I\), for which
\(V(q;\hat L) = \sup\{\operatorname{tr}(\hat L(\rho-\sigma)) : \mathcal{D}(\rho\|\sigma) \le q\}\)
is the maximal deviation from the baseline expectation.

\label{rem-inf}

\end{refremark}

\subsection{Strong duality}\label{sec-strong-duality}

Fix an observable \(L \in M_n^{\text{sa}}\) and consider the
\textbf{value function} \(V(q)=V(q;L)\) as a function of the uncertainty
level \(q\). The following result is the standard duality for divergence
balls \citep{ben-tal-etal2013, shapiro2017, breuer-csiszar2016}, whose
penalty form goes back to \citep{bental-benisrael-teboulle1991}, stated
for the trace pairing: strong duality reduces the computation of the
value function --- a priori an optimization over \(\mathcal{S}_n\) ---
to a one-dimensional optimization.

\begin{proposition}[Properties of the value function and strong
duality]\protect\hypertarget{prp-dro-duality}{}\label{prp-dro-duality}

For a fixed observable \(L \in M_n^{\text{sa}}\), the value function
\(V(q)=V(q;L)\) satisfies the following properties.

\begin{enumerate}
\def\labelenumi{(\roman{enumi})}
\item
  \textbf{Monotonicity.} \(V(q)\) is nondecreasing on \([0,\infty)\).
\item
  \textbf{Concavity.} \(V(q)\) is concave on \([0,\infty)\).
\item
  \textbf{Bounds.}
  \(\operatorname{tr}(L\sigma) \le V(q) \le \lambda_{\max}(L)\) for all
  \(q \ge 0\).
\item
  \textbf{Strong duality.} We have
\end{enumerate}

\begin{equation}\phantomsection\label{eq-strong-dual}{
V(q) = \inf_{s \ge 0}\bigl\{s\, q +
  G(s)\bigr\},
}\end{equation}

where the \textbf{free energy} \(G\) is \[
G(s) = \sup_{\rho \in \mathcal{S}_n}
\{\operatorname{tr}(L\rho) -
s\,\mathcal{D}(\rho\|\sigma)\}\,,
\] the supremum being over the effective domain
\(\{\mathcal{D}(\cdot\|\sigma) < \infty\}\). \(G\) is convex and
nonincreasing, with
\(G(0) = \sup\{\operatorname{tr}(L\rho) : \mathcal{D}(\rho\|\sigma) < \infty\}\),
which equals \(\lambda_{\max}(L)\) whenever
\(\mathcal{D}(\cdot\|\sigma)\) is finite on the top eigenspace of \(L\).

\end{proposition}

\begin{proof}
(i) Larger \(q\) gives a larger feasible set. (ii) If \(\rho_i\) is
optimal for \(V(q_i)\) and \(\lambda \in [0,1]\), convexity of
\(\mathcal{D}(\cdot\|\sigma)\) makes
\(\lambda\rho_1 + (1-\lambda)\rho_2\) feasible for
\(V(\lambda q_1 + (1-\lambda)q_2)\), with objective
\(\lambda V(q_1) + (1-\lambda)V(q_2)\). (iii) \(\rho = \sigma\) is
feasible for every \(q \ge 0\) by (D2), and
\(\operatorname{tr}(L\rho) \le \lambda_{\max}(L)\) for every state.

(iv) Define \(g(q) := -V(q)\) for \(q \ge 0\) and \(g := +\infty\) on
\((-\infty,0)\): proper by (iii), convex by (ii), and lower
semicontinuous because \(V\) is continuous on \([0,\infty)\) --- on
\((0,\infty)\) as a finite concave function, and at \(0\) because if
\(q_k \downarrow 0\) and \(\rho_k\) is optimal for \(V(q_k)\), every
limit point \(\rho\) satisfies
\(\mathcal{D}(\rho\|\sigma) \le \liminf_k \mathcal{D}(\rho_k\|\sigma) = 0\),
so \(\rho\) is feasible at \(q = 0\) and
\(\lim_k V(q_k) = \operatorname{tr}(L\rho) \le V(0) \le \lim_k V(q_k)\)
by (i). Hence \(g = g^{**}\) (Fenchel--Moreau). The conjugate
\(g^*(\mu) = \sup_{q \ge 0}\{\mu q + V(q)\}\) is \(+\infty\) for
\(\mu > 0\) by (iii); for \(\mu = -s < 0\), writing \(V(q)\) as a
supremum over \(\{\mathcal{D}(\rho\|\sigma) \le q\}\) and exchanging the
two suprema, the decreasing map \(q \mapsto -sq\) is maximized at
\(q = \mathcal{D}(\rho\|\sigma)\), so
\(g^*(-s) = \sup_\rho\{\operatorname{tr}(L\rho) - s\mathcal{D}(\rho\|\sigma)\} = G(s)\);
and \(g^*(0) = \sup_q V(q) = G(0)\), a state being feasible for some
finite \(q\) iff its divergence is finite. Then
\(V(q) = -g^{**}(q) = \inf_{\mu \le 0}\{g^*(\mu) - \mu q\} = \inf_{s \ge 0}\{sq + G(s)\}\).
\(\square\)
\end{proof}

\subsection{The rate function}\label{sec-duality-structure}

The value function \(V(q)\) answers the question ``given a divergence
budget \(q\), how large can \(\operatorname{tr}(\hat{L}\,\rho)\) be?''
We introduce next the \textbf{rate function} which asks the reverse:
given a target expectation \(x\), what is the minimal divergence cost to
achieve it?

\begin{definition}[Rate
function]\protect\hypertarget{def-ratefunction}{}\label{def-ratefunction}

Given \(L \in M_n^{\text{sa}}\), \(x \in \mathbb{R}\), and a baseline
model \(\sigma\), the rate function is
\begin{equation}\phantomsection\label{eq-rate-function}{
R(x, L) := \inf\bigl\{\mathcal{D}(\rho\|\sigma) :
  \rho \in \mathcal{S}_n,\;
  \operatorname{tr}(L\,\rho) \ge x\bigr\}.
}\end{equation}

\end{definition}

The names follow the large-deviation analogy --- for the classical
relative entropy \(R\) is the rate function of Cramér's theorem ---
whose convexity structure survives even where the probabilistic
interpretation does not. Alongside the \textbf{free energy} \(G\) of
Proposition~\ref{prp-dro-duality} we introduce the \textbf{abstract
cumulant generating function}
\begin{equation}\phantomsection\label{eq-abstract-cgf}{
F(t) := \sup_{\rho \in \mathcal{S}_n}
\{ t \operatorname{tr}(L\rho) -\,\mathcal{D}(\rho\|\sigma)\}, \qquad t \in \mathbb{R},
}\end{equation} so that \(G\) is the perspective of \(F\):
\(G(s) = s\,F(1/s)\) for \(s > 0\), equivalently \(F(t) = t\,G(1/t)\).
\(F\) is convex, \(F(0) = 0\), \(F(t) \ge t\operatorname{tr}(L\sigma)\),
and \(F(t;L) = F(-t;-L)\). The following proposition connects the rate
function to the value function and identifies \(R\) with the
Fenchel--Legendre transform \(F^*(x) = \sup_{t \ge 0}\{xt - F(t)\}\) of
\(F\). Throughout, \(L\) is fixed and \(R(x) = R(x,L)\). Part (iv) and
the perspective formula (\ref{eq-DRO-perspective}) in its proof are the
standard inversion of a Legendre transform
\citep{boucheron-lugosi-massart2013}; for the relative entropy see
\citep{ahmadi-javid2012, breuer-csiszar2013, chowdhary-dupuis2013}, for
Rényi divergences \citep{atar-chowdhary-dupuis2015}, and for the same
inversion as a goal-oriented uncertainty bound
\citep{dupuis-katsoulakis-pantazis-plechac2016}.

\begin{proposition}[Constraint-objective
exchange]\protect\hypertarget{prp-exchange}{}\label{prp-exchange}

For all \(q \ge 0\) and \(x \in \mathbb{R}\),
\begin{equation}\phantomsection\label{eq-exchange}{
V(q) \ge x \quad\Longleftrightarrow\quad R(x) \le q.
}\end{equation} Consequently, \(V\) and \(R\) are generalized inverses:
\[
V(q) = \sup\{x \in \mathbb{R} : R(x) \le q\} \quad  \text{ and } \quad R(x) = \inf\{q \ge 0 : V(q) \ge x\}\,.
\] Furthermore:

\begin{enumerate}
\def\labelenumi{(\roman{enumi})}
\item
  \textbf{Monotonicity.} \(R\) is nondecreasing on
  \([\operatorname{tr}(L \sigma),\infty)\).
\item
  \textbf{Convexity.} \(R\) is convex on
  \([\operatorname{tr}(L \sigma),\infty)\).
\item
  \textbf{Bounds.} \(R(\operatorname{tr}(L \sigma)) = 0\) and
  \(R(x) = \infty\) for \(x > \lambda_{\max}(L)\).
\item
  \textbf{Strong duality.} \(R(x) = F^*(x)\) for all
  \(x \ge \operatorname{tr}(L\sigma)\).
\end{enumerate}

\end{proposition}

\begin{proof}
\emph{The exchange.} Both extrema are attained (the feasible sets are
compact and the objectives are continuous, resp. lower semicontinuous),
so \(V(q) \ge x\) and \(R(x) \le q\) each say that some
\(\rho \in \mathcal{S}_n\) satisfies both
\(\mathcal{D}(\rho\|\sigma) \le q\) and
\(\operatorname{tr}(L \rho) \ge x\). This proves (\ref{eq-exchange}) and
the generalized-inverse relation.

\emph{Properties (i)--(iii).} Through the exchange, monotonicity of
\(R\) is monotonicity of \(V\), and convexity of \(R\) is concavity of
\(V\): if \(R(x_i) \le q_i\) then
\(V(\lambda q_1 + (1-\lambda)q_2) \ge \lambda V(q_1) + (1-\lambda)V(q_2) \ge \lambda x_1 + (1-\lambda)x_2\).
The bounds follow from (D2) and
\(\operatorname{tr}(L\rho) \le \lambda_{\max}(L)\).

\emph{Strong duality: \(R = F^*\).} By Proposition~\ref{prp-dro-duality}
and the substitution \(s = 1/t\) (the \(s = 0\) term never wins, \(G\)
being nonincreasing), the perspective formula holds:
\begin{equation}\phantomsection\label{eq-DRO-perspective}{
V(q) = \inf_{s > 0}\left\{s\, q +
  G(s)\right\} = \inf_{t > 0}\left\{ \frac{q + F(t)}{t} \right\}.
}\end{equation} Now, by the exchange (\ref{eq-exchange}), \[
R(x) = \inf\{ q \ge 0 : V(q) \ge x \}.
\] Using (\ref{eq-DRO-perspective}): \(V(q) \ge x\) if and only if
\((q + F(t))/t \ge x\) for all \(t > 0\), i.e., \(q \ge xt - F(t)\) for
all \(t > 0\). Therefore \[
R(x) = \inf\bigl\{ q \ge 0 : q \ge xt - F(t) \;\text{for all}\; t > 0 \bigr\} = \sup_{t > 0}\bigl\{ xt - F(t) \bigr\} = F^*(x),
\] the last equality because the \(t = 0\) term of \(F^*\) is
\(0 \le \sup_{t>0}\{xt - F(t)\}\) when
\(x \ge \operatorname{tr}(L\sigma)\) (let \(t \to 0^+\)). \(\square\)
\end{proof}

\textbf{The saturation budget.} Since
\(\operatorname{tr}(L\rho) \le \lambda_{\max}(L)\) with equality iff
\(\rho\) is supported on the top eigenspace of \(L\), we set
\begin{equation}\phantomsection\label{eq-qmax-def}{
q_{\max} \;:=\; R(\lambda_{\max}(L)) \;=\; \inf\bigl\{\mathcal{D}(\rho\|\sigma) : \rho \in \mathcal{S}_n,\ \operatorname{tr}(L\rho) = \lambda_{\max}(L)\bigr\},
}\end{equation} attained by compactness of the feasible set and lower
semicontinuity of \(\mathcal{D}\), and finite for every divergence in
this paper since \(\sigma > 0\); by (\ref{eq-exchange}),
\(V(q) = \lambda_{\max}(L)\) iff \(q \ge q_{\max}\).

\begin{figure}[htbp]

\centering{

\pandocbounded{\includegraphics[keepaspectratio]{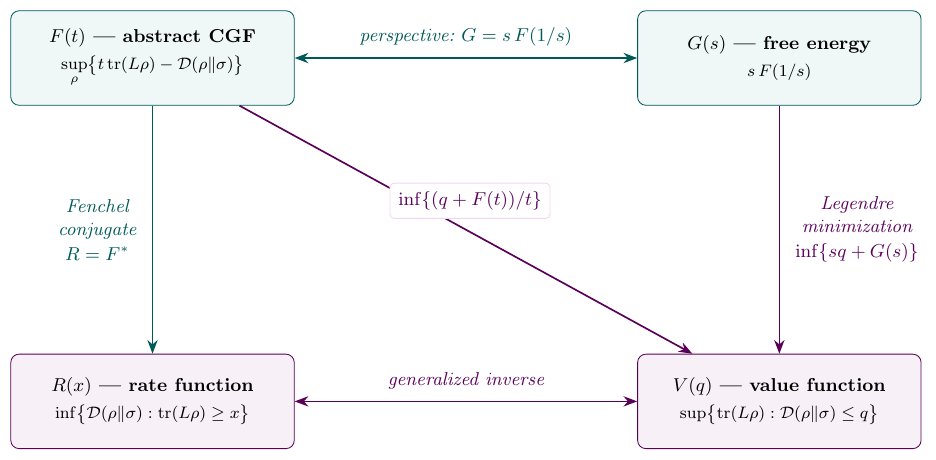}}

}

\caption{\label{fig-dro-square}The DRO square: \(F\) and \(G\) are
related by the perspective \(G(s) = sF(1/s)\), \(R\) and \(V\) are
generalized inverses, and the two paths from \(F\) to \(V\) --- Fenchel
conjugation \(R = F^*\) then inversion, or perspective then
\(V(q) = \inf_{s}\{sq + G(s)\}\) --- commute
(Proposition~\ref{prp-exchange}); the diagonal is
\(V(q) = \inf_{t>0}\{(q + F(t))/t\}\).}

\end{figure}

The four functions form the commutative square of
Figure~\ref{fig-dro-square}, and the entire DRO strategy reduces to a
single question: \textbf{can you compute \(F(t)\) or \(G(s)\)
explicitly?} This is the only step that depends on \(\mathcal{D}\); the
rest --- concavity and strong duality --- is automatic.

\begin{refremark}[Illustration: relative entropy and the BMV bridge]
For the classical relative entropy the Gibbs variational principle gives
\(F(t) = \log\mathbb{E}_Q[e^{tL}]\), the classical cumulant generating
function, and the square is the classical duality between risk-sensitive
integrals and relative entropy
\citep{petersen-james-dupuis2000, hansen-sargent2008, chowdhary-dupuis2013}.
For the Umegaki relative entropy it gives
\(F(t) = \log\operatorname{tr}\,e^{\log\sigma + tL}\), which is
\emph{not} the log-moment generating function
\(\log\operatorname{tr}[\sigma e^{tL}]\) when \([\sigma,L] \neq 0\) ---
yet, by Stahl's theorem \citep{stahl2013}, is the cumulant generating
function of a probability measure \(\omega_1\) on the spectral interval
of \(L\) (Section~\ref{sec-example-alpha-one}): the \textbf{BMV bridge}
that carries classical large-deviation tools to the quantum
\(\alpha = 1\) problem.

\label{rem-classical-KL}

\end{refremark}

\subsection{The rate function picture}\label{sec-rate-function-picture}

Applied to \(\hat L\) and to \(-\hat L\), the perspective formula
(\ref{eq-DRO-perspective}) reads both branches of \(F\): the line
through \((0,-q)\) tangent to the graph of \(F(\cdot;\hat L)\) on the
right has slope \(x_+ = V(q;\hat{L})\), the tangent on the left has
slope \(x_- = -V(q;-\hat{L})\), and
\(\{x : R(x;\hat L) \le q,\ R(-x;-\hat L) \le q\} = [x_-, x_+]\) is the
\textbf{two-sided uncertainty interval}:
\begin{equation}\phantomsection\label{eq-two-sided}{
-V(q;-\hat{L}) \;\le\; \operatorname{tr}\bigl(\hat{L}(\rho - \sigma)\bigr) \;\le\; V(q;\hat{L})
\qquad \text{for all } \rho \text{ with } \mathcal{D}(\rho\|\sigma) \le q,
}\end{equation} asymmetric in general.

\begin{figure}[htbp]

\centering{

\pandocbounded{\includegraphics[keepaspectratio]{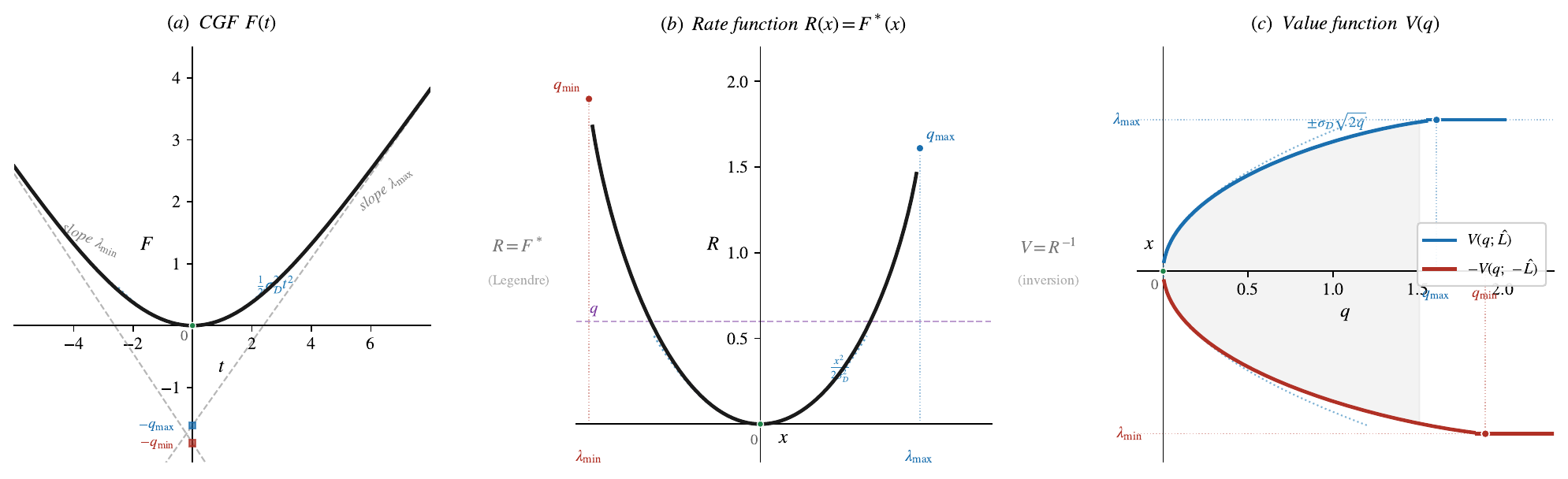}}

}

\caption{\label{fig-triptych}The convex duality chain. \emph{(a)}
\(F(t)\), convex with \(F(0) = F'(0) = 0\), linear asymptotes of slopes
\(\lambda_{\min}(\hat L)\), \(\lambda_{\max}(\hat L)\) and tangent
parabola \(\tfrac12 s_{\mathcal{D}}^2t^2\) at the origin. \emph{(b)}
\(R = F^*\) on \([\lambda_{\min},\lambda_{\max}]\), with
\(R(\lambda_{\max}(\hat L)) = q_{\max}\). \emph{(c)} The two branches
\(V(q;\hat L)\) and \(-V(q;-\hat L)\), diverging as
\(\pm s_{\mathcal{D}}\sqrt{2q}\) and saturating at \(\lambda_{\max}\),
\(\lambda_{\min}\); the shaded region is (\ref{eq-two-sided}).}

\end{figure}

The full convex duality chain \(F \to R \to V\) is illustrated in
Figure~\ref{fig-triptych}.

Read as the inverse of \(R\), \(V\) has three regimes: a
\textbf{square-root onset} as \(q \to 0^+\), governed by the curvature
of \(R\) at the origin; a \textbf{strict increase} on \((0, q_{\max})\),
where the constraint is active; and \textbf{saturation} at
\(\lambda_{\max}(L)\) for \(q \ge q_{\max}\). The onset is the classical
square-root law
\citep{dupuis-katsoulakis-pantazis-plechac2016, lam2016robust, gotoh-kim-lim2018, duchi-namkoong2019},
governed by a single number, here called the \textbf{DRO variance}
\(s_{\mathcal{D}}^2(\hat L)\):

\begin{proposition}[Square-root
law]\protect\hypertarget{prp-square-root}{}\label{prp-square-root}

Suppose that \(F(t) = \tfrac12s_{\mathcal{D}}^2\,t^2 + o(t^2)\) as
\(t \to 0^+\) for some
\(s_{\mathcal{D}} = s_{\mathcal{D}}(\hat L) > 0\). Then \[
V(q;\hat{L}) = s_{\mathcal{D}}(\hat{L})\,\sqrt{2q} + o(\sqrt{q}) \qquad\text{as } q \to 0^+.
\]

\end{proposition}

\begin{proof}
Write \(s := s_{\mathcal{D}}\) and use the perspective formula
\(V(q;\hat L) = \inf_{t>0}(q + F(t))/t\) of (\ref{eq-DRO-perspective})
for the centered observable. \emph{Upper bound.} Taking
\(t = \sqrt{2q}/s\) gives \(V(q;\hat L) \le s\sqrt{2q} + o(\sqrt q)\).
\emph{Lower bound.} Fix \(\varepsilon \in (0,s^2)\) and \(\delta > 0\)
with \(F(t) \ge \tfrac12(s^2 - \varepsilon)t^2\) on \((0,\delta]\). For
\(t \le \delta\),
\((q + F(t))/t \ge \inf_{t>0}\bigl(q + \tfrac12(s^2-\varepsilon)t^2\bigr)/t = \sqrt{2q(s^2-\varepsilon)}\).
For \(t > \delta\): \(F\) is convex with \(F(0) = 0\), so \(F(t)/t\) is
nondecreasing and
\((q + F(t))/t \ge F(\delta)/\delta \ge \tfrac12(s^2 - \varepsilon)\delta\),
which exceeds \(\sqrt{2q(s^2-\varepsilon)}\) once \(q\) is small. Hence
\(V(q;\hat L) \ge \sqrt{2q(s^2 - \varepsilon)}\) for small \(q\), and
\(\varepsilon\) is arbitrary. \(\square\)
\end{proof}

For \(\mathcal{D} = H_\alpha\), \(\alpha \in [1/2,\infty)\), the
hypothesis is verified in Section~\ref{sec-linearization} for every
\(L\) that is not a multiple of the identity, with
\(s_{\mathcal{D}}^2(\hat L) = \langle \hat L, \mathcal{H}_\alpha^{-1}\hat L\rangle\)
the inverse Hessian of \(H_\alpha(\cdot\|\sigma)\) at \(\sigma\)
(Theorem~\ref{thm-hessian-bridge},
Corollary~\ref{cor-dro-variance-sandwiched}).

\begin{refremark}[Connection to risk measures]
For fixed \(q \ge 0\), the map \(L \mapsto V(q;L)\) is a
\textbf{coherent risk measure} in the sense of Artzner, Delbaen, Eber,
and Heath \citep{artzner1999} --- monotone, translation equivariant,
positively homogeneous and subadditive, all immediate from the supremum
form; at \(\alpha = 1\) it is the quantum counterpart of the entropic
value-at-risk \citep{ahmadi-javid2012}, with the Rényi extension of
\citep{pichler-schlotter2020} at the other orders, at
\(\alpha = \infty\) of the conditional value-at-risk of Rockafellar and
Uryasev \citep{rockafellar-uryasev2000}
(Section~\ref{sec-alpha-infty-dro}), and in the dual form of
Theorem~\ref{thm-main} an optimized certainty equivalent
\citep{ben-tal-teboulle2007}.

\label{rem-risk-measures}

\end{refremark}

\section{\texorpdfstring{DRO for
\(\alpha\)-divergences}{DRO for \textbackslash alpha-divergences}}\label{sec-main-theorem}

\subsection{Definitions and notations}\label{definitions-and-notations}

This section computes \(V_\alpha(q)\) on the whole range
\(\alpha \in [\tfrac12, \infty]\). The answer (Theorem~\ref{thm-main})
is a one-dimensional dual, the quantum form of the optimized certainty
equivalent of Ben-Tal and Teboulle \citep{ben-tal-teboulle2007}, with
the worst-case state explicit in terms of the dual minimizer
(Corollary~\ref{cor-optimizer-large}); \(\alpha = 1\) is the Gibbs
variational principle and \(\alpha = \infty\) a quantum conditional
value-at-risk. Everything rests on convexity of
\(H_\alpha(\cdot\|\sigma)\) in \(\rho\), so the duality of
Section~\ref{sec-framework} applies verbatim.

Quantum refinements of the classical Rényi divergence
\citep{renyi1961, vanerven-harremoes2014} are not unique. Alongside the
sandwiched family, Petz's quasi-entropies
\citep{petz1986, ohya-petz2004}, the \(\alpha\)-\(z\) divergences
\citep{jaksic-ogata-pautrat-pillet2012, audenaert-datta2015}, and the
measured and maximal \(f\)-divergences
\citep{hiai2011, hiai-mosonyi2017} all agree with it on the commuting
locus but differ elsewhere; the measured Rényi divergences, the largest
classical divergence obtainable by a measurement, have
Donsker--Varadhan-type variational formulas of their own
\citep{berta2017}. The sandwiched family is monotone under quantum
channels for \(\alpha
\geq \tfrac12\) \citep{frank-lieb2013} (for \(\alpha > 1\) also
\citep{beigi2013}) and not below \citep{berta2017}, and monotone in
\(\alpha\) \citep{muller-lennert2013}; its \(\alpha = 1\) member is the
Umegaki relative entropy, monotone under channels by
\citep{lindblad1975, uhlmann1977}, and its \(\alpha = \infty\) endpoint
the max-relative entropy \citep{datta2009}.

Throughout this section, \(\sigma > 0\) denotes a faithful state on
\(\mathbb{C}^n\), \(L = L^*\) a Hermitian observable, and
\(\alpha \in [\tfrac12,\infty]\setminus\{1\}\) carries the conjugate
exponent and sandwiching data
\begin{equation}\phantomsection\label{eq-exponent-conventions}{
\alpha' = \frac{\alpha}{\alpha-1},
\qquad
\beta = \frac{\alpha-1}{2\alpha} = \frac{1}{2\alpha'},
\qquad
\gamma := -\beta = \frac{1-\alpha}{2\alpha},
\qquad
p := -\alpha' \big|_{\alpha<1} = \frac{\alpha}{1-\alpha} \ge 1 .
}\end{equation} So \(\beta>0\) for \(\alpha>1\) and \(\gamma>0\) for
\(\alpha \in [\tfrac12,1)\), where \(\alpha' = -p\): \(\beta\) is used
on the first range and \(\gamma = -\beta\) on the second, so that the
sandwiching exponents \(\sigma^\beta\), \(\sigma^\gamma\) read as
positive powers. The \textbf{sandwiched Rényi power} is given for
\(\alpha \ne 1\) by
\begin{equation}\phantomsection\label{eq-Q-alpha-def}{
Q_\alpha(\rho\|\sigma) := \operatorname{tr}\!\left[\bigl(\sigma^{\frac{1-\alpha}{2\alpha}}\,\rho\,\sigma^{\frac{1-\alpha}{2\alpha}}\bigr)^{\alpha}\right]
= \|\sigma^{\frac{1-\alpha}{2\alpha}}\,\rho\,\sigma^{\frac{1-\alpha}{2\alpha}}\|_\alpha^\alpha.
}\end{equation}

The \textbf{sandwiched \(\alpha\)-divergence} is then defined by

\begin{definition}[Sandwiched
\(\alpha\)-divergence]\protect\hypertarget{def-sandwiched-hellinger}{}\label{def-sandwiched-hellinger}

\[
H_\alpha(\rho\|\sigma)
  :=
  \left\{
  \begin{array}{cl}
  \frac{Q_\alpha(\rho\|\sigma) - 1}{\alpha - 1} & \alpha \in [\frac{1}{2},1) \cup (1,\infty) \\
  \operatorname{tr}(\rho(\log\rho - \log\sigma)) & \alpha =1 \\
  \|\sigma^{-1/2}\rho\,\sigma^{-1/2}\|_\infty -1 & \alpha =\infty
  \end{array}
  \right.  
\]

\end{definition}

Only convexity of \(H_\alpha(\cdot\|\sigma)\) in \(\rho\) is used below;
at \(\alpha = \infty\) it is the only convexity available.

The \textbf{DRO primal} with \(H_\alpha\)-budget \(q \ge 0\) is
\begin{equation}\phantomsection\label{eq-renyi-dro-primal}{
V_\alpha(q) = \sup\bigl\{\operatorname{tr}(L\rho) :
  \rho \in \mathcal{S}_n,\;
  H_\alpha(\rho\|\sigma) \le q\bigr\}.
}\end{equation} The logarithmic sandwiched Rényi divergence is the
monotone reparametrization
\(D_\alpha = \tfrac{1}{\alpha-1}\log Q_\alpha = \tfrac{1}{\alpha-1}\log\bigl(1 + (\alpha-1)H_\alpha\bigr)\)
(with \(D_1 = H_1\) and \(D_\infty = \log(1 + H_\infty)\)), so the
\(H_\alpha\)-ball of radius \(q\) \emph{is} the \(D_\alpha\)-ball of
radius \begin{equation}\phantomsection\label{eq-budget-dictionary}{
r = \frac{\log\bigl(1 + (\alpha-1)q\bigr)}{\alpha-1},
\qquad\text{equivalently}\qquad
q = \frac{e^{(\alpha-1)r} - 1}{\alpha-1}
}\end{equation} (\(r = q\) at \(\alpha = 1\), \(r = \log(1+q)\) at
\(\alpha = \infty\)), and at each fixed order every result below
transfers verbatim to the logarithmic convention through this
dictionary. We dualize \(H_\alpha\) rather than \(D_\alpha\) for a
convexity reason: \(H_\alpha(\cdot\|\sigma)\) is convex (axiom (D3) of
Section~\ref{sec-framework}), which is what the DRO duality rests on,
whereas \(D_\alpha\) is jointly convex only for
\(\alpha \in [\tfrac12, 1]\) (as in the classical case
\citep{vanerven-harremoes2014}) and for \(\alpha > 1\), including
\(\alpha = \infty\), is not even convex in \(\rho\) --- already for
commuting pairs.

\begin{refremark}[Monotonicity in \(\alpha\) and the limit
\(\alpha \to \infty\)]
For every pair, \(\alpha \mapsto D_\alpha(\rho\|\sigma)\) is
nondecreasing on \([\tfrac12,\infty)\) \citetext{\citealp[Thm.
7]{muller-lennert2013}; \citealp{beigi2013}}, and
\(x \mapsto (e^{(\alpha-1)x}-1)/(\alpha-1)\) is increasing in both
\(\alpha\) and \(x\); hence \(\alpha \mapsto H_\alpha(\rho\|\sigma)\) is
nondecreasing as well, the \(H_\alpha\)-balls of a fixed radius are
nested downward in \(\alpha\), and \[
\alpha \mapsto V_\alpha(q) \ \text{ is nonincreasing on } [\tfrac12,\infty) \text{ for every } q \geq 0
\] (visible in Figure~\ref{fig-risk-family}). The same holds classically
for \(H_\alpha^{\rm cl}\) \citep[Thm. 3]{vanerven-harremoes2014}, which
is how it is used in Proposition~\ref{prp-explicit-bernstein}.

The limit \(\alpha \to \infty\), however, belongs to the Rényi scale. At
fixed \(q\) the radius \(r\) of (\ref{eq-budget-dictionary}) tends to
\(0\), so the balls shrink to \(\{\sigma\}\) and
\(V_\alpha(q) \downarrow \operatorname{tr}(L\sigma)\). At fixed \(r\)
the balls \(\{D_\alpha \le r\}\) decrease to
\(\{D_\infty \le r\} = \{H_\infty \le e^r - 1\}\), and the values
\(V_\alpha(q)\) at \(q = (e^{(\alpha-1)r}-1)/(\alpha-1)\) decrease to
\(V_\infty(e^r - 1)\). Thus \(H_\infty\) in
Definition~\ref{def-sandwiched-hellinger} is the endpoint of the Rényi
scale, not a limit of the \(H_\alpha\).

\label{rem-alpha-monotone}

\end{refremark}

\begin{refremark}[Minimality of the sandwiched family]
Any quantum Rényi divergence \(D_\alpha\) that is additive under tensor
products, monotone under quantum channels and equal to the classical
Rényi divergence on commuting pairs dominates the sandwiched one:
pinching \(\rho^{\otimes n}\) in the eigenbasis of
\(\sigma^{\otimes n}\) gives
\(D_\alpha(\rho\|\sigma) \geq \tfrac1n D_\alpha\bigl(\mathcal{P}_{\sigma^{\otimes n}}(\rho^{\otimes n})\,\big\|\,\sigma^{\otimes n}\bigr)\),
a classical quantity whose limit as \(n \to \infty\) is the sandwiched
divergence \citep[§4.2.2 and Prop.~4.12]{tomamichel2016}; for this
reason Tomamichel calls the latter the \emph{minimal} quantum Rényi
divergence. Since the ball of such a \(D_\alpha\) is contained in the
sandwiched ball of the same radius, in either normalization, its robust
value is at most \(V_\alpha(q)\): the value of Theorem~\ref{thm-main} is
the largest, and every upper bound of Section~\ref{sec-concentration}
remains valid, for every admissible quantum Rényi divergence at the same
order and budget. Petz's family dominates it at every order, by the
Araki--Lieb--Thirring inequality; the measured Rényi divergence, which
lies below the sandwiched one \citep{berta2017}, is not additive and is
not covered.

\label{rem-minimality}

\end{refremark}

\subsection{\texorpdfstring{Relative entropy DRO
(\(\alpha = 1\))}{Relative entropy DRO (\textbackslash alpha = 1)}}\label{sec-umegaki-dro}

At \(\alpha = 1\) the generalized cumulant is the Gibbs free energy
\(\log\operatorname{tr}\,e^{\log\sigma + tL}\), in closed form, whereas
for \(\alpha \neq 1\) it carries a one-dimensional optimization. The
quantum Gibbs variational principle below is a classical result
\citep{petz1988variational, ohya-petz2004, donald1986}; we include a
short, self-contained proof.

\begin{theorem}[Gibbs Variational
Principle]\protect\hypertarget{thm-quantum-dv}{}\label{thm-quantum-dv}

For \(\sigma > 0\) and \(L = L^*\), \[
\sup_{\rho \in \mathcal{S}_n}\bigl\{\operatorname{tr}(L\rho) - H_1(\rho\|\sigma)\bigr\}
= \log \operatorname{tr}\bigl(e^{\log \sigma + L}\bigr),
\] attained uniquely at the Gibbs state
\(\tau := e^{\log\sigma + L}/\operatorname{tr}(e^{\log\sigma + L})\).

\end{theorem}

\begin{proof}
A direct computation gives the key identity \[
H_1(\rho\|\tau) = H_1(\rho\|\sigma) - \operatorname{tr}(L\rho) + \log \operatorname{tr}(e^{\log\sigma + L}),
\] so
\(\operatorname{tr}(L\rho) - H_1(\rho\|\sigma) = \log\operatorname{tr}(e^{\log\sigma+L}) - H_1(\rho\|\tau)\).
Non-negativity and faithfulness of \(H_1\) give the supremum and the
uniqueness of the maximizer \(\rho = \tau\). \(\square\)
\end{proof}

The Gibbs cumulant
\(F_1(t) := \log \operatorname{tr}\bigl(e^{\log\sigma + tL}\bigr)\),
\(t > 0\), identifies a one-parameter family of tilted Gibbs states \[
\sigma_t := \frac{e^{\log\sigma + tL}}{\operatorname{tr}(e^{\log\sigma + tL})},
\qquad F_1'(t) = \operatorname{tr}(L\sigma_t),
\] and the Legendre identity
\(H_1(\sigma_t\|\sigma) = tF_1'(t) - F_1(t)\) follows by the same
computation as in the proof of Theorem~\ref{thm-quantum-dv}.

The next theorem is the quantum counterpart of the classical
relative-entropy DRO
\citep{petersen-james-dupuis2000, ahmadi-javid2012, breuer-csiszar2013}.
In quantum risk-sensitive control, the expectation under a state at
finite relative entropy from a nominal state is bounded through the
Gibbs variational principle and the Golden--Thompson inequality, that is
with \(\log\operatorname{tr}(\sigma e^{tL})\) in place of \(F_1(t)\)
\citep[Lemma 2.1 and eq. (13)]{yamamoto-bouten2009}, and the worst case
over a relative-entropy ball is bounded by the corresponding infimum
over the risk parameter \citep[Thm. 4]{vladimirov-petersen-james2018}.
The theorem gives the exact value, with \(F_1\), and the worst-case
state.

\begin{theorem}[Relative entropy
robustness]\protect\hypertarget{thm-DRO-1}{}\label{thm-DRO-1}

Let \(\sigma > 0\) and let \(L = L^*\) not be a multiple of the
identity. For every \(q \ge 0\),
\begin{equation}\phantomsection\label{eq-dro-1}{
V_1(q) \;=\; \inf_{t > 0}\;\frac{q + F_1(t)}{t}.
}\end{equation} For \(q \in (0, q_{\max})\) (\(q_{\max}\) as in
(\ref{eq-qmax-def})) the infimum is attained at the unique \(t^* > 0\)
satisfying \(H_1(\sigma_{t^*}\|\sigma) = q\), and
\begin{equation}\phantomsection\label{eq-dro-1-primal}{
V_1(q) \;=\; F_1'(t^*) \;=\; \operatorname{tr}(L\,\sigma_{t^*}).
}\end{equation}

\end{theorem}

\begin{proof}
Equation (\ref{eq-dro-1}) is the perspective formula
(\ref{eq-DRO-perspective}), since by Theorem~\ref{thm-quantum-dv}
applied to \(tL\) the abstract CGF of (\ref{eq-abstract-cgf}) is
\(F = F_1\). Setting the derivative in \(t\) to zero, \[
\frac{d}{dt}\frac{q + F_1(t)}{t} = \frac{tF_1'(t) - F_1(t) - q}{t^2},
\] gives the stationarity condition \(tF_1'(t) - F_1(t) = q\), which by
the Legendre identity above is exactly \(H_1(\sigma_t\|\sigma) = q\).
Existence and uniqueness of \(t^*\) follow from strict convexity of
\(F_1\) --- its second derivative \(F_1''(t)\) is the Kubo--Mori
(Bogoliubov) variance of \(L\) in the tilted state \(\sigma_t\), hence
strictly positive since \(L\) is not a multiple of the identity
\citep{petz1993, hiai-petz2014} --- as follows:
\(\tfrac{d}{dt}H_1(\sigma_t\|\sigma) = tF_1''(t) > 0\), so
\(t \mapsto H_1(\sigma_t\|\sigma)\) is continuous and strictly
increasing on \((0,\infty)\), starting from \(0\); and its limit as
\(t \to \infty\) is
\(\lim_t\{\lambda_{\max}(L)\,t - F_1(t)\} = F_1^*(\lambda_{\max}(L)) = q_{\max}\)
by Proposition~\ref{prp-exchange}(iv) and (\ref{eq-qmax-def}). More
explicitly, if \(P_n\) is the spectral projection of \(L\) at
\(l_n=\lambda_{\max}(L)\), finite-dimensional analytic perturbation of
\(L+t^{-1}\log\sigma\) gives \[
F_1(t)=t l_n+\log\operatorname{tr}_{\operatorname{Ran}P_n}\exp(P_n\log\sigma P_n)+O(t^{-1}),
\] and differentiation (or the same perturbation expansion one order
further) yields \(t(l_n-F_1'(t))\to0\). Notice that in the noncommuting
case the spectral subspace itself need only approach
\(\operatorname{Ran}P_n\) at order \(t^{-1}\); exponential convergence
of \(\sigma_t\) to that fixed subspace is not required and is generally
false.

At \(t^*\), (\ref{eq-dro-1}) gives
\(V_1(q) = (q + F_1(t^*))/t^* = t^*F_1'(t^*)/t^* = F_1'(t^*)\), which is
(\ref{eq-dro-1-primal}). \(\square\)
\end{proof}

\subsection{\texorpdfstring{DRO for finite
\(\alpha \ne 1\)}{DRO for finite \textbackslash alpha \textbackslash ne 1}}\label{sec-finite-alpha-dro}

For \(\alpha \in [\tfrac12,\infty) \setminus \{1\}\) the sandwiched
divergence \(H_\alpha\) is convex in its first argument --- for
\(\alpha>1\) because \(Q_\alpha\) itself is convex in \(\rho\), and for
\(\alpha \in [\tfrac12,1)\) because \(Q_\alpha\) is \emph{concave} and
\(H_\alpha = (Q_\alpha-1)/(\alpha-1)\) divides by the negative constant
\(\alpha-1\) --- so the abstract DRO machinery of
Section~\ref{sec-framework} applies throughout. The entire calculation
reduces to computing the generalized cumulant
\(G_\alpha(s) = \sup_\rho\{\operatorname{tr}(L\rho) - sH_\alpha(\rho\|\sigma)\}\),
which Theorem~\ref{thm-variational} below obtains uniformly on both
sides of \(\alpha=1\) --- via the signed variable \(s=1/(\alpha-1)\)
tracking the convexity/concavity of \(Q_\alpha\) --- from the convex
duality of trace functions, a matrix Fenchel--Young inequality; see
\citep{carlen2010} for this circle of ideas. (The formulas below hold
for all \(\alpha \in (0,1)\); the range \(\alpha \ge \tfrac12\) is where
\(H_\alpha\) is monotone under quantum channels and where the
classicalizing measure of Section~\ref{sec-spectral-measure} is
positive.)

Frank and Lieb \citep[Lemma 4]{frank-lieb2013} write \(Q_\alpha\) as a
supremum, respectively an infimum, of functions affine in \(\rho\); the
variational formula below is the dual companion of that representation,
derived here directly from the Fenchel--Young inequality for trace
functions.

\subsubsection{Variational formula and its
consequences}\label{variational-formula-and-its-consequences}

\begin{theorem}[Quantum variational formula for
\(H_\alpha\)]\protect\hypertarget{thm-variational}{}\label{thm-variational}

For \(\sigma > 0\), \(L = L^*\) and
\(\alpha \in [\tfrac12,\infty)\setminus\{1\}\), let
\(\widehat G(s) := \sup_{\rho \in \mathcal{S}_n}\{\operatorname{tr}(L\rho) - s\,Q_\alpha(\rho\|\sigma)\}\).

\textbf{Case (i), \(\alpha > 1\):} for every \(s > 0\),
\begin{equation}\phantomsection\label{eq-ghat-general}{
\widehat G(s) = \inf_{c \in \mathbb{R}}\left\{c + \frac{\alpha-1}{\alpha^{\alpha'}\,s^{\alpha'-1}}\bigl\|\bigl(\sigma^\beta(L-cI)\sigma^\beta\bigr)_+\bigr\|_{\alpha'}^{\alpha'}\right\}.
}\end{equation}

\textbf{Case (ii), \(\alpha \in [\tfrac12, 1)\):} for every
\(\zeta > 0\),
\begin{equation}\phantomsection\label{eq-gtilde-mu-general}{
\widehat G(-\zeta) = \inf_{c\,>\,\lambda_{\max}(L)}\left\{c + (1-\alpha)\,\alpha^p\,\zeta^{1+p}\,\|\sigma^\gamma(cI-L)^{-1}\sigma^\gamma\|_p^p\right\}.
}\end{equation}

\end{theorem}

\begin{proof}
Both cases are proved by the same four steps, for \(s\) with
\(\operatorname{sign}(s) = \operatorname{sign}(\alpha-1)\) --- the sign,
verified in Step 1, that makes \(\widehat G\) the supremum of a concave
function; only Step 3 depends on it. Recall
\(Q_\alpha(\rho\|\sigma) = \|\sigma^{-\beta}\rho\sigma^{-\beta}\|_\alpha^\alpha\).

\textbf{Step 1 --- Infimal-convolution split.} On \(M_n^{\text{sa}}\)
with pairing \(\langle A,\rho\rangle = \operatorname{tr}(A\rho)\),
define \[
M(\rho) = \begin{cases} s\,Q_\alpha(\rho\|\sigma) & \rho \ge 0 \\ +\infty & \text{else} \end{cases},
\qquad
N(\rho) = \begin{cases} 0 & \operatorname{tr}\rho = 1 \\ +\infty & \text{else} \end{cases}.
\] \(N\) is proper, convex, and lsc. So is \(M\):
\(\rho \mapsto Q_\alpha(\rho\|\sigma)\) is convex on the positive cone
for \(\alpha>1\) and concave for \(\alpha \in [\tfrac12,1)\) --- since
\(X \mapsto \operatorname{tr}X^\alpha\) is convex, respectively concave,
on positive matrices, as \(\operatorname{tr}f\) inherits the convexity
of the scalar function \(f\) \citep{carlen2010} --- so \(sQ_\alpha\) is
convex exactly when
\(\operatorname{sign}(s) = \operatorname{sign}(\alpha-1)\), the case at
hand. Since
\(I/n \in \operatorname{ri}(\operatorname{dom} M) \cap \operatorname{ri}(\operatorname{dom} N)\),
the Moreau--Rockafellar / Fenchel duality theorem
\citep{rockafellar1970} gives \[
\widehat G(s) = (M+N)^*(L) = \inf_{A}\{M^*(L-A) + N^*(A)\} = \inf_{c \in \mathbb{R}}\{c + M^*(L-cI)\},
\] the last step because \(N^*(A) = c\) for \(A = cI\) and \(+\infty\)
otherwise, collapsing the \(A\)-infimum onto that line; here
\(M^*(B) = \sup_{\rho \ge 0}\{\operatorname{tr}(B\rho) - sQ_\alpha(\rho\|\sigma)\}\).

\textbf{Step 2 --- Change of variables.} Set
\(X = \sigma^{-\beta}\rho\sigma^{-\beta}\) (so
\(\rho = \sigma^\beta X\sigma^\beta\)), a bijection between positive
cones --- the same substitution on both sides of \(\alpha = 1\). By
cyclicity,
\(\operatorname{tr}(B\rho) = \operatorname{tr}(\widetilde B X)\) with
\(\widetilde B := \sigma^\beta B\sigma^\beta\), and
\(M^*(B) = \Psi(\widetilde B)\) where \[
\Psi(A) := \sup_{X \ge 0}\bigl\{\operatorname{tr}(AX) - s\|X\|_\alpha^\alpha\bigr\}.
\]

\textbf{Step 3 --- Computation of \(\Psi(A)\): the sign split.} Let
\(a_1 \ge \cdots \ge a_n\) and \(x_1 \ge \cdots \ge x_n \ge 0\) be the
eigenvalues of \(A\), \(X\). By von Neumann's trace inequality
\citep{bhatia1997},
\(\operatorname{tr}(AX) - s\|X\|_\alpha^\alpha \le \sum_i(a_ix_i - sx_i^\alpha)\),
with equality when \(A,X\) share a co-ordered eigenbasis, reducing
\(\Psi(A)\) to \(n\) copies of the scalar problem
\(\sup_{x\ge0}(ax-sx^\alpha)\).

\emph{(a) \(\alpha > 1\) (\(s>0\)).} The optimum is \(0\) at \(x^*=0\)
for \(a \le 0\); for \(a>0\), \(x^*(a) = (a/(s\alpha))^{1/(\alpha-1)}\),
with value
\(\tfrac{\alpha-1}{\alpha^{\alpha'}s^{\alpha'-1}}a^{\alpha'}\). Since
\(a \mapsto x^*(a_+)\) is nondecreasing, the optimal \(x_i^*\) inherit
the order of \(a_i\), and equality in von Neumann is attained by
diagonalizing \(X\) in the eigenbasis of \(A\). Summing,
\begin{equation}\phantomsection\label{eq-psi-formula}{
\Psi(A) = \frac{\alpha-1}{\alpha^{\alpha'}\,s^{\alpha'-1}}\,\|A_+\|_{\alpha'}^{\alpha'}.
}\end{equation}

\emph{(b) \(\alpha \in [\tfrac12,1)\) (\(s<0\); write
\(\zeta := -s>0\)).} Here \(ax-sx^\alpha=ax+\zeta x^\alpha\); since
\(x\mapsto x^\alpha\) is sublinear, the supremum is \(+\infty\) whenever
\(a \ge 0\), so finiteness of \(\Psi(A)\) forces \(A<0\). For \(a<0\),
\(x^*(a) = (\zeta\alpha/|a|)^{1/(1-\alpha)}\), with value
\((1-\alpha)\zeta(\zeta\alpha/|a|)^p\); again \(a \mapsto x^*(a)\) is
nondecreasing on \(\{a<0\}\) (it decreases in \(|a|\)), so the
rearrangement is non-binding and equality is attained by diagonalizing
\(X\) in the eigenbasis of \(A\). Summing,
\begin{equation}\phantomsection\label{eq-psi-formula-small}{
\Psi(A) = \begin{cases}
(1-\alpha)\,\alpha^p\,\zeta^{1+p}\,\bigl\|(-A)^{-1}\bigr\|_p^p & A < 0 \\
+\infty & \text{otherwise.}
\end{cases}
}\end{equation}

\textbf{Step 4 --- Reassembly.} Combining Steps 1--3 with
\(B = L - cI\), \(\widetilde B = \sigma^\beta(L-cI)\sigma^\beta\): for
\(s > 0\), (\ref{eq-psi-formula}) gives (\ref{eq-ghat-general}); for
\(s = -\zeta < 0\), finiteness of \(\Psi(\widetilde B)\) forces
\(\widetilde B < 0\), i.e.~\(c > \lambda_{\max}(L)\), and
\((-\widetilde B)^{-1} = \sigma^{-\beta}(cI-L)^{-1}\sigma^{-\beta} = \sigma^\gamma(cI-L)^{-1}\sigma^\gamma\)
in (\ref{eq-psi-formula-small}) gives (\ref{eq-gtilde-mu-general}).
\(\square\)
\end{proof}

\subsubsection{DRO value}\label{dro-value}

For commuting \(\sigma\) and \(L\), case (i) of the following theorem is
the dual formula of Duchi and Namkoong \citep[Lemma
1]{duchi-namkoong2021}, constant included, in their normalization of the
radius (\(q/\alpha\) in place of \(q\)); see
Lemma~\ref{lem-classical-dual}. The norm
\(X \mapsto \|\sigma^\beta X\sigma^\beta\|_{\alpha'}\) is the
\(\sigma\)-weighted noncommutative \(L_{\alpha'}\) norm of
\citep[§II]{beigi2013}; in the theorem it is applied to a positive part
taken after sandwiching.

\begin{theorem}[DRO for sandwiched
\(\alpha\)-divergence]\protect\hypertarget{thm-main}{}\label{thm-main}

Let \(\sigma > 0\), \(L = L^*\), and \(q \ge 0\).

\textbf{Case (i), \(\alpha > 1\):}
\begin{equation}\phantomsection\label{eq-main-theorem}{
V_\alpha(q) = \inf_{c \in \mathbb{R}}\left\{c + \bigl(1+(\alpha-1)q\bigr)^{1/\alpha}\left\|\bigl(\sigma^\beta(L-cI)\sigma^\beta\bigr)_+\right\|_{\alpha'}\right\},
}\end{equation} and for \(q > 0\) the infimum is attained.

\textbf{Case (ii), \(\alpha \in [\tfrac12, 1)\):}
\begin{equation}\phantomsection\label{eq-main-theorem-small}{
V_\alpha(q) = \inf_{c\,>\,\lambda_{\max}(L)}\left\{c - \frac{\bigl(1+(\alpha-1)q\bigr)_+^{1/\alpha}}{\bigl\|\sigma^\gamma(cI-L)^{-1}\sigma^\gamma\bigr\|_p}\right\},
}\end{equation} and for \(q \in (0, q_{\max})\) the infimum is attained.

\end{theorem}

The positive part in the prefactor of (ii) matters only for
\(q \ge 1/(1-\alpha)\), where \(H_\alpha \le 1/(1-\alpha)\) makes the
ball all of \(\mathcal{S}_n\) and both sides equal
\(\lambda_{\max}(L)\).

\begin{proof}
Both cases are proved by the same Step 1; the inner minimization and the
attainment argument then differ genuinely between the two ranges, so
Steps 2--3 are given separately for each.

\textbf{Step 1 --- Rewrite in terms of \(Q_\alpha\).} Since
\(H_\alpha = (Q_\alpha-1)/(\alpha-1)\), the substitution
\(\tau := s/(\alpha-1)\) turns (\ref{eq-DRO-perspective}) into \[
V_\alpha(q) = \inf_{s>0}\{qs + G_\alpha(s)\} = \inf_{\operatorname{sign}(\tau)=\operatorname{sign}(\alpha-1)}\bigl\{\widetilde q\,\tau + \widehat G(\tau)\bigr\},
\qquad \widetilde q := 1+(\alpha-1)q,
\] where \(\widehat G\) is the signed \(Q_\alpha\)-cumulant of
Theorem~\ref{thm-variational}, given in closed form by
(\ref{eq-ghat-general}) (\(\tau>0\)) and (\ref{eq-gtilde-mu-general})
(\(\tau<0\)).

\textbf{Case (i), \(\alpha>1\) (\(\tau=t>0\)).}

\textbf{Step 2 --- Inner \(t\)-minimization.} Fix \(c\), set
\(\widetilde L(c) := \sigma^\beta(L-cI)\sigma^\beta\) and
\(T := \|\widetilde L(c)_+\|_{\alpha'}^{\alpha'}\). Elementary calculus,
using \(\alpha'-1 = 1/(\alpha-1)\), gives \[
\inf_{t>0}\left\{\widetilde q\,t + \frac{(\alpha-1)T}{\alpha^{\alpha'}}\,t^{1-\alpha'}\right\}
= \widetilde q^{\,1/\alpha}\,T^{1/\alpha'} = \widetilde q^{\,1/\alpha}\,\|\widetilde L(c)_+\|_{\alpha'}.
\] Substituting into Step 1, \[
V_\alpha(q) = \inf_{c\in\mathbb{R}}\left\{c + \bigl(1+(\alpha-1)q\bigr)^{1/\alpha}\|\widetilde L(c)_+\|_{\alpha'}\right\}.
\]

\textbf{Step 3 --- Attainment.} Denote the objective by \(\Phi(c)\). It
is convex (composition of the affine map
\(c \mapsto \widetilde L(c) = \sigma^\beta L\sigma^\beta - c\sigma^{2\beta}\)
with
\(A \mapsto \|A_+\|_{\alpha'} = \sup\{\operatorname{tr}(AX) : X \ge 0,\ \|X\|_\alpha \le 1\}\),
a supremum of linear functionals), finite and continuous on
\(\mathbb{R}\). For \(c \ge \lambda_{\max}(L)\) we have
\(\widetilde L(c)_+ = 0\), so \(\Phi(c) = c\); in particular
\(\Phi(\lambda_{\max}(L)) = \lambda_{\max}(L)\), which is the value
\(V_\alpha(q)\) when \(q \ge q_{\max}\), so the infimum is attained
there. For \(0 < q < q_{\max}\) we have \(\widetilde q > 1\), and
\(\Phi\) is coercive: \(\Phi(c) = c \to +\infty\) as \(c \to +\infty\),
while for \(c < \lambda_{\min}(L)\), \(\widetilde L(c) \succ 0\), so
\(\widetilde L(c)_+ = \widetilde L(c)\) and, using
\(\|\sigma^{2\beta}\|_{\alpha'} = 1\) (since
\(\operatorname{tr}\sigma = 1\) and \(2\beta\alpha' = 1\)), the triangle
inequality gives
\(\Phi(c) \ge (\widetilde q^{\,1/\alpha} - 1)\,|c| - \widetilde q^{\,1/\alpha}\|\sigma^\beta L\sigma^\beta\|_{\alpha'} \to +\infty\)
as \(c \to -\infty\).

\textbf{Case (ii), \(\alpha\in[\tfrac12,1)\) (\(\tau=-\zeta\),
\(\zeta>0\)).}

\textbf{Step 2 --- Inner \(\zeta\)-minimization.} Fix
\(c>\lambda_{\max}(L)\) and set
\(T := \|\sigma^\gamma(cI-L)^{-1}\sigma^\gamma\|_p > 0\). For
\(\widetilde q > 0\), elementary calculus, using \(1 + 1/p = 1/\alpha\)
and \((1-\alpha)(1+p) = 1\), gives \[
\inf_{\zeta > 0}\bigl\{-\widetilde q\,\zeta + (1-\alpha)\,\alpha^p\,T^p\,\zeta^{1+p}\bigr\} = -\widetilde q^{\,1/\alpha}/T.
\] Substituting into Step 1, \[
V_\alpha(q) = \inf_{c\,>\,\lambda_{\max}(L)}\left\{c - \widetilde q^{\,1/\alpha}/T(c)\right\}.
\]

\textbf{Step 3 --- Attainment.} Denote the objective by \(\Phi(c)\),
continuous on \((\lambda_{\max}(L), +\infty)\). As
\(c \to \lambda_{\max}(L)^+\), \(T(c) \to +\infty\) and
\(\Phi(c) \to \lambda_{\max}(L)\). As \(c \to +\infty\), expanding
\((cI - L)^{-1} = c^{-1}(I + L/c + O(c^{-2}))\) and using
\(\|\sigma^{2\gamma}\|_p = 1\) (since \(2\gamma p = 1\)) together with
the derivative of the Schatten norm,
\(T(c) = c^{-1}\bigl(1 + \operatorname{tr}(\sigma L)/c + O(c^{-2})\bigr)\),
hence \[
\Phi(c) = \bigl(1 - \widetilde q^{\,1/\alpha}\bigr)\,c + \widetilde q^{\,1/\alpha}\operatorname{tr}(\sigma L) + O(1/c),
\] which tends to \(+\infty\) exactly when \(\widetilde q < 1\),
i.e.~\(q > 0\). For \(0 < q < q_{\max}\) the infimum \(V_\alpha(q)\) is
strictly below the left-end limit \(\lambda_{\max}(L)\) (by
(\ref{eq-exchange})), so it is attained at some
\(c^* \in (\lambda_{\max}(L), \infty)\). \(\square\)
\end{proof}

\subsubsection{Optimal state and primal--dual
identity}\label{optimal-state-and-primaldual-identity}

The dual characterization in Theorem~\ref{thm-main} pins down the
worst-case state in each case, by the same stationarity argument, and
yields a clean decomposition of the robust value.

\begin{corollary}[Worst-case state and primal--dual
identity]\protect\hypertarget{cor-optimizer-large}{}\label{cor-optimizer-large}

In the active regime \(q \in (0, q_{\max})\), let \(c^*\) be a minimizer
in the corresponding case of Theorem~\ref{thm-main} and set
\(\widetilde q := 1 + (\alpha-1)q\).

\textbf{Case (i), \(\alpha > 1\):} with
\(\widetilde L^* := \sigma^\beta(L - c^* I)\sigma^\beta\), the supremum
in \(V_\alpha(q)\) is attained at
\begin{equation}\phantomsection\label{eq-optimizer-large}{
\rho^* \;=\; \widetilde q^{\,1/\alpha}\,\frac{\sigma^\beta\,(\widetilde L^*_+)^{\alpha'-1}\,\sigma^\beta}{\|\widetilde L^*_+\|_{\alpha'}^{\alpha'-1}},
\qquad \alpha'-1 = \tfrac{1}{\alpha-1} > 0,
}\end{equation} the divergence constraint is tight,
\(H_\alpha(\rho^*\|\sigma) = q\), and
\begin{equation}\phantomsection\label{eq-primal-dual-identity}{
V_\alpha(q) \;=\; \operatorname{tr}(L\rho^*) \;=\; c^* + \widetilde q^{\,1/\alpha}\,\|\widetilde L^*_+\|_{\alpha'}.
}\end{equation}

\textbf{Case (ii), \(\alpha \in [\tfrac12,1)\):} with
\(\widetilde M^* := \sigma^{-\gamma}(c^* I - L)\sigma^{-\gamma} > 0\),
the supremum in \(V_\alpha(q)\) is attained at
\begin{equation}\phantomsection\label{eq-optimizer-small}{
\rho^* \;=\; \widetilde q^{\,1/\alpha}\,\frac{\sigma^{-\gamma}\,(\widetilde M^*)^{\alpha'-1}\,\sigma^{-\gamma}}{\|(\widetilde M^*)^{-1}\|_p^{1+p}},
\qquad \alpha'-1 = \tfrac{1}{\alpha-1} = -(1+p) < 0,
}\end{equation} the divergence constraint is tight,
\(H_\alpha(\rho^*\|\sigma) = q\), and
\begin{equation}\phantomsection\label{eq-primal-dual-identity-small}{
V_\alpha(q) \;=\; \operatorname{tr}(L\rho^*) \;=\; c^* - \frac{\widetilde q^{\,1/\alpha}}{\|(\widetilde M^*)^{-1}\|_p}.
}\end{equation}

\end{corollary}

\begin{proof}
\textbf{Case (i), \(\alpha>1\).}

\textbf{Form of \(\rho^*\).} Define \(\rho^*\) by
(\ref{eq-optimizer-large}), the form read off Step 3(a) of the proof of
Theorem~\ref{thm-variational}. Since \(\|A_+\|_{\alpha'}\) is
differentiable in \(A\) wherever \(A_+ \neq 0\) (\(\alpha' > 1\)), and
\(\widetilde L^*_+ \neq 0\) because
\(\Phi(c^*) = V_\alpha(q) < \lambda_{\max}(L) \le \Phi(c)\) for every
\(c \ge \lambda_{\max}(L)\), stationarity of \(\Phi\) at \(c^*\) reads
\[
0 = \Phi'(c^*) = 1 - \widetilde q^{\,1/\alpha}\,\frac{\operatorname{tr}\!\bigl[\sigma^{2\beta}(\widetilde L^*_+)^{\alpha'-1}\bigr]}{\|\widetilde L^*_+\|_{\alpha'}^{\alpha'-1}},
\] which is exactly \(\operatorname{tr}\rho^* = 1\) for the state
(\ref{eq-optimizer-large}); and \(\rho^* \ge 0\) is clear.

\textbf{Feasibility and tightness.} With
\(X^* = \sigma^{-\beta}\rho^*\sigma^{-\beta} = \widetilde q^{\,1/\alpha}(\widetilde L^*_+)^{\alpha'-1}/\|\widetilde L^*_+\|_{\alpha'}^{\alpha'-1}\)
and \((\alpha'-1)\alpha = \alpha'\), \[
Q_\alpha(\rho^*\|\sigma) = \|X^*\|_\alpha^\alpha = \widetilde q\,\frac{\operatorname{tr}\bigl[(\widetilde L^*_+)^{\alpha'}\bigr]}{\|\widetilde L^*_+\|_{\alpha'}^{\alpha'}} = \widetilde q,
\]
i.e.~\(H_\alpha(\rho^*\|\sigma) = (\widetilde q - 1)/(\alpha-1) = q\):
\(\rho^*\) is feasible and the constraint is tight.

\textbf{Primal--dual identity and optimality.} Cyclicity and
(\ref{eq-optimizer-large}) give \[
\operatorname{tr}((L-c^*I)\rho^*) = \operatorname{tr}(\widetilde L^* X^*) = \widetilde q^{\,1/\alpha}\,\frac{\|\widetilde L^*_+\|_{\alpha'}^{\alpha'}}{\|\widetilde L^*_+\|_{\alpha'}^{\alpha'-1}} = \widetilde q^{\,1/\alpha}\,\|\widetilde L^*_+\|_{\alpha'},
\] so
\(\operatorname{tr}(L\rho^*) = c^* + \operatorname{tr}(\widetilde L^*X^*) = \Phi(c^*) = V_\alpha(q)\)
by Theorem~\ref{thm-main} (case (i)). Since \(\rho^*\) is feasible, it
is optimal, which is (\ref{eq-primal-dual-identity}).

\textbf{Case (ii), \(\alpha\in[\tfrac12,1)\).}

\textbf{Form of \(\rho^*\).} Define \(\rho^*\) by
(\ref{eq-optimizer-small}), the form read off Step 3(b) of the proof of
Theorem~\ref{thm-variational}. Stationarity of \(\Phi\) at \(c^*\),
\(0 = \Phi'(c^*) = 1 - \widetilde q^{\,1/\alpha}\operatorname{tr}[\sigma^{-2\gamma}(\widetilde M^*)^{-1-p}]/\|(\widetilde M^*)^{-1}\|_p^{1+p}\),
reads \[
\operatorname{tr}\!\bigl[\sigma^{-\gamma}\,(\widetilde M^*)^{\alpha'-1}\,\sigma^{-\gamma}\bigr] \;=\; \widetilde q^{\,-1/\alpha}\,\|(\widetilde M^*)^{-1}\|_p^{1+p},
\] which is \(\operatorname{tr}\rho^* = 1\) for
(\ref{eq-optimizer-small}); \(\rho^* \ge 0\) is clear.

\textbf{Feasibility and tightness.} With
\(X^* = \sigma^\gamma\rho^*\sigma^\gamma = \widetilde q^{\,1/\alpha}(\widetilde M^*)^{-(1+p)}/\|(\widetilde M^*)^{-1}\|_p^{1+p}\),
\((1+p)\alpha = p\) and
\(\|X^*\|_\alpha^\alpha = \widetilde q\,\operatorname{tr}[(\widetilde M^*)^{-p}]/\|(\widetilde M^*)^{-1}\|_p^{p} = \widetilde q\),
so \(Q_\alpha(\rho^*\|\sigma) = \widetilde q\) and
\(H_\alpha(\rho^*\|\sigma) = (\widetilde q - 1)/(\alpha-1) = q\):
\(\rho^*\) is feasible and the constraint is tight.

\textbf{Primal--dual identity and optimality.} With \(\alpha' = -p\),
cyclicity and (\ref{eq-optimizer-small}) give \[
\operatorname{tr}\!\bigl((c^*I - L)\rho^*\bigr) = \operatorname{tr}(\widetilde M^* X^*) = \widetilde q^{\,1/\alpha}\,\frac{\operatorname{tr}\bigl((\widetilde M^*)^{\alpha'}\bigr)}{\|(\widetilde M^*)^{-1}\|_p^{1+p}}  = \widetilde q^{\,1/\alpha}\,\frac{\|(\widetilde M^*)^{-1}\|_p^p}{\|(\widetilde M^*)^{-1}\|_p^{1+p}} = \frac{\widetilde q^{\,1/\alpha}}{\|(\widetilde M^*)^{-1}\|_p},
\] so
\(\operatorname{tr}(L\rho^*) = c^* - \operatorname{tr}\bigl((c^*I-L)\rho^*\bigr) = \Phi(c^*) = V_\alpha(q)\)
by Theorem~\ref{thm-main} (case (ii)); since \(\rho^*\) is feasible it
is optimal, which is (\ref{eq-primal-dual-identity-small}). \(\square\)
\end{proof}

\subsubsection{\texorpdfstring{The Hilbert--Schmidt endpoint
\(\alpha = 2\)}{The Hilbert--Schmidt endpoint \textbackslash alpha = 2}}\label{the-hilbertschmidt-endpoint-alpha-2}

At \(\alpha = 2\) the Schatten norm in (\ref{eq-main-theorem}) is the
Hilbert--Schmidt norm and the \(c\)-minimization closes in elementary
form, as long as the positive part is inactive. With \(\hat L\) the
centered observable, write
\begin{equation}\phantomsection\label{eq-s2-def}{
s_2^2 \;:=\; \tfrac{1}{2}\operatorname{tr}\bigl(\sigma^{1/2}\hat L\,\sigma^{1/2}\hat L\bigr),
}\end{equation} identified in Section~\ref{sec-linearization} as the DRO
variance \(s_2^2(\hat L)\). At this order
\(H_2(\rho\|\sigma) = \operatorname{tr}(\sigma^{-1/2}X\sigma^{-1/2}X)\),
\(X = \rho - \sigma\), is the geometric-mean member of the quantum
\(\chi^2\) family of \citep{temme-etal2010}, and the corollary below is
the Cauchy--Schwarz inequality for it (\ref{eq-hcr}), with equality as
long as the worst-case state stays positive.

\begin{corollary}[The Hilbert--Schmidt case \(\alpha = 2\): exact
square-root
law]\protect\hypertarget{cor-alpha-two-exact}{}\label{cor-alpha-two-exact}

Let \(\hat L \neq 0\) and \(q_0 := 2s_2^2/\lambda_{\min}(\hat L)^2\).
For \(0 < q \le q_0\) the infimum in (\ref{eq-main-theorem}) is attained
at \(c^* = \operatorname{tr}(L\sigma) - s_2\sqrt{2/q}\), and
\begin{equation}\phantomsection\label{eq-alpha-two-exact}{
V_2(q) \;=\; \operatorname{tr}(L\sigma) + s_2\sqrt{2q},
\qquad
\rho^* \;=\; \sigma^{1/2}\Bigl(I + \sqrt{\tfrac{q}{2}}\,\frac{\hat L}{s_2}\Bigr)\sigma^{1/2}.
}\end{equation} For \(q > q_0\),
\(V_2(q) < \operatorname{tr}(L\sigma) + s_2\sqrt{2q}\).

\end{corollary}

\begin{proof}
At \(\alpha = 2\), \(\beta = 1/4\), \(\alpha' = 2\) and
\(\widetilde q = 1 + q\). For \(c < \lambda_{\min}(L)\) the matrix
\(\widetilde L(c) = \sigma^{1/4}(L - cI)\sigma^{1/4}\) is positive, so
with \(u := \operatorname{tr}(L\sigma) - c > 0\), \[
\|\widetilde L(c)_+\|_2^2 = \operatorname{tr}\bigl[\sigma^{1/2}(\hat L + uI)\sigma^{1/2}(\hat L + uI)\bigr] = 2s_2^2 + u^2,
\] the cross term vanishing because
\(\operatorname{tr}(\sigma\hat L) = 0\). The objective of
(\ref{eq-main-theorem}) becomes
\(\Phi(c) = \operatorname{tr}(L\sigma) - u + \sqrt{1+q}\sqrt{2s_2^2 + u^2}\),
which is strictly convex in \(u\) with stationary point
\(u_*^2(1+q) = 2s_2^2 + u_*^2\), i.e.~\(u_* = s_2\sqrt{2/q}\), and value
\(\operatorname{tr}(L\sigma) - u_* + (1+q)u_* = \operatorname{tr}(L\sigma) + s_2\sqrt{2q}\).
The condition
\(c^* = \operatorname{tr}(L\sigma) - u_* < \lambda_{\min}(L)\) reads
\(u_* > -\lambda_{\min}(\hat L)\) (note \(\lambda_{\min}(\hat L) < 0\)
as \(\operatorname{tr}(\sigma\hat L) = 0\), \(\hat L \neq 0\)),
i.e.~\(q < q_0\). Since \(\Phi\) is convex on \(\mathbb{R}\) (Step 3 of
the proof of Theorem~\ref{thm-main}, case (i)) and coincides with the
function above on \((-\infty, \lambda_{\min}(L)]\), its stationary point
\(c^*\) in that half-line is its global minimizer (for \(q = q_0\),
\(c^* = \lambda_{\min}(L)\) is the endpoint, and \(\Phi\) is
nondecreasing to the right of a point where its left derivative
vanishes). Then (\ref{eq-optimizer-large}) --- whose verification in
Corollary~\ref{cor-optimizer-large} uses only that \(c^*\) is a
minimizer with \(\widetilde L^*_+ \neq 0\), both exhibited here --- with
\(\widetilde L^*_+ = \widetilde L^*\), \(\alpha' - 1 = 1\), gives
\(\rho^* = \sqrt{1+q}\,\sigma^{1/2}(\hat L + u_*I)\sigma^{1/2}/\sqrt{2s_2^2+u_*^2} = \sigma^{1/2}(I + \hat L/u_*)\sigma^{1/2}\),
which is (\ref{eq-alpha-two-exact}). For \(q > q_0\) the point
\(c_u := \operatorname{tr}(L\sigma) - u_*\) lies above
\(\lambda_{\min}(L)\), so \(L - c_uI\) has a negative eigenvalue and
\(\|\widetilde L(c_u)_+\|_2 < \|\widetilde L(c_u)\|_2\) strictly; hence
\(V_2(q) \le \Phi(c_u) < \operatorname{tr}(L\sigma) - u_* + \sqrt{1+q}\sqrt{2s_2^2+u_*^2} = \operatorname{tr}(L\sigma) + s_2\sqrt{2q}\).
\(\square\)
\end{proof}

The worst-case state is a sandwiched linear tilt of \(\sigma\), and
\(q_0\) is the budget at which it loses positivity --- the quantum
counterpart of the classical \(\chi^2\)-ball DRO
\citep[§2.1]{duchi-namkoong2019}. Section~\ref{sec-linearization}
establishes the square-root law for every \(\alpha\); at \(\alpha = 2\)
it is exact on an interval.

\subsubsection{\texorpdfstring{The fidelity endpoint
\(\alpha = 1/2\)}{The fidelity endpoint \textbackslash alpha = 1/2}}\label{sec-fidelity-boundary}

At \(\alpha = 1/2\), the parameters of Theorem~\ref{thm-main} (case
(ii)) specialize to \(\gamma = 1/2\), \(p = 1\), and
\(Q_{1/2}(\rho\|\sigma) = \mathrm{Fid}(\rho, \sigma)\), the \textbf{root
fidelity}
\(\mathrm{Fid}(\rho,\sigma) := \operatorname{tr}\bigl|\sigma^{1/2}\rho^{1/2}\bigr|\),
giving \(H_{1/2} = 2(1 - \mathrm{Fid}(\rho,\sigma))\). Since \(p = 1\)
and \(\sigma^{1/2}(cI-L)^{-1}\sigma^{1/2} \ge 0\), the Schatten
\(1\)-norm reduces to the trace, which gives the corollary below. The
inequality \(\le\) in it is Alberti's variational formula
\(\mathrm{Fid}(\rho,\sigma)^2 \le \operatorname{tr}(\rho X)\operatorname{tr}(\sigma X^{-1})\)
\citep{alberti1983} at \(X = cI - L\). Bounds on expectation values over
a fidelity ball that use only the mean of the observable, or its mean
and variance, are in \citep[Thms. 1 and 2]{weber-etal2022} (see also
\citep{weber-etal2021}); the corollary gives the value for a given
observable.

\begin{corollary}[Fidelity
DRO]\protect\hypertarget{cor-fidelity-dro}{}\label{cor-fidelity-dro}

For \(\sigma > 0\), \(L = L^*\), and \(0 \le q \le 2\)
(\(H_{1/2} = 2(1-\mathrm{Fid}) < 2\) on \(\mathcal{S}_n\)), \[
V_{1/2}(q) = \inf_{c\,>\,\lambda_{\max}(L)}\left\{c - \frac{(1-q/2)^2}{\operatorname{tr}\bigl[\sigma(cI-L)^{-1}\bigr]}\right\}.
\]

\end{corollary}

\begin{proof}
Set \(\alpha = 1/2\) in Theorem~\ref{thm-main} (case (ii)). Cyclicity of
the trace absorbs the sandwiching:
\(\|\sigma^{1/2}(cI-L)^{-1}\sigma^{1/2}\|_1 = \operatorname{tr}[\sigma(cI-L)^{-1}]\).
The prefactor specializes to
\((1 + (\alpha-1)q)^{1/\alpha} = (1 - q/2)^2\). \(\square\)
\end{proof}

\begin{refremark}[Stieltjes structure at \(\alpha = 1/2\)]
The denominator \(f(c) := \operatorname{tr}[\sigma(cI-L)^{-1}]\) is the
\textbf{operator Stieltjes transform} of \(L\) weighted by \(\sigma\),
reappearing in the proof of the Fuchs--van de Graaf inequality in
Section~\ref{sec-state-discrimination}; the collapse of the sandwiched
norm to a \(\sigma\)-weighted trace is the spectral face of the
coincidence \(H_{1/2}^{\rm meas} = H_{1/2}\) of the measured and
sandwiched divergences at \(\alpha = \tfrac12\) \citep{berta2017}.

\label{rem-stieltjes}

\end{refremark}

\subsection{\texorpdfstring{DRO for
\(\alpha = \infty\)}{DRO for \textbackslash alpha = \textbackslash infty}}\label{sec-alpha-infty-dro}

As \(\alpha \to \infty\) the sandwiched Rényi divergence converges to
the \textbf{max-relative entropy} \citep{datta2009}
\(D_\infty(\rho\|\sigma) = \inf\{\lambda : \rho \le e^\lambda\sigma\} = \log Q_\infty(\rho\|\sigma)\),
and we set \[
H_\infty(\rho\|\sigma) := Q_\infty(\rho\|\sigma) - 1, \qquad Q_\infty(\rho\|\sigma) := \|\sigma^{-1/2}\rho\,\sigma^{-1/2}\|_\infty = \inf\{\mu \ge 0 : \rho \le \mu\sigma\}.
\] Since \(Q_\infty\) is convex in \(\rho\), the duality of
Section~\ref{sec-framework} applies, and the constraint
\(H_\infty(\rho\|\sigma) \le q\) is the operator inequality
\(\rho \le (1+q)\sigma\), making (\ref{eq-renyi-dro-primal}) a
semidefinite program:
\begin{equation}\phantomsection\label{eq-cvar-primal}{
V_\infty(q) = \sup\bigl\{\operatorname{tr}(L\rho) : \rho \in \mathcal{S}_n,\; \rho \le (1+q)\sigma\bigr\}.
}\end{equation}

The Fenchel strategy of Theorem~\ref{thm-variational} applies verbatim
at \(\alpha = \infty\), with \(\beta = \tfrac12\):
\(Q_\infty = \|Y\|_\infty\) is \(1\)-homogeneous, so its conjugate is an
indicator, of \(\{\operatorname{tr}B_+ \le s\}\), and the
\(s\)-minimization of (\ref{eq-strong-dual}) returns the trace of the
positive part directly. The resulting formula
(Theorem~\ref{thm-quantum-cvar}) is the dual of a quantum
Neyman--Pearson problem --- for \(L \geq 0\), the dual form of the
quantum relative Lorenz curve of Buscemi and Gour
\citep{buscemi-gour2017}: in the variable
\(T = \sigma^{-1/2}\rho\,\sigma^{-1/2}/(1+q)\), (\ref{eq-cvar-primal})
maximizes \((1+q)\operatorname{tr}(\sigma^{1/2}L\sigma^{1/2}\,T)\) over
tests \(0 \le T \le I\) with \(\operatorname{tr}(\sigma T) = 1/(1+q)\).

\subsubsection{DRO value}\label{dro-value-1}

\begin{theorem}[Quantum
CVaR]\protect\hypertarget{thm-quantum-cvar}{}\label{thm-quantum-cvar}

For \(\sigma > 0\), \(L = L^*\), and \(q \ge 0\),
\begin{equation}\phantomsection\label{eq-quantum-cvar}{
V_\infty(q) = \inf_{c \in \mathbb{R}}\bigl\{c + (1+q)\,\operatorname{tr}\bigl[\bigl(\sigma^{1/2}(L - cI)\sigma^{1/2}\bigr)_+\bigr]\bigr\},
}\end{equation} and for \(q > 0\) the infimum is attained.

\end{theorem}

\begin{proof}
Steps 1--2 of the proof of Theorem~\ref{thm-variational} apply verbatim
with \(Q_\infty\) in place of \(Q_\alpha\) --- \(Q_\infty\) is convex
and continuous on the positive cone --- and \(\beta = \tfrac12\): for
\(s > 0\), \[
\begin{aligned}
\widehat G_\infty(s) &:= \sup_{\rho \in \mathcal{S}_n}\{\operatorname{tr}(L\rho) - sQ_\infty(\rho\|\sigma)\}
= \inf_{c \in \mathbb{R}}\bigl\{c + \Psi_\infty\bigl(\sigma^{1/2}(L-cI)\sigma^{1/2}\bigr)\bigr\},
\\
\Psi_\infty(A) &:= \sup_{X \ge 0}\bigl\{\operatorname{tr}(AX) - s\|X\|_\infty\bigr\}.
\end{aligned}
\] By von Neumann's trace inequality,
\(\operatorname{tr}(AX) \le \|X\|_\infty\operatorname{tr}A_+\) for
\(X \ge 0\), with equality at \(X = \lambda\mathbf{1}_{(0,\infty)}(A)\),
\(\lambda \ge 0\); so
\(\Psi_\infty(A) = \sup_{\lambda \ge 0}\lambda(\operatorname{tr}A_+ - s)\)
is \(0\) if \(\operatorname{tr}A_+ \le s\) and \(+\infty\) otherwise,
and
\(\widehat G_\infty(s) = \inf\{c : \operatorname{tr}[(\sigma^{1/2}(L-cI)\sigma^{1/2})_+] \le s\}\).
Since \(H_\infty = Q_\infty - 1\), the strong duality
(\ref{eq-strong-dual}) reads \[
V_\infty(q) = \inf_{s \ge 0}\bigl\{(1+q)s + \widehat G_\infty(s)\bigr\}
= \inf_{c \in \mathbb{R}}\ \inf_{s \ge \operatorname{tr}[(\sigma^{1/2}(L-cI)\sigma^{1/2})_+]}\bigl\{c + (1+q)s\bigr\},
\] which is (\ref{eq-quantum-cvar}). For \(q > 0\) the infimum is
attained: the objective
\(\Phi(c) := c + (1+q)\operatorname{tr}[(\sigma^{1/2}(L-cI)\sigma^{1/2})_+]\)
is convex and continuous, \(\Phi(c) \ge c \to +\infty\) as
\(c \to +\infty\), and for \(c \le \lambda_{\min}(L)\) it equals
\(c + (1+q)(\operatorname{tr}(L\sigma) - c) \to +\infty\) as
\(c \to -\infty\). \(\square\)
\end{proof}

\subsubsection{Optimal state and primal--dual
identity}\label{optimal-state-and-primaldual-identity-1}

\begin{corollary}[Worst-case state and primal--dual identity
(\(\alpha = \infty\))]\protect\hypertarget{cor-optimizer-infty}{}\label{cor-optimizer-infty}

In the active regime \(q \in (0, q_{\max})\), let \(c^*\) be a minimizer
in (\ref{eq-quantum-cvar}),
\(\widetilde L^* := \sigma^{1/2}(L - c^*I)\sigma^{1/2}\), and
\(P_+ := \mathbf{1}_{(0,\infty)}(\widetilde L^*)\). Assume
\(\ker \widetilde L^* = \{0\}\). Then the supremum in \(V_\infty(q)\) is
attained at the unique
\begin{equation}\phantomsection\label{eq-optimizer-infty}{
\rho^* = (1+q)\,\sigma^{1/2}P_+\sigma^{1/2},
}\end{equation} \(c^*\) satisfies
\((1+q)\operatorname{tr}(\sigma P_+) = 1\),
\((1+q)\sigma - \rho^* = (1+q)\sigma^{1/2}(I - P_+)\sigma^{1/2}\), and
\begin{equation}\phantomsection\label{eq-primal-dual-identity-infty}{
V_\infty(q) = \operatorname{tr}(L\rho^*) = c^* + (1+q)\,\|\widetilde L^*_+\|_1.
}\end{equation}

\end{corollary}

\begin{proof}
In sandwiched coordinates \(Y = \sigma^{-1/2}\rho\,\sigma^{-1/2}\),
(\ref{eq-cvar-primal}) reads
\(c^* + \max\{\operatorname{tr}(\widetilde L^* Y) : 0 \le Y \le (1+q)I,\; \operatorname{tr}(\sigma Y) = 1\}\).
For feasible \(Y\),
\(\operatorname{tr}(\widetilde L^* Y) \le \operatorname{tr}(\widetilde L^*_+ Y) \le (1+q)\operatorname{tr}\widetilde L^*_+\),
with equality in the first step iff \(Y\) vanishes on the range of
\(\widetilde L^*_-\) and in the second iff \(Y = (1+q)I\) on the range
of \(P_+\); and
\(c^* + (1+q)\operatorname{tr}\widetilde L^*_+ = V_\infty(q)\) by
Theorem~\ref{thm-quantum-cvar}. When \(\ker\widetilde L^* = \{0\}\) the
two conditions force \(Y^* = (1+q)P_+\), and stationarity of the
objective at \(c^*\), \(0 = 1 - (1+q)\operatorname{tr}(\sigma P_+)\)
(the objective is differentiable there, \(\widetilde L^*\) having no
zero eigenvalue), is the normalization
\(\operatorname{tr}(\sigma Y^*) = 1\); so \(Y^*\) is feasible and
optimal, and unique. Undoing the sandwich gives
(\ref{eq-optimizer-infty}); (\ref{eq-primal-dual-identity-infty}) is
(\ref{eq-quantum-cvar}) at \(c^*\). \(\square\)
\end{proof}

\begin{refremark}[Neyman--Pearson structure, and the degenerate case]
The worst-case state (\ref{eq-optimizer-infty}) is a threshold test,
with \(P_+\) the Holevo--Helstrom optimal test
\citep{holevo1973, helstrom1976} --- the projection onto the positive
part of the noncommutative likelihood ratio --- and \(c^*\) the
Neyman--Pearson level. It is \(\widetilde L^*\), not the spectral cut
\(\mathbf{1}_{\{L > c^*\}}\), that appears: the adversary tests the
\emph{sandwiched} observable, not \(L\) itself.

The hypothesis \(\ker\widetilde L^* = \{0\}\) in
Corollary~\ref{cor-optimizer-infty} fails exactly when \(c^*\) is an
eigenvalue of \(L\): these are the exceptional budgets, in general whole
intervals of \(q\), on which \(c^*\) is constant and \(V_\infty\) is
affine in \(q\). The two equality conditions still pin \(Y^*\) off the
kernel \(\Pi_0 := \mathbf{1}_{\{0\}}(\widetilde L^*)\), but leave it
free there subject only to the normalization
\(\operatorname{tr}(\sigma Y^*) = 1\); when \(c^*\) is a simple
eigenvalue of \(L\), \(\Pi_0\) has rank one and the normalization fixes
\(Y_0 = y\,\Pi_0\), so the worst-case state is still unique, while at a
multiple eigenvalue (\(\operatorname{rank}\Pi_0 \ge 2\)) it is
non-unique, \[
\rho^* = (1+q)\,\sigma^{1/2}P_+\sigma^{1/2} + \sigma^{1/2}Y_0\,\sigma^{1/2},
\qquad 0 \le Y_0 \le (1+q)\Pi_0,
\] while \(V_\infty(q)\) and (\ref{eq-primal-dual-identity-infty}) are
unaffected, since \(\widetilde L^*\) vanishes on \(\Pi_0\). This is the
quantum face of the randomization at the Value-at-Risk quantile in the
classical CVaR \citep{rockafellar-uryasev2000}: when the spectral
measure carries an atom at the threshold, the optimal tail allocation
may split it, and only the split, within a multiple eigenspace, is
undetermined --- never the value.

\label{rem-neyman-pearson}

\end{refremark}

\subsection{Reading the dual as a quantum risk
measure}\label{sec-schatten-structure}

\begin{figure}[htbp]

\centering{

\includegraphics[width=1\linewidth,height=\textheight,keepaspectratio]{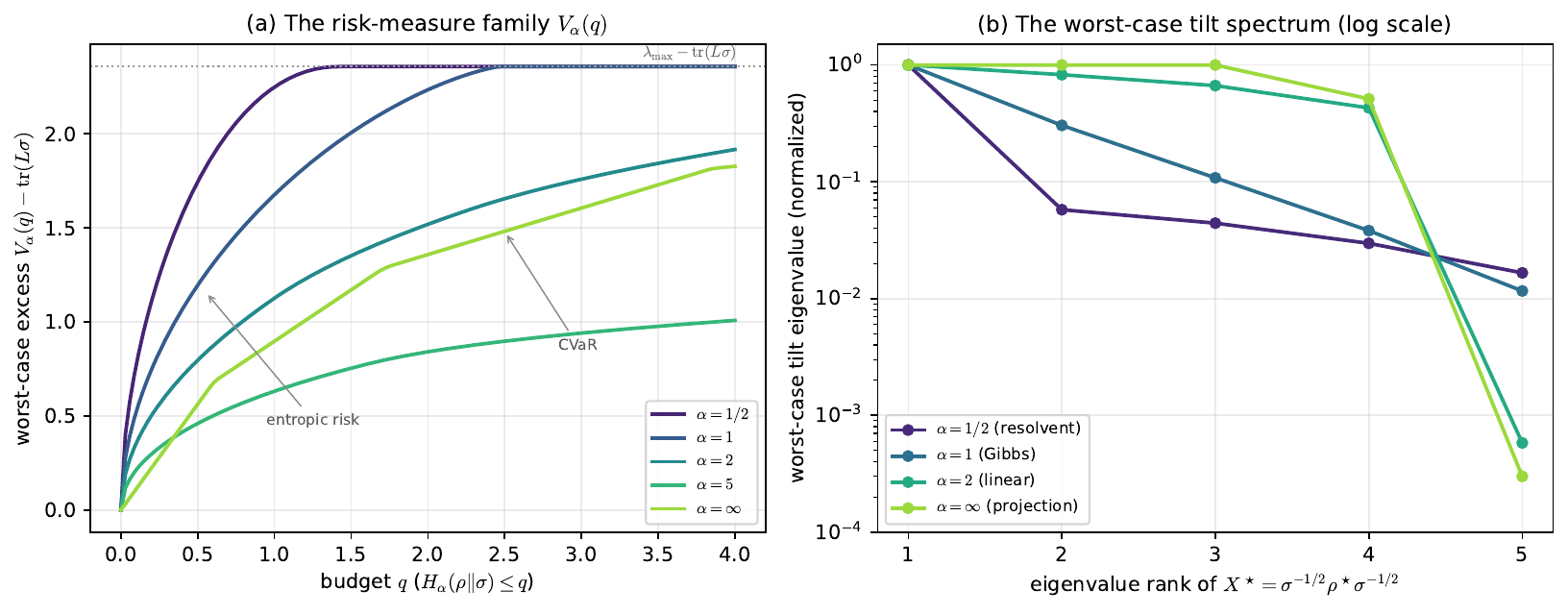}

}

\caption{\label{fig-risk-family}The quantum risk-measure family for a
noncommuting pair on \(\mathbb{C}^5\)
(\(\sigma = \operatorname{diag}(0.40,0.25,0.15,0.12,0.08)\), \(L\)
tridiagonal with diagonal \((0,0.8,1.6,2.4,3.2)\) and coupling
\(0.35\)). \emph{(a)} Worst-case excess
\(V_\alpha(q) - \operatorname{tr}(L\sigma)\) against the budget \(q\)
for \(\alpha \in \{1/2,1,2,5,\infty\}\), from the entropic risk measure
(\(\alpha = 1\)) to the quantum CVaR (\(\alpha = \infty\)), each
saturating at \(\lambda_{\max} -
\operatorname{tr}(L\sigma)\) (beyond the plotted range for
\(\alpha \ge 2\)); the affine segments of the \(\alpha = \infty\) curve
are the exceptional-budget intervals of Remark~\ref{rem-neyman-pearson}.
\emph{(b)} Spectrum of the worst-case tilt
\(X^* = \sigma^{-1/2}\rho^*\sigma^{-1/2}\) at \(q = 1\) (sorted,
normalized, log scale): full support for \(\alpha \le 1\) (resolvent
weighting at \(\alpha = \tfrac12\), Gibbs tilt at \(\alpha = 1\)),
truncation \(\propto (\widetilde L^*_+)^{\alpha'-1}\) for
\(\alpha > 1\), down to the Neyman--Pearson projection at \(\alpha =
\infty\), where this budget is exceptional and the fourth eigenvalue
shows the fractional split on \(\ker\widetilde L^*\).}

\end{figure}

Both faces of the family --- the value functions and the optimizer ---
are shown in Figure~\ref{fig-risk-family}.

When \([\sigma, L] = 0\), (\ref{eq-main-theorem}) reduces to the
classical \(\alpha\)-divergence DRO of the Born law
\(\sum_i p_i\,\delta_{l_i}\) (Lemma~\ref{lem-classical-dual});
Section~\ref{sec-spectral-measure} shows that the quantum DRO is a
classical DRO for every pair, of a measure \(\omega_\alpha\) that
encodes the noncommutativity.

\section{The quantum spectral measures}\label{sec-spectral-measure}

The dual formulas of Section~\ref{sec-main-theorem} express the DRO
value through Schatten norms of sandwiched operators. The main result of
this section is that this apparently quantum object is \emph{exactly} a
classical DRO: for each \(\alpha\) there is a probability measure
\(\omega_\alpha\) on the spectral interval \([\lambda_{\min}(L),
\lambda_{\max}(L)]\), the \textbf{quantum spectral measure}, with
\(V_\alpha(q) = V_\alpha^{\mathrm{cl}}(q; \omega_\alpha)\) for every
budget; we call this reduction \textbf{classicalization}. The measure
comes from one transform identity per regime: stop-loss for \(\alpha >
1\) from Heinävaara's tracial joint spectral measure
\citep{heinavaara2024tracial}, Stieltjes for \(\alpha < 1\) from the
companion note \citep{reybellet2026stieltjes}, Laplace at \(\alpha = 1\)
from Stahl's theorem.

\subsection{The classicalization theorem}\label{sec-classicalization}

Fix \(\sigma > 0\) and \(L\) Hermitian on \(\mathbb{C}^n\) with
eigenvalues \(l_1 \leq \cdots \leq l_n\) and an orthonormal eigenbasis
\(|1\rangle,
\ldots, |n\rangle\). With \(\beta = (\alpha-1)/2\alpha\),
\(\alpha' = \alpha/(\alpha-1)\) and \(p = -\alpha'\) as in
(\ref{eq-exponent-conventions}) (\(2\beta\alpha' = 1\); here \(\beta\)
is used on both ranges, \(\beta < 0\) for \(\alpha < 1\)), the
\textbf{sandwiched pair} is
\begin{equation}\phantomsection\label{eq-pair-def}{
A = \sigma^\beta L\,\sigma^\beta, \qquad B = \sigma^{2\beta} \succ 0,
}\end{equation} and we write
\(N_i := \langle i|\sigma^{-2\beta}|i\rangle > 0\). The dual formulas of
Section~\ref{sec-main-theorem} are traces of functions of the pencil
combination \(\sigma^\beta(L - cI)\sigma^\beta = A - cB\): a
positive-part power \(\operatorname{tr}(A - cB)_+^{\alpha'}\) for
\(\alpha
> 1\), a resolvent power \(\operatorname{tr}(cB - A)^{-p}\) for
\(\alpha <
1\). \emph{Standing convention.} In the pencil analysis of
Section~\ref{sec-heinavaara} and in the structure statements of this
section \(L\) has \emph{simple} spectrum, \(l_1 < \cdots < l_n\); the
existence part of Theorem~\ref{thm-spectral-rep} needs no such
assumption (Remark~\ref{rem-multiplicities}).

\textbf{The classical DRO value.} For a probability measure \(\nu\) on a
compact interval \(J = [\ell_{\min}, \ell_{\max}] \subset \mathbb{R}\)
and a budget \(q \geq 0\),
\begin{equation}\phantomsection\label{eq-classical-dro-def}{
V_\alpha^{\mathrm{cl}}(q;\,\nu)
:= \sup\Bigl\{\textstyle\int \ell\,dP :
P \in \mathcal{P}(J),\; H_\alpha^{\mathrm{cl}}(P\|\nu) \leq q\Bigr\},
}\end{equation} where \(H_\alpha^{\mathrm{cl}}\) is the classical
Hellinger divergence of order \(\alpha\), the commutative case of
Definition~\ref{def-sandwiched-hellinger}:
\(H_\alpha^{\mathrm{cl}}(P\|\nu) = \frac{1}{\alpha-1}\bigl(\int
(dP_{\mathrm{ac}}/d\nu)^\alpha\,d\nu - 1\bigr)\) with
\(P_{\mathrm{ac}}\) the \(\nu\)-absolutely continuous part of \(P\) and
\(H_\alpha^{\mathrm{cl}}(P\|\nu) = +\infty\) for \(\alpha > 1\) unless
\(P
\ll \nu\); at \(\alpha = 1\) the relative entropy \(D(P\|\nu) = \int
\log(dP/d\nu)\,dP\); at \(\alpha = \infty\),
\(H_\infty^{\mathrm{cl}}(P\|\nu) = \|dP/d\nu\|_{L^\infty(\nu)} - 1\)
(\(+\infty\) unless \(P \ll \nu\)).

\begin{lemma}[Classical
duals]\protect\hypertarget{lem-classical-dual}{}\label{lem-classical-dual}

Let \(\nu \in \mathcal{P}(J)\) and \(q \geq 0\). Then
\begin{equation}\phantomsection\label{eq-classical-duals}{
V_\alpha^{\mathrm{cl}}(q;\,\nu) =
\begin{cases}
\displaystyle\inf_{c \in \mathbb{R}}\Bigl\{c +
\bigl(1+(\alpha-1)q\bigr)^{1/\alpha}\,
\bigl\|(\ell - c)_+\bigr\|_{L^{\alpha'}(\nu)}\Bigr\},
& \alpha \in (1, \infty),\\[3mm]
\displaystyle\inf_{c \in \mathbb{R}}\Bigl\{c +
(1+q)\textstyle\int(\ell - c)_+\,d\nu\Bigr\},
& \alpha = \infty,\\[3mm]
\displaystyle\inf_{t \,>\, 0}\;\frac{1}{t}\Bigl\{q +
\log\textstyle\int e^{t\ell}\,d\nu\Bigr\},
& \alpha = 1,\\[3mm]
\displaystyle\inf_{c \,>\, \ell_{\max}}\Bigl\{c -
\bigl(1+(\alpha-1)q\bigr)_+^{1/\alpha}\,
\bigl\|(c-\ell)^{-1}\bigr\|_{L^{p}(\nu)}^{-1}\Bigr\},
& \alpha \in [\tfrac12, 1).
\end{cases}
}\end{equation}

\end{lemma}

The proof is in \hyperref[sec-app-moment-tools]{Appendix A}.

\begin{theorem}[Classicalization]\protect\hypertarget{thm-spectral-rep}{}\label{thm-spectral-rep}

Let \(\sigma > 0\) and let \(L\) be Hermitian on \(\mathbb{C}^n\).

\begin{enumerate}
\def\labelenumi{(\roman{enumi})}
\item
  For every \(\alpha \in [\tfrac12, \infty]\) there is a probability
  measure \(\omega_\alpha\) on
  \([\lambda_{\min}(L), \lambda_{\max}(L)]\) with mean
  \(\operatorname{tr}(L\sigma)\) such that
  \begin{equation}\phantomsection\label{eq-spectral-dro}{
  \boxed{\,V_\alpha(q) = V_\alpha^{\mathrm{cl}}(q;\,\omega_\alpha),
  \qquad q \geq 0.\,}
  }\end{equation} It is the unique measure on that interval satisfying
  the transform identity of its regime ---
  (\ref{eq-classicalization-alpha-big}) for \(\alpha
  \in (1,\infty]\), (\ref{eq-classicalization-alpha-less-one}) for
  \(\alpha \in
  [\tfrac12, 1)\), (\ref{eq-bmv-explicit}) for \(\alpha = 1\) --- and if
  \([L,
  \sigma] = 0\) it is the Born measure \(\nu_L\) for every \(\alpha\).
\item
  If \(L\) has simple spectrum, \(\omega_\alpha = \sum_i
  \omega_\alpha(\{l_i\})\,\delta_{l_i} + \omega_\alpha^{\mathrm{ac}}\)
  with atoms \(\omega_\alpha(\{l_i\}) = N_i^{-\alpha'}\) for
  \(\alpha \neq
  1\) and \(e^{\langle i|\log\sigma|i\rangle}\) for \(\alpha = 1\), and
  an absolutely continuous part on \((l_1, l_n)\), given in
  Proposition~\ref{prp-explicit-density} (\(\alpha \ne 1\)) and
  (\ref{eq-bmv-density}) (\(\alpha = 1\)) as an integral of Heinävaara's
  density, which vanishes if and only if \([L, \sigma] = 0\) or
  \(\alpha =
  \tfrac12\).
\end{enumerate}

\end{theorem}

\emph{Related work.} The existence of a positive measure with the
transform identity is not ours in any regime. It is Heinävaara's theorem
for \(\alpha \in (1, \infty)\) \citep{heinavaara2024tracial}, the
companion note for \(\alpha \in (\tfrac12, 1)\)
\citep{reybellet2026stieltjes}, which is built on Heinävaara's measure,
and Stahl's theorem at \(\alpha = 1\) \citep{stahl2013}; at
\(\alpha = \infty\) it needs only the convexity of
\(c \mapsto \operatorname{tr}(A - cB)_+\), and at \(\alpha = \tfrac12\)
only trace cyclicity. At \(\alpha = \infty\) the measure
\(\omega_\infty\) is the layer-cake measure of \citetext{\citealp[Eqs.
(15)--(16)]{gour2026}; \citealp[Eq. (17), Lemma 2.2, Prop.
6.1]{liu-hirche-cheng2025}}, taken here for the pair
\((\sigma^{1/2}L\sigma^{1/2}, \sigma)\), and for \(L \geq 0\) the value
\(V_\infty\) is, up to normalization, the quantum relative Lorenz curve
of \citep{buscemi-gour2017}. Concurrently and independently, Schmitt
\citep{schmitt2026} obtains the same measure as the push-forward of
Heinävaara's measure under \((a,b) \mapsto a/b\), with the stop-loss
identity \citep[Prop.~6.2]{schmitt2026} and the dichotomy between the
commuting and the non-commuting case \citep[Thm.~6.4]{schmitt2026}. What
the theorem adds is the identification with the DRO value at every
budget and for every finite \(\alpha\), with the normalization (mass
one, mean \(\operatorname{tr}(L\sigma)\)) and the atoms.

\begin{refremark}[Degenerate spectra]
Simple spectrum enters only through the explicit atoms and the ray
structure of (ii); existence and the transform identities in (i) hold
for arbitrary multiplicities, since Heinävaara's identity, the Stieltjes
representation of Theorem~\ref{thm-note-stieltjes} and Stahl's theorem
need no non-degeneracy. The analogue of (ii) for repeated eigenvalues
follows by the same arguments; for \(\alpha < 1\) it is carried out in
\citep[Cor.~A.1]{reybellet2026stieltjes}.

\label{rem-multiplicities}

\end{refremark}

\textbf{The fidelity endpoint.} At \(\alpha = \tfrac12\) (\(\beta =
-\tfrac12\), \(p = 1\)) trace cyclicity collapses the sandwich,
\(\operatorname{tr}[\sigma^{1/2}(cI-L)^{-1}\sigma^{1/2}] =
\operatorname{tr}[\sigma(cI-L)^{-1}] = \int (c-\ell)^{-1}\,d\nu_L\), so
\begin{equation}\phantomsection\label{eq-omega-half}{
\omega_{1/2} = \nu_L = \sum_i \langle i|\sigma|i\rangle\,\delta_{l_i}
}\end{equation} for every \(L\), and the DRO dual
Corollary~\ref{cor-fidelity-dro} is the classical \(\alpha = \tfrac12\)
DRO of the Born measure. It is the only order at which \(\omega_\alpha\)
is the Born measure for every \(L\)
(Theorem~\ref{thm-variance-ordering}), the spectral trace of the
coincidence \(H_{1/2}^{\mathrm{meas}} = H_{1/2}\)
(Remark~\ref{rem-stieltjes}).

\subsection{Tools: the pencil and Heinävaara's
measure}\label{sec-heinavaara}

A \emph{matrix pencil} \((A,B)\) is the family \(\{\ell B - A : \ell \in
\mathbb{R}\}\); its \emph{roots} are the zeros of \(\det(\ell B - A)\),
the eigenvalues of \(B^{-1}A\) when \(B\) is invertible, and the pencil
is \emph{non-degenerate} if some real combination of \(A, B\) is
invertible.

\begin{lemma}[Pencil of the sandwiched
pair]\protect\hypertarget{lem-pencil}{}\label{lem-pencil}

Let \(L\) have simple spectrum \(l_1 < \cdots < l_n\) with unit
eigenvectors \(|1\rangle, \ldots, |n\rangle\), and let \(\beta \in
\mathbb{R}\). The pencil \((A,B)\) of (\ref{eq-pair-def}) is
non-degenerate with roots \(l_1, \ldots, l_n\). At \(l_i\) its unit
eigenvector is \[
|v_i\rangle = \frac{\sigma^{-\beta}|i\rangle}{N_i^{1/2}},
\qquad N_i = \langle i|\sigma^{-2\beta}|i\rangle,
\] with \(\langle v_i|A|v_i\rangle = l_i/N_i\) and
\(\langle v_i|B|v_i\rangle = 1/N_i\). Moreover, for \(b > 0\) and
\(\ell = a/b \notin \operatorname{spec}(L)\),
\begin{equation}\phantomsection\label{eq-pencil-resolvent}{
(bA - aB)^{-1} = b^{-1}\,\sigma^{-\beta}(L - \ell I)^{-1}\sigma^{-\beta}.
}\end{equation}

\end{lemma}

\begin{proof}
\(\ell B - A = \sigma^\beta(\ell I - L)\sigma^\beta\) with
\(\sigma^\beta\) invertible, so the roots are the \(l_i\), \(B\) is
invertible, and
\(\ker(l_iB - A) = \operatorname{span}\{\sigma^{-\beta}|i\rangle\}\);
normalizing gives \(|v_i\rangle\) and the two expectations, and
inverting the three factors of \(bA - aB = b\,\sigma^\beta(L - \ell
I)\sigma^\beta\) gives (\ref{eq-pencil-resolvent}). \(\square\)
\end{proof}

\begin{theorem}[Tracial joint spectral measure
\citep{heinavaara2024tracial}]\protect\hypertarget{thm-heinavaara}{}\label{thm-heinavaara}

Let \(A, B\) be Hermitian on \(\mathbb{C}^n\).

\begin{enumerate}
\def\labelenumi{(\roman{enumi})}
\item
  There is a positive measure \(\mu_{A,B}\) on \(\mathbb{R}^2\) with
  \(\mu_{A,B}(\{0\}) = 0\) such that, for every measurable
  \(f : \mathbb{R}
  \to \mathbb{R}\) with \(f(0) = 0\) and \(\int_{-M}^{M} |f(t)/t|\,dt <
  \infty\) for all \(M > 0\), and every \((x, y) \in \mathbb{R}^2\),
  \begin{equation}\phantomsection\label{eq-heinavaara}{
  \operatorname{tr}\,H(f)(xA + yB)
  = \int_{\mathbb{R}^2} f(ax + by)\,d\mu_{A,B}(a,b),
  \qquad H(f)(x) := \int_0^1 \frac{1-t}{t}\,f(xt)\,dt.
  }\end{equation}
\item
  Let \(\mu_{A,B} = \mu_c + \mu_s\) be the Lebesgue decomposition. If
  the pencil \((A,B)\) is non-degenerate, \(\mu_c\) has density
  \begin{equation}\phantomsection\label{eq-heinavaara-density}{
  \mathfrak{h}_{A,B}(a,b)
  = \frac{1}{2\pi}\sum_{i=1}^n
  \bigl|\operatorname{Im}\lambda_i(C_1(a,b))\bigr|,
  \qquad
  C_1(a,b) = \Bigl(I - \tfrac{aA + bB}{a^2 + b^2}\Bigr)
  (bA - aB)^{-1},
  }\end{equation} and if moreover \(A\) is invertible and \(A^{-1}B\)
  has \(n\) distinct eigenvalues, then for
  \(\varphi \in C_c^\infty(\mathbb{R}^2 \setminus \{0\})\)
  \begin{equation}\phantomsection\label{eq-heinavaara-singular}{
  \mu_s(\varphi) = \sum_{v \in E(A,B)}
  \int_0^1 \frac{1-t}{t}\,
  \varphi\bigl(t\langle v|A|v\rangle,\, t\langle v|B|v\rangle\bigr)\,dt,
  }\end{equation} \(E(A,B)\) containing one unit eigenvector of the
  pencil per real root.
\item
  If \(xA + yB \succeq 0\) for some real \((x,y)\), then
  \(\operatorname{supp}\mu_{A,B} \subseteq \{xa + yb \geq 0\}\); \(\mu_c
  \equiv 0\) if and only if \([A,B] = 0\); and
  \(\operatorname{supp}\mu_c\) is contained in the joint numerical range
  \(W(A,B) := \{(\langle
  v|A|v\rangle, \langle v|B|v\rangle) : \|v\| = 1\}\).
\end{enumerate}

\end{theorem}

Item (i) is \citep[Thm.~1.3]{heinavaara2024tracial} (uniqueness away
from the origin is its Prop.~3.9) and (ii) is
\citep[Thm.~1.6]{heinavaara2024tracial}; in (iii) the half-plane
statement is \citep[Lemma 1.5]{heinavaara2024tracial}, stated there for
\((x,y) = (1,0)\) and valid in every direction by the same proof, and
the dichotomy and the inclusion
\(\operatorname{supp}\mu_c \subseteq W(A,B)\) are \citep[
Prop.~4]{heinavaara-thesis}. The specialization we use is the one-sided
power \(f(t) = t_+^{s}\), \(s >
0\), which Heinävaara uses for integer \(s\) in the proof of
\citep[Cor.~1.2]{heinavaara2024tracial} and which is stated for every
\(s >
0\) in \citep[Prop.~2.2]{schmitt2026}; it is admissible since
\(\int_{-M}^{M}|f(t)/t|\,dt = M^s/s\), with \(H(f)(x)
= x_+^s\int_0^1(1-t)t^{s-1}dt = x_+^s/(s(s+1))\):
\begin{equation}\phantomsection\label{eq-heinavaara-schatten}{
\operatorname{tr}\,(xA + yB)_+^{\,s} = s(s+1)\int (ax + by)_+^{\,s}\,d\mu_{A,B}(a,b),
\qquad s > 0 .
}\end{equation}

\begin{lemma}[Support for the sandwiched
pair]\protect\hypertarget{lem-support}{}\label{lem-support}

For the pair (\ref{eq-pair-def}), any \(\beta \in \mathbb{R}\) and
arbitrary multiplicities of \(L\): \(b > 0\) and
\(\ell := a/b \in [l_1, l_n]\) for \(\mu_{A,B}\)-a.e.~\((a,b)\).

\end{lemma}

\begin{proof}
The matrices \(B = \sigma^{2\beta}\), \(A - l_1 B = \sigma^\beta(L -
l_1 I)\sigma^\beta\) and \(l_n B - A = \sigma^\beta(l_n I -
L)\sigma^\beta\) are positive semidefinite, so by the half-plane
statement of Theorem~\ref{thm-heinavaara}(iii) the measure \(\mu_{A,B}\)
is carried by \(\{b \geq 0,\ l_1 b \leq a \leq l_n b\}\). On this set
\(b = 0\) forces \(a = 0\), and the origin carries no mass. \(\square\)
\end{proof}

\begin{lemma}[Transform
uniqueness]\protect\hypertarget{lem-transform-unique}{}\label{lem-transform-unique}

A finite measure \(\nu\) on \([l_1, l_n]\) is determined by any one of:
its stop-loss transform \(c \mapsto \int(\ell - c)_+^{\alpha'}d\nu\) on
\(\mathbb{R}\) (\(\alpha' \geq 1\)); its Stieltjes transform \(c \mapsto
\int (c-\ell)^{-p}d\nu\) on \((l_n, \infty)\) (\(p > 0\)); its Laplace
transform \(t \mapsto \int e^{t\ell}d\nu\) on \([0,\infty)\).

\end{lemma}

The proof is in \hyperref[sec-app-moment-tools]{Appendix A}.

For \(\alpha < 1\) the classicalization rests on a companion result ---
a Stieltjes representation of \(\operatorname{tr}[(A + tB)^{-p}]\) built
from Heinävaara's measure --- which we import; it is Theorem 1.1 there,
stated in the note's convention, in which the positive-definite matrix
comes first; the pair (\ref{eq-pair-def}) enters below with the roles
exchanged.

\begin{theorem}[Stieltjes-\(p\) representation
\citep{reybellet2026stieltjes}]\protect\hypertarget{thm-note-stieltjes}{}\label{thm-note-stieltjes}

Let \(A \succ 0\) and \(B \succeq 0\) be Hermitian on \(\mathbb{C}^n\),
and let \(p \geq 1\). With \(\lambda_1 < \cdots < \lambda_r\)
(\(r \leq n\)) the \emph{distinct} eigenvalues of \(A^{-1/2}BA^{-1/2}\)
and \(P_1, \ldots, P_r\) its spectral projections, there is a unique
positive finite measure \(\rho_p\) on \([\lambda_1, \lambda_r]\) with
\begin{equation}\phantomsection\label{eq-note-stieltjes-p}{
\operatorname{tr}\bigl[(A + tB)^{-p}\bigr]
= \int (1 + t\ell)^{-p}\,d\rho_p(\ell),
\qquad t \geq 0.
}\end{equation} Its atoms are \(\rho_p(\{\lambda_s\}) =
\operatorname{tr}[(P_s A^{-1} P_s)^p]\), \(s = 1, \ldots, r\), and its
continuous part has density
\begin{equation}\phantomsection\label{eq-note-density}{
\frac{d\rho_p^{\mathrm{ac}}}{d\ell}(\ell)
= p(p-1)\int_0^\infty a^{1-p}\,\mathfrak{h}_{A,B}(a,\, \ell a)\,da,
}\end{equation} with \(\mathfrak{h}_{A,B}\) the density
(\ref{eq-heinavaara-density}); the continuous part vanishes iff
\(p = 1\) or \([A,B] = 0\).

\end{theorem}

The representation is obtained in \citep[§4.2]{reybellet2026stieltjes}
from (\ref{eq-heinavaara}) by inverting Heinävaara's transform on
\(x^{-p}\) with a cutoff at the origin, whose boundary terms cancel
against the singular rays of \(\mu_{A,B}\); the weight \(a^{1-p}\) is
integrable because \(a \ge
\lambda_{\min}(A)\) on \(\operatorname{supp}\mu_c\), a by-product of the
same argument.

\subsection{The three transform identities}\label{sec-construction}

Each regime is one identity expressing the trace functional of the
quantum dual as a classical transform of a measure that the identity
itself defines.

\subsubsection{\texorpdfstring{Stop-loss identity,
\(\alpha \in (1, \infty]\)}{Stop-loss identity, \textbackslash alpha \textbackslash in (1, \textbackslash infty{]}}}\label{sec-stoploss}

The identity of this regime is Heinävaara's, namely his
(\ref{eq-heinavaara-schatten}) read in the spectral variable \(\ell =
a/b\); what the lemma adds is the normalization and the atoms for the
sandwiched pair. At \(\alpha = \infty\) and for a pair of states, the
same push-forward and the same identity are
\citep[Prop.~6.2]{schmitt2026}, obtained independently.

\begin{lemma}[Stop-loss
identity]\protect\hypertarget{lem-stoploss}{}\label{lem-stoploss}

Let \(\alpha \in (1, \infty]\), so \(\alpha' \in [1, \infty)\)
(\(\alpha'
= 1\), \(\beta = \tfrac12\) at \(\alpha = \infty\)), and define
\(\omega_\alpha\) by
\begin{equation}\phantomsection\label{eq-omega-alpha-def}{
\int \varphi(\ell)\,d\omega_\alpha(\ell)
:= \alpha'(\alpha'+1)\int \varphi(a/b)\,b^{\alpha'}\,d\mu_{A,B}(a,b).
}\end{equation} Then \(\omega_\alpha\) is a probability measure on
\([l_1, l_n]\) and, for every \(c \in \mathbb{R}\),
\begin{equation}\phantomsection\label{eq-classicalization-alpha-big}{
\operatorname{tr}\bigl[\sigma^\beta(L - cI)\sigma^\beta\bigr]_+^{\alpha'}
= \int (\ell - c)_+^{\alpha'}\,d\omega_\alpha(\ell).
}\end{equation} If \(L\) has simple spectrum,
\(\omega_\alpha = \sum_i N_i^{-\alpha'}
\delta_{l_i} + \omega_\alpha^{\mathrm{ac}}\), where
\(\omega_\alpha^{\mathrm{ac}}\), the image of \(b^{\alpha'}\mu_c\), is
absolutely continuous and vanishes iff \([L, \sigma] = 0\).

\end{lemma}

\begin{proof}
By Lemma~\ref{lem-support}, \(\ell = a/b\) is defined
\(\mu_{A,B}\)-a.e.~with values in \([l_1, l_n]\), so
(\ref{eq-omega-alpha-def}) defines a positive measure there. Identity
(\ref{eq-heinavaara-schatten}) with \(s = \alpha'\) at \((x,y)
= (1, -c)\) gives
\(\operatorname{tr}[\sigma^\beta(L - cI)\sigma^\beta]_+^{\alpha'} =
\alpha'(\alpha'+1)\int (a - cb)_+^{\alpha'}\,d\mu_{A,B} =
\alpha'(\alpha'+1)\int b^{\alpha'}(a/b - c)_+^{\alpha'}\,d\mu_{A,B}\),
which is (\ref{eq-classicalization-alpha-big}); at \((x,y) = (0,1)\) it
gives \(\omega_\alpha(\mathbb{R}) = \operatorname{tr} B^{\alpha'} =
\operatorname{tr}\sigma^{2\beta\alpha'} = \operatorname{tr}\sigma = 1\).
For simple spectrum (and \(0 \notin \operatorname{spec}L\), as
(\ref{eq-heinavaara-singular}) requires; otherwise apply what follows to
\(L + \kappa I\) and translate by \(\kappa\), which is legitimate by
Lemma~\ref{lem-transform-unique} since the transform of \(L + \kappa I\)
is that of \(L\) shifted in \(c\)), (\ref{eq-heinavaara-singular}) and
Lemma~\ref{lem-pencil} describe \(\mu_s\) as the radial measures
\((1-t)t^{-1}dt\) on the segments
\(\{t(l_i/N_i, 1/N_i) : 0 < t \leq 1\}\), on which \(a/b = l_i\) and
\(b^{\alpha'} = (t/N_i)^{\alpha'}\), so
\begin{equation}\phantomsection\label{eq-atom-weights}{
\omega_\alpha(\{l_i\})
= \alpha'(\alpha'+1)\,N_i^{-\alpha'}\int_0^1 (1-t)\,t^{\alpha'-1}\,dt
= N_i^{-\alpha'} .
}\end{equation} The image of \(b^{\alpha'}\mu_c\) under
\((a,b) \mapsto a/b\) is absolutely continuous because
\((\ell, b) \mapsto (\ell b, b)\) is a diffeomorphism of
\(\mathbb{R} \times (0,\infty)\) with Jacobian \(b\) (its density is
computed in Proposition~\ref{prp-explicit-density}); it vanishes iff
\(\mu_c
\equiv 0\) iff \([A,B] = 0\) (Theorem~\ref{thm-heinavaara}(iii)) iff
\([L, \sigma] =
0\), since \([A,B] = \sigma^\beta[L, \sigma^{2\beta}]\sigma^\beta\) and
\(\sigma > 0\). \(\square\)
\end{proof}

\textbf{The CVaR endpoint \(\alpha = \infty\).} At \(\alpha = \infty\)
(\(\beta = \tfrac12\), \(\alpha' = 1\)) the pair is
\((A, B) = (\sigma^{1/2}L\sigma^{1/2}, \sigma)\) and
Lemma~\ref{lem-stoploss} reads \[
\operatorname{tr}\bigl[\bigl(\sigma^{1/2}(L - cI)\sigma^{1/2}\bigr)_+\bigr]
= \int_c^{l_n}(\ell - c)\,d\omega_\infty ,
\] the order-one stop-loss transform of \(\omega_\infty\). The left side
is convex in \(c\), so at this endpoint the existence of
\(\omega_\infty\) needs nothing from Theorem~\ref{thm-heinavaara};
\(\omega_\infty\) is the layer-cake measure of \citetext{\citealp[Eq.
(15)]{gour2026}; \citealp[Eq.
(17)]{liu-hirche-cheng2025}; \citealp[Prop.~6.2]{schmitt2026}} for this
pair (see the related work after Theorem~\ref{thm-spectral-rep}).
Inserted into the max-divergence dual (\ref{eq-quantum-cvar}), it makes
the \(\alpha =
\infty\) case of Theorem~\ref{thm-spectral-rep} the Rockafellar--Uryasev
conditional value-at-risk \citep{rockafellar-uryasev2000} at tail level
\(u =
(1+q)^{-1}\): \begin{equation}\phantomsection\label{eq-cvar-classical}{
V_\infty(q) = V_\infty^{\mathrm{cl}}(q;\omega_\infty)
= \operatorname{CVaR}_u(\omega_\infty)
= \inf_{c \in \mathbb{R}}\left\{c + (1+q)\int_c^{l_n}(\ell - c)\,d\omega_\infty\right\}.
}\end{equation} The quantum CVaR is dominated by its Born-measure
shadow, \(V_\infty(q) \leq \operatorname{CVaR}_u(\nu_L)\), with equality
for all \(q\) iff \([L,\sigma] = 0\): this is the \(\alpha = \infty\)
case of the stop-loss domination Proposition~\ref{prp-born-domination}.

\subsubsection{\texorpdfstring{Stieltjes identity,
\(\alpha \in [\tfrac12, 1)\)}{Stieltjes identity, \textbackslash alpha \textbackslash in {[}\textbackslash tfrac12, 1)}}\label{sec-alpha-small}

\begin{lemma}[Stieltjes identity
\citep{reybellet2026stieltjes}]\protect\hypertarget{lem-stieltjes}{}\label{lem-stieltjes}

Let \(\alpha \in [\tfrac12, 1)\), \(p = \alpha/(1-\alpha)\), fix \(c_0 >
\lambda_{\max}(L)\), and let \(\rho_p\) be the measure of
Theorem~\ref{thm-note-stieltjes} for the pair \[
A_0 := \sigma^\beta(c_0 I - L)\sigma^\beta \succ 0,
\qquad B_0 := \sigma^{2\beta} \succ 0 .
\] With \(m(\ell') := c_0 - 1/\ell'\), define \(\omega_\alpha :=
m_*\bigl(\ell'^{\,-p}\,d\rho_p\bigr)\). Then \(\omega_\alpha\) is a
probability measure on \([l_1, l_n]\), independent of \(c_0\), and for
every \(c > \lambda_{\max}(L)\)
\begin{equation}\phantomsection\label{eq-classicalization-alpha-less-one}{
\operatorname{tr}\bigl[\sigma^\beta(cI-L)\sigma^\beta\bigr]^{-p}
= \int (c - \ell)^{-p}\,d\omega_\alpha(\ell).
}\end{equation} Its atoms sit at the eigenvalues of \(L\), with weight
\(N_i^{-\alpha'}\) at a simple eigenvalue \(l_i\); its continuous part,
the image of \(\ell'^{-p}\rho_p^{\mathrm{ac}}\), vanishes iff \(p = 1\)
or \([L,\sigma]
= 0\).

\end{lemma}

\begin{proof}
This is the pencil form of the Stieltjes-\(p\) representation,
\citep[Cor.~A.1]{reybellet2026stieltjes}, for \(P = \sigma^{2\beta}\)
and \(Q
= \sigma^\beta L\sigma^\beta\). Indeed \(cP - Q = \sigma^\beta(cI -
L)\sigma^\beta\), \(P^{-1/2}QP^{-1/2} = L\) and
\((c_0P - Q,\, P) = (A_0,
B_0)\), so the measure of the corollary is \(\omega_\alpha\). It is
carried by \([l_1, l_n]\), it is positive since \(p \geq 1\), and the
corollary gives (\ref{eq-classicalization-alpha-less-one}) for every
\(c >
\lambda_{\max}(L)\); by Lemma~\ref{lem-transform-unique} it is the
unique measure with this property, hence independent of \(c_0\). Its
mass is
\(\operatorname{tr}P^{-p} = \operatorname{tr}\sigma^{-2\beta p} =
\operatorname{tr}\sigma = 1\) (as \(-2\beta p = 1\)). Its atom at a
simple eigenvalue \(l_i\) is \(\langle i|P^{-1}|i\rangle^{p} = N_i^{p} =
N_i^{-\alpha'}\). Its continuous part is the image of
\(\ell'^{-p}\rho_p^{\mathrm{ac}}\) and vanishes iff \(p = 1\) or
\([P, Q] =
\sigma^\beta[\sigma^{2\beta}, L]\sigma^\beta = 0\), that is, iff \(p =
1\) or \([L, \sigma] = 0\). \(\square\)
\end{proof}

At \(\alpha = \tfrac12\) the lemma returns (\ref{eq-omega-half}):
\(p = 1\) kills the continuous part, and the weights
\(N_i^{-\alpha'} = \langle
i|\sigma|i\rangle\) are the Born weights.

\subsubsection{\texorpdfstring{Laplace identity, \(\alpha = 1\): the
Stahl
measure}{Laplace identity, \textbackslash alpha = 1: the Stahl measure}}\label{sec-example-alpha-one}

At \(\alpha = 1\) the sandwiching exponent is \(\beta = 0\); the pair
degenerates and the state-dependence sits in the Gibbs cumulant
\(F_1(t) = \log\operatorname{tr}\,e^{\log\sigma + tL}\) of
Theorem~\ref{thm-DRO-1}.

\begin{lemma}[Laplace identity (Stahl;
Heinävaara)]\protect\hypertarget{lem-laplace}{}\label{lem-laplace}

There is a unique probability measure \(\omega_1\) on \([l_1, l_n]\)
with \begin{equation}\phantomsection\label{eq-bmv-explicit}{
\operatorname{tr}\,e^{\log\sigma + tL}
= \int e^{t\ell}\,d\omega_1(\ell),
\qquad t \in \mathbb{R},
\qquad\quad
\omega_1
= \sum_{i=1}^n e^{\langle i|\log\sigma|i\rangle}\,\delta_{l_i}
+ \mathfrak{h}_1(\ell)\,d\ell
\quad\text{(simple spectrum)},
}\end{equation} where
\begin{equation}\phantomsection\label{eq-bmv-density}{
\mathfrak{h}_1(\ell) = \frac{1}{2\pi}\int_{\mathbb{R}} e^{a}
\sum_{j=1}^n \bigl|\operatorname{Im}\lambda_j\bigl(
(aI - \log\sigma)(L - \ell I)^{-1}\bigr)\bigr|\,da,
\qquad \ell \in (l_1, l_n),
}\end{equation} \(\mathfrak{h}_1 = 0\) off \([l_1, l_n]\), and
\(\mathfrak{h}_1 \equiv 0\) iff \([L, \sigma] = 0\). The mean of
\(\omega_1\) is \(\operatorname{tr}(L\sigma)\).

\end{lemma}

\begin{proof}
The atoms and the density are those of \citep[§1.4, Thm.
24]{heinavaara-thesis}, and the derivation is Heinävaara's \citep[proof
of Thm. 24]{heinavaara-thesis}: it is his identity (\ref{eq-heinavaara})
for the pair \((\log\sigma,\ \lambda_{\max}(L)I - L)\) with the test
function \(f(u) =
u^{-2}e^{-1/u}\mathbf{1}_{u > 0}\), for which \(H(f)(x) =
e^{-1/x}\mathbf{1}_{x > 0}\): a reciprocal spectral variable produces
the exponential, and the shift \(\ell = \lambda_{\max}(L) - b\) is
absorbed. The support lies in \([l_1, l_n]\) because \(\mu_{A,B}\) is
carried by the joint numerical range \(W(A,B)\) and the segments from
the origin to its points (Theorem~\ref{thm-heinavaara}(ii)--(iii), for
simple spectrum; in general the support statement is part of Stahl's
theorem \citep{stahl2013}), whose \(b\)-projection is
\([0, l_n - l_1]\). Mass and mean are the transform and its derivative
at \(t = 0\), uniqueness is Lemma~\ref{lem-transform-unique}, and the
dichotomy is that of Theorem~\ref{thm-heinavaara}(iii). \(\square\)
\end{proof}

The existence of a positive representing measure is Stahl's theorem
\citep{stahl2013}, the former BMV conjecture \citep{bessis1975}, whose
proof (and Eremenko's \citep{eremenko2015, clivaz2016}) yields the
density implicitly, as a contour integral whose positivity is not
manifest; in Heinävaara's formula (\ref{eq-bmv-density}) it is read off
the absolute values. By Jensen's inequality each atom \(e^{\langle
i|\log\sigma|i\rangle}\) is dominated by the Born weight \(\langle
i|\sigma|i\rangle\), the deficit being carried by \(\mathfrak{h}_1\).

\subsection{Proof of the classicalization
theorem}\label{sec-proof-classicalization}

\emph{Step 1 (dual coincidence).} For \(\alpha \in (1,\infty)\) insert
(\ref{eq-classicalization-alpha-big}) into Theorem~\ref{thm-main}(i):
\(\|(\sigma^\beta(L-cI)\sigma^\beta)_+\|_{\alpha'} = \bigl(\int(\ell -
c)_+^{\alpha'}d\omega_\alpha\bigr)^{1/\alpha'} = \|(\ell -
c)_+\|_{L^{\alpha'}(\omega_\alpha)}\). At \(\alpha = \infty\) insert it
(\(\alpha' = 1\)) into Theorem~\ref{thm-quantum-cvar}. For
\(\alpha \in [\tfrac12,1)\) insert
(\ref{eq-classicalization-alpha-less-one}) into
Theorem~\ref{thm-main}(ii), where
\(\|\sigma^\gamma(cI-L)^{-1}\sigma^\gamma\|_p^{-1} =
\bigl(\operatorname{tr}[\sigma^\beta(cI-L)\sigma^\beta]^{-p}\bigr)^{-1/p}
= \|(c-\ell)^{-1}\|_{L^p(\omega_\alpha)}^{-1}\). At \(\alpha = 1\)
insert (\ref{eq-bmv-explicit}) into Theorem~\ref{thm-DRO-1}. In each
case the quantum dual becomes, verbatim, the corresponding row of
Lemma~\ref{lem-classical-dual} at \(\nu = \omega_\alpha\); both hold for
every \(q \geq 0\), which is (\ref{eq-spectral-dro}).

\emph{Step 2 (mean).} \(V_\alpha(0) = \operatorname{tr}(L\sigma)\)
(Proposition~\ref{prp-dro-duality}) while
\(V_\alpha^{\mathrm{cl}}(0;\nu) = \int
\ell\,d\nu\), since at \(q = 0\) the classical ball is \(\{\nu\}\).

\emph{Step 3 (commuting case).} If \([L,\sigma] = 0\), in a joint
eigenbasis
\(\operatorname{tr}[\sigma^\beta(L-cI)\sigma^\beta]_+^{\alpha'}
= \sum_i \langle i|\sigma|i\rangle (l_i - c)_+^{\alpha'}\),
\(\operatorname{tr}[\sigma^\beta(cI-L)\sigma^\beta]^{-p} = \sum_i
\langle i|\sigma|i\rangle (c - l_i)^{-p}\) and
\(\operatorname{tr}e^{\log\sigma + tL} = \sum_i \langle
i|\sigma|i\rangle e^{tl_i}\) (using \(2\beta\alpha' = -2\beta p = 1\)):
the transforms of \(\nu_L\). By Lemma~\ref{lem-transform-unique},
\(\omega_\alpha =
\nu_L\).

\emph{Step 4 (uniqueness).} This is Lemma~\ref{lem-transform-unique}
applied to the identity of each regime.

\emph{Step 5 (structure).} Under simple spectrum,
Lemma~\ref{lem-stoploss}, Lemma~\ref{lem-stieltjes} and
Lemma~\ref{lem-laplace} give the atoms of (ii) and exhibit the remainder
as absolutely continuous, vanishing iff \([L,\sigma] = 0\) or \(p = 1\),
i.e.~\(\alpha = \tfrac12\). \(\square\)

\subsection{Structure of the spectral
measure}\label{sec-spectral-structure}

\begin{proposition}[The spectral measure,
explicitly]\protect\hypertarget{prp-explicit-density}{}\label{prp-explicit-density}

For \(\sigma>0\), \(L=L^*\) with simple spectrum \(l_1<\cdots<l_n\), and
\(\alpha \in [\tfrac12,\infty]\setminus\{1\}\) (\(\alpha' = 1\),
\(\beta = \tfrac12\) at \(\alpha = \infty\)), the spectral measure of
Theorem~\ref{thm-spectral-rep} has atoms
\(\omega_\alpha(\{l_i\}) = N_i^{-\alpha'}\) and continuous density, for
\(\ell \in (l_1,l_n)\),

\textbf{Case (i), \(\alpha>1\):}
\begin{equation}\phantomsection\label{eq-explicit-density-big}{
\frac{d\omega_\alpha^{\mathrm{ac}}}{d\ell}(\ell)
= \alpha'(\alpha'+1)\int_0^\infty b^{\alpha'+1}\,
\mathfrak{h}_{A,B}(\ell b,\,b)\,db,
}\end{equation}

\textbf{Case (ii), \(\alpha \in [\tfrac12,1)\):}
\begin{equation}\phantomsection\label{eq-explicit-density-small}{
\frac{d\omega_\alpha^{\mathrm{ac}}}{d\ell}(\ell)
= p(p-1)\,\ell'^{\,2-p}\int_0^\infty a^{1-p}\,
\mathfrak{h}_{A_0,B_0}(a,\,\ell' a)\,da,
\qquad \ell' := \frac{1}{c_0-\ell},
}\end{equation}

for any fixed \(c_0 > \lambda_{\max}(L)\), with \(\mathfrak{h}_{A,B}\)
(resp.~\(\mathfrak{h}_{A_0,B_0}\)) the continuous Heinävaara density
(\ref{eq-heinavaara-density}) of the pair \((A,B)\) of
(\ref{eq-pair-def}) (resp.~\((A_0,B_0)\) of Lemma~\ref{lem-stieltjes}).

\end{proposition}

\begin{proof}
The atoms are (\ref{eq-atom-weights}). In case (i), restrict
(\ref{eq-omega-alpha-def}) to \(\mu_c\) and substitute \(a = \ell b\) at
fixed \(b > 0\) (so \(da = b\,d\ell\)):
\(\int\varphi\,d\omega_\alpha^{\rm ac} =
\int\varphi(\ell)\bigl[\alpha'(\alpha'+1)\int_0^\infty
b^{\alpha'+1}\mathfrak{h}_{A,B}(\ell b,b)\,db\bigr]\,d\ell\), which is
(\ref{eq-explicit-density-big}). In case (ii), push the density
(\ref{eq-note-density}) of \(\rho_p\) forward through
\(\ell = c_0 - 1/\ell'\) (so \(d\ell'/d\ell = \ell'^2\)):
\(d\omega_\alpha^{\rm ac}/d\ell(\ell) =
\ell'^{-p}\,(d\rho_p^{\rm ac}/d\ell')(\ell')\cdot\ell'^2\), which is
(\ref{eq-explicit-density-small}). \(\square\)
\end{proof}

\begin{figure}[htbp]

\centering{

\includegraphics[width=1\linewidth,height=\textheight,keepaspectratio]{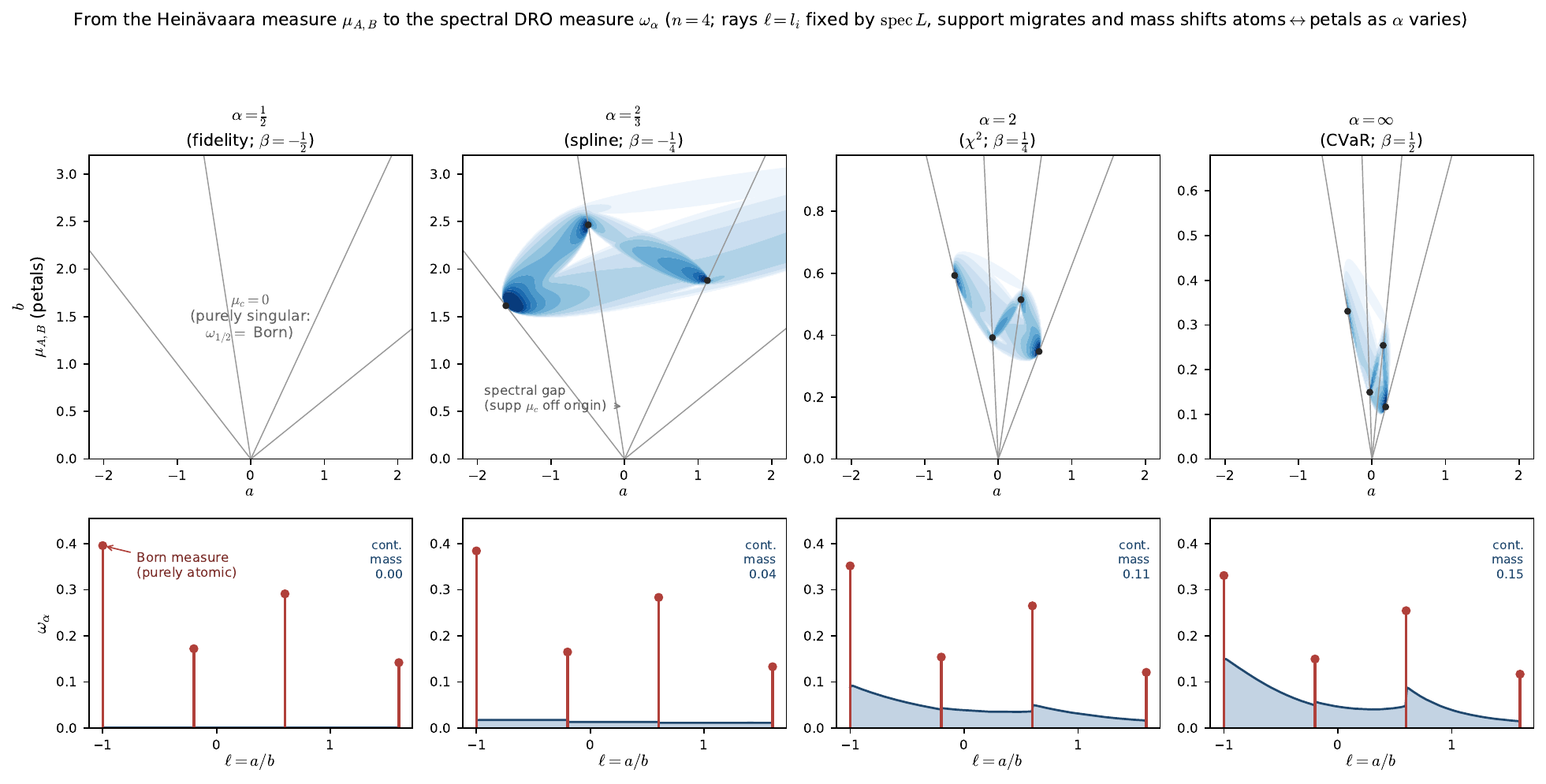}

}

\caption{\label{fig-omega-sweep}From the tracial measure to the spectral
measure (\(n = 4\); \(\sigma =
\operatorname{diag}(0.45,0.30,0.15,0.10)\), \(L\) with eigenvalues
\(\{-1,-0.2,0.6,1.6\}\) rotated so that \([\sigma,L] \neq 0\)).
\emph{Top:} the continuous part of \(\mu_{A,B}\) in the \((a,b)\)
half-plane; the rays \(\ell = l_i\) are fixed by
\(\operatorname{spec}L\), while the singular points \((l_i/N_i, 1/N_i)\)
and the continuous density move with \(\alpha\). \emph{Bottom:}
\(\omega_\alpha\) on \([l_1,l_n]\). As \(\alpha\) grows from
\(\tfrac12\) to \(\infty\), mass migrates from the atoms into the
continuous part; at \(\alpha = \tfrac12\) the measure is the purely
atomic Born measure.}

\end{figure}

For \(n = 2\) and \(\alpha < 1\), closed forms follow from
\citep[§5]{reybellet2026stieltjes}; Figure~\ref{fig-omega-sweep} shows a
non-commuting \(n = 4\) example.

\section{Linearization and quantum Fisher
metrics}\label{sec-linearization}

A distributionally robust objective is often used through its
\emph{linearization}: at a small budget the worst-case value is the
nominal value plus a variance term, the variance-regularized surrogate
that makes DRO tractable in machine learning
\citep{duchi-namkoong2019, lam2016robust, gotoh-kim-lim2018}, and the
linearized goal-oriented divergence bounds of uncertainty quantification
\citep{dupuis-katsoulakis-pantazis-plechac2016}. For the quantum
sandwiched DRO the square-root law of Proposition~\ref{prp-square-root}
gives \begin{equation}\phantomsection\label{eq-linearization}{
V_\alpha(q) = \operatorname{tr}(L\sigma) + s_\alpha(\hat L)\sqrt{2q} + o(\sqrt q),
\qquad q \to 0^+,
}\end{equation} with \(\hat L\) the centered observable; the coefficient
\(s_\alpha^2(\hat L)\), the \textbf{DRO variance}, is the quantum
variance-regularizer, and computing it is the business of this section.

\begin{theorem}[DRO
variance]\protect\hypertarget{thm-dro-variance}{}\label{thm-dro-variance}

Let \(\sigma > 0\), \(L\) Hermitian and not a multiple of the identity,
and \(\alpha \in [1/2, \infty)\). The DRO value linearizes as in
(\ref{eq-linearization}), and the \textbf{DRO variance} has the two
forms \begin{equation}\phantomsection\label{eq-dro-variance-two-forms}{
s_\alpha^2(\hat L)
\;=\; \langle \hat L, \mathcal{H}_\alpha^{-1}(\hat L)\rangle
\;=\; \tfrac{1}{\alpha}\operatorname{Var}_{\omega_\alpha}(\ell),
}\end{equation} where \(\mathcal{H}_\alpha\) is the Hessian of
\(H_\alpha(\cdot\|\sigma)\) at \(\sigma\) and \(\omega_\alpha\) the
spectral measure of Theorem~\ref{thm-spectral-rep}. At matched curvature
the variance
\(\operatorname{Var}_{\omega_\alpha}(\ell) = \alpha\,s_\alpha^2(\hat L)\)
is non-increasing in \(\alpha\): constant, equal to
\(\operatorname{Var}_\sigma(L)\), if \([\sigma, L] = 0\), and strictly
decreasing otherwise.

\end{theorem}

The theorem is proved by combining Theorem~\ref{thm-hessian-bridge} (the
linearization), Corollary~\ref{cor-dro-variance-sandwiched} and
Corollary~\ref{cor-spectral-variance} (the two forms of the variance)
and Theorem~\ref{thm-variance-ordering} (the ordering), established in
that order below.

\subsection{The DRO variance bridge}\label{sec-hessian-bridge}

For a divergence \(\mathcal{D}\) with \(\mathcal{D}(\cdot\|\sigma)\)
twice continuously differentiable on a relative neighborhood of
\(\sigma\), the second variation defines a symmetric bilinear form on
\(T_\sigma\mathcal{S}_n := \{A = A^* \in M_n(\mathbb{C}) : \operatorname{tr} A = 0\}\),
whose self-adjoint operator is the \textbf{Hessian superoperator}
\(\mathcal{H}\): \begin{equation}\phantomsection\label{eq-hessian-def}{
\langle A,\mathcal{H}(A)\rangle \;:=\; \frac{d^2}{d\epsilon^2}\,\mathcal{D}(\sigma+\epsilon A\|\sigma)\Big|_{\epsilon=0},
\qquad A \in T_\sigma\mathcal{S}_n.
}\end{equation} We extend \(\mathcal{H}\) to all Hermitian matrices by
\(\mathcal{H}(I) = 0\); equivalently \(\mathcal{H}^{-1}\) is the
Moore--Penrose pseudoinverse, and
\(\mathcal{H}^{-1}(A) = \mathcal{H}^{-1}(A|_T)\) where
\(A|_T := A - (\operatorname{tr} A/n)I\).

This curvature is what the DRO linearization sees:
Theorem~\ref{thm-hessian-bridge} below is the classical fact that the
Hessian of a Legendre conjugate is the inverse Hessian
\citep{crouzeix1977}, on the tangent space of
\(\{\operatorname{tr}\rho = 1\}\), combined with the square-root law of
Proposition~\ref{prp-square-root}; its classical counterpart is the
variance expansion of \citetext{\citealp[Thm.
3.1]{lam2016robust}; \citealp{duchi-namkoong2019}; \citealp{duchi-glynn-namkoong2021}; \citealp{gotoh-kim-lim2018}}.
Recall \(\operatorname{tr}(\hat L\sigma) = 0\); trace pairings
\(\operatorname{tr}(\hat L\,X)\) for \(X \in T_\sigma\mathcal{S}_n\)
depend only on \(\hat L|_T\) since \(\operatorname{tr}(I\cdot X) = 0\).

\begin{theorem}[Hessian
bridge]\protect\hypertarget{thm-hessian-bridge}{}\label{thm-hessian-bridge}

Let \(\mathcal{D}\) satisfy (D1)--(D3), let
\(\mathcal{D}(\cdot\|\sigma)\) be \(C^2\) on a relative neighborhood of
\(\sigma\) with Hessian \(\mathcal{H}\) positive definite on
\(T_\sigma\mathcal{S}_n\), and let \(L\) not be a multiple of the
identity. Then
\(F''(0) = \operatorname{tr}(\hat L\,\mathcal{H}^{-1}(\hat L)) > 0\),
and \begin{equation}\phantomsection\label{eq-sqrt-law-hessian}{
V_\mathcal{D}(q;\hat L)
\;=\; \sqrt{2q\,\operatorname{tr}\bigl(\hat L\,\mathcal{H}^{-1}(\hat L)\bigr)}
  \;+\; o(\sqrt{q}),
\qquad q \to 0^+.
}\end{equation}

\end{theorem}

\begin{proof}
Write \[
\varphi(t,\rho) := t\operatorname{tr}(\hat L\rho) - \mathcal{D}(\rho\|\sigma),
\qquad
F(t) := \sup_{\rho \in \mathcal{S}_n}\varphi(t,\rho),
\] so that \(F\) is the abstract CGF of (\ref{eq-abstract-cgf}) for the
observable \(\hat L\). Since \(\mathcal{H} \succ 0\) and
\(\mathcal{D}(\cdot\|\sigma)\) is \(C^2\), there is a relatively open
convex neighborhood \(U\) of \(\sigma\) in \(\mathcal{S}_n\), contained
in the relative interior (as \(\sigma \succ 0\)), on which
\(\mathcal{D}(\cdot\|\sigma)\) is strictly convex. Maximizers of
\(\varphi(t,\cdot)\) exist by compactness and lower semicontinuity, and
any maximizer \(\rho\) satisfies
\(\mathcal{D}(\rho\|\sigma) \le t\operatorname{tr}(\hat L(\rho - \sigma)) \le 2|t|\,\|\hat L\|\)
(compare with \(\rho = \sigma\), using
\(\operatorname{tr}(\hat L\sigma) = 0\) and
\(\|\rho - \sigma\|_1 \le 2\)); by (D2), lower semicontinuity and
compactness, all maximizers therefore lie in \(U\) for \(|t| < \delta\),
where \(\varphi(t,\cdot)\) is strictly concave, so the maximizer
\(\rho^*(t)\) is unique and interior for \(|t| < \delta\). We locate
\(\rho^*(t)\) by the implicit function theorem and then differentiate
\(F\) by the envelope theorem, on \((-\delta,\delta)\).

\emph{Step 1 --- first-order optimality and the multiplier.} Write
\(g(\rho) := \nabla_1\mathcal{D}(\rho\|\sigma)\) for the differential of
\(\mathcal{D}(\cdot\|\sigma)\) in its first argument, a Hermitian-matrix
valued map that is \(C^1\) near \(\sigma\) by the \(C^2\) hypothesis and
satisfies the tangent projection \(g(\sigma)|_T = 0\) since \(\sigma\)
minimizes \(\mathcal{D}(\cdot\|\sigma)\) subject to
\(\operatorname{tr}\rho=1\) (the ambient gradient may be a scalar
multiple of \(I\)). The only constraint active near the interior
maximizer is \(\operatorname{tr}\rho = 1\), with Lagrangian
\(\varphi(t,\rho) - \zeta(\operatorname{tr}\rho - 1)\); stationarity in
\(\rho\) reads \begin{equation}\phantomsection\label{eq-foc-raw}{
t\,\hat L \;=\; g(\rho^*(t)) + \zeta(t)\,I .
}\end{equation} Project onto the tangent space
\(T_\sigma\mathcal{S}_n = \{X = X^* :
\operatorname{tr} X = 0\}\) by
\(X \mapsto X|_T := X - (\operatorname{tr}
X/n)I\). The multiplier term drops out, \((\zeta I)|_T = 0\), leaving
\begin{equation}\phantomsection\label{eq-first-order}{
G(\rho^*(t)) \;=\; t\,\hat L\big|_T,
\qquad G(\rho) := g(\rho)\big|_T .
}\end{equation} Strict convexity on \(U\) makes \(\sigma\) the unique
zero of the tangent first-order condition \(G(\rho)=0\) in \(U\), so at
\(t = 0\) (\ref{eq-first-order}) forces \(\rho^*(0) = \sigma\).

\emph{Step 2 --- implicit function theorem and \(\dot\rho^*(0)\).}
Define \(\Phi(t,\rho) := t\,\hat L|_T - G(\rho)\) on a relative
neighborhood of \((0,\sigma)\), valued in \(T_\sigma\mathcal{S}_n\). It
is \(C^1\) with \(\Phi(0,\sigma) = 0\), and its partial derivative in
\(\rho\) there is \(\partial_\rho\Phi|_{(0,\sigma)} = -\,DG|_\sigma\).
Now \(DG|_\sigma\) is the Fréchet derivative of \(G\) at \(\sigma\),
i.e.~for \(A \in
T_\sigma\mathcal{S}_n\), \[
DG|_\sigma(A) = \bigl(\partial_\rho g|_\sigma (A)\bigr)\big|_T = \mathcal{H}(A),
\] the second differential of \(\mathcal{D}(\cdot\|\sigma)\) at
\(\sigma\) of (\ref{eq-hessian-def}) (the projection \((\cdot)|_T\) is
redundant because
\(\operatorname{tr}\mathcal{H}(A) = \langle I, \mathcal{H}(A)\rangle =
\langle \mathcal{H}(I), A\rangle = 0\) by self-adjointness and
\(\mathcal{H}(I) = 0\)). So \(\partial_\rho\Phi|_{(0,\sigma)} =
-\mathcal{H}\), invertible on \(T_\sigma\mathcal{S}_n\) by hypothesis,
and the implicit function theorem yields a unique \(C^1\) curve
\(t \mapsto \tilde\rho(t)\) with \(\tilde\rho(0) = \sigma\) solving
\(\Phi(t,\tilde\rho(t)) = 0\); it is the maximizer,
\(\tilde\rho(t) = \rho^*(t)\) for small \(|t|\), because \(G\) is
injective on the convex set \(U\) (strict convexity gives
\(\langle G(\rho_1) - G(\rho_2), \rho_1 - \rho_2\rangle > 0\) for
\(\rho_1 \neq \rho_2\) in \(U\)), so (\ref{eq-first-order}) has at most
one solution in \(U\), and both \(\rho^*(t)\) and \(\tilde\rho(t)\) lie
in \(U\) and solve it. Differentiating \(G(\rho^*(t)) = t\,\hat L|_T\)
at \(t = 0\) gives \(\mathcal{H}(\dot\rho^*(0)) = \hat L|_T\), whence
\begin{equation}\phantomsection\label{eq-rho-dot}{
\dot\rho^*(0) \;=\; \mathcal{H}^{-1}(\hat L\big|_T) \;=\; \mathcal{H}^{-1}(\hat L) \;\in\; T_\sigma\mathcal{S}_n
}\end{equation} by the extension convention
\(\mathcal{H}^{-1}(\hat L) = \mathcal{H}^{-1}(\hat L|_T)\).

\emph{Step 3 --- envelope theorem.} The feasible set \(\mathcal{S}_n\)
is compact and independent of \(t\), \(\partial_t\varphi(t,\rho) =
\operatorname{tr}(\hat L\rho)\) is jointly continuous, and the maximizer
\(\rho^*(t)\) is unique for \(|t| < \delta\). Danskin's theorem
\citep[Thm.~4.13]{bonnans-shapiro2000} therefore gives, with no
differentiability of \(\rho^*\) required, \[
F'(t) \;=\; \partial_t\varphi(t,\rho^*(t)) \;=\; \operatorname{tr}(\hat L\,\rho^*(t))
\qquad\text{for } |t| < \delta .
\] In particular \(F'(0) = \operatorname{tr}(\hat L\sigma) = 0\) because
\(\hat L\) is \(\sigma\)-centered, and since \(\rho^*\) is \(C^1\) near
\(0\) by Step 2, (\ref{eq-rho-dot}) yields \[
F''(0) \;=\; \operatorname{tr}(\hat L\,\dot\rho^*(0)) \;=\; \operatorname{tr}\bigl(\hat L\,\mathcal{H}^{-1}(\hat L)\bigr) \;=\; \langle \hat L|_T, \mathcal{H}^{-1}(\hat L|_T)\rangle \;>\; 0,
\] the \(I\)-component of \(\hat L\) not contributing since
\(\mathcal{H}^{-1}(\hat L) \in T_\sigma\mathcal{S}_n\), and positivity
holding because \(\hat L|_T \neq 0\) when \(L\) is not a multiple of
\(I\). Feeding \(F(0) = \varphi(0,\sigma) = 0\), \(F'(0) = 0\) and
\(F''(0)\) into the square-root law of Proposition~\ref{prp-square-root}
gives (\ref{eq-sqrt-law-hessian}). \(\square\)
\end{proof}

\subsection{The sandwiched Hessian}\label{sec-hessian-sandwiched}

We now recompute this Hessian for \(\mathcal{D} = H_\alpha\): it is the
metric of Takahashi and Fujiwara \citep[Lemma
4]{takahashi-fujiwara2017}, obtained by the same divided-difference
computation in \citep[Cor. 21]{wilde2025fisher}. The short computation
fixes the DRO normalization (\(H_\alpha\) rather than \(D_\alpha\)) and
produces the mean functions \(m_\alpha = 1/h_\alpha\) used throughout.

\begin{proposition}[Sandwiched Fisher kernel
\citep{takahashi-fujiwara2017, wilde2025fisher}]\protect\hypertarget{prp-hessian-sandwiched}{}\label{prp-hessian-sandwiched}

Let \(\sigma = \sum_i p_i\,|i\rangle\langle i| > 0\) and
\(\alpha \in [1/2, \infty) \setminus \{1\}\). For
\(A \in T_\sigma\mathcal{S}_n\),
\begin{equation}\phantomsection\label{eq-hessian-explicit}{
\frac{d^2}{d\epsilon^2}\,
H_\alpha(\sigma+\epsilon A\,\|\,\sigma)\Big|_{\epsilon=0}
\;=\; \sum_{i,j} h_\alpha(p_i,p_j)\,|A_{ij}|^2,
}\end{equation} where, for \(a, b > 0\) with \(a \neq b\),
\begin{equation}\phantomsection\label{eq-fisher-kernel}{
h_\alpha(a,b) \;=\; \frac{\alpha}{\alpha-1}\cdot
  \frac{a^{(\alpha-1)/\alpha} - b^{(\alpha-1)/\alpha}}
       {a^{1/\alpha} - b^{1/\alpha}}\cdot
  (ab)^{(1-\alpha)/\alpha},
}\end{equation} with \(h_\alpha(a,a) = \alpha/a\). The formula extends
continuously at \(\alpha = 1\) to
\(h_1(a,b) = (\log a - \log b)/(a-b)\), agreeing with the Hessian of the
Umegaki relative entropy \(H_1\).

\end{proposition}

\begin{proof}
Set \(\gamma := (1-\alpha)/(2\alpha) = -\beta\) as in
(\ref{eq-exponent-conventions}), \(X_0 := \sigma^{1/\alpha}\)
(eigenvalues \(\pi_i := p_i^{1/\alpha}\)), and
\(Y := \sigma^\gamma A\,\sigma^\gamma\). The identity
\(2\gamma + 1 = 1/\alpha\) gives
\(\sigma^\gamma\,\sigma\,\sigma^\gamma = \sigma^{1/\alpha} = X_0\),
hence \[
\sigma^\gamma(\sigma+\epsilon A)\sigma^\gamma \;=\; X_0 + \epsilon\,Y,
\qquad
Q_\alpha(\sigma+\epsilon A\|\sigma) \;=\; \operatorname{tr}\bigl[(X_0 + \epsilon Y)^\alpha\bigr].
\] Since \(H_\alpha = (Q_\alpha-1)/(\alpha-1)\) and
\(Q_\alpha(\sigma\|\sigma) = 1\), it suffices to expand the right side
to second order in \(\epsilon\). The first variation is
\(\alpha\operatorname{tr}[X_0^{\alpha-1}Y] = \alpha\operatorname{tr}[\sigma^{(\alpha-1)/\alpha + 2\gamma}A] = \alpha\operatorname{tr}A = 0\).
For the second, differentiate
\(\alpha\operatorname{tr}[(X_0+\epsilon Y)^{\alpha-1}Y]\) once more via
the Daleckiĭ--Krein formula \citep[Thm.~V.3.3]{bhatia1997} for
\(f'(x) = \alpha x^{\alpha-1}\) (smooth near the spectrum of
\(X_0 > 0\)): \[
\frac{d^2}{d\epsilon^2}\operatorname{tr}\bigl[(X_0+\epsilon Y)^\alpha\bigr]\Big|_0
\;=\; \sum_{i,j} f^{[1,1]}(\pi_i,\pi_j)\,|Y_{ij}|^2,
\qquad
f^{[1,1]}(a,b) := \frac{f'(a)-f'(b)}{a-b},\quad f^{[1,1]}(a,a) = f''(a).
\] With \(\pi_i^{\alpha-1} = p_i^{(\alpha-1)/\alpha}\) and, in the
eigenbasis of \(\sigma\),
\(|Y_{ij}|^2 = (p_ip_j)^{2\gamma}|A_{ij}|^2 = (p_ip_j)^{(1-\alpha)/\alpha}|A_{ij}|^2\),
dividing by \(\alpha - 1\) gives \[
\frac{d^2 H_\alpha}{d\epsilon^2}\bigg|_0
\;=\; \frac{\alpha}{\alpha-1}\sum_{i,j}\frac{p_i^{(\alpha-1)/\alpha} - p_j^{(\alpha-1)/\alpha}}
                             {p_i^{1/\alpha} - p_j^{1/\alpha}}\,(p_ip_j)^{(1-\alpha)/\alpha}\,|A_{ij}|^2
\;=\; \sum_{i,j} h_\alpha(p_i,p_j)\,|A_{ij}|^2,
\] the diagonal \(h_\alpha(a,a) = \alpha/a\) coming from
\(f''(\pi) = \alpha(\alpha-1)\pi^{\alpha-2}\) at \(\pi = a^{1/\alpha}\).
At \(\alpha = 1\), L'Hôpital in (\ref{eq-fisher-kernel}) gives
\(h_1(a,b) = (\log a - \log b)/(a-b)\), and the same computation for
\(\operatorname{tr}\rho\log\rho\) (Daleckiĭ--Krein for \(\log\); the
term \(-\operatorname{tr}\rho\log\sigma\) is linear) gives the
Bogoliubov--Kubo--Mori Hessian \(\sum_{ij}h_1(p_i,p_j)|A_{ij}|^2\)
\citep[Ex.~7.18]{petz1993, hiai-petz2014}. \(\square\)
\end{proof}

In detail, \(h_\alpha/\alpha\) is the Takahashi--Fujiwara metric
\citep{takahashi-fujiwara2017}, to which the \(\alpha\)-\(z\)
information matrices of Wilde \citep{wilde2025fisher} specialize at
\(z = \alpha\); the difference of normalizations (\(\log Q_\alpha\)
against \(Q_\alpha - 1\)) vanishes to second order at \(\rho = \sigma\).
The reciprocal kernel is a classical object: multiplying numerator and
denominator of (\ref{eq-fisher-kernel}) by \((ab)^{(\alpha-1)/\alpha}\)
shows \begin{equation}\phantomsection\label{eq-power-difference}{
\frac{\alpha}{h_\alpha(a,b)} \;=\; M_{1/\alpha}(a,b),
\qquad
M_r(a,b) := \frac{r-1}{r}\cdot\frac{a^{r}-b^{r}}{a^{r-1}-b^{r-1}},
}\end{equation} the \emph{power-difference mean} of Hiai and Kosaki
\citep[§2]{hiai-kosaki1999} (limits at \(r \in \{0,1\}\) and on the
diagonal understood by continuity).

Write \(m_\alpha := 1/h_\alpha\) for the reciprocal kernel, the
\textbf{DRO mean function} of the sandwiched Rényi family.

\begin{corollary}[Inverse of the sandwiched
Hessian]\protect\hypertarget{cor-hessian-superop}{}\label{cor-hessian-superop}

For \(\alpha \in [1/2, \infty)\) the Hessian \(\mathcal{H}_\alpha\) of
\(H_\alpha(\cdot\|\sigma)\) at \(\sigma\) is positive definite on
\(T_\sigma\mathcal{S}_n\), and for every Hermitian \(B\) with
\(\operatorname{tr}(\sigma B) = 0\), in the eigenbasis of \(\sigma\),
\begin{equation}\phantomsection\label{eq-hessian-inverse-diagonal}{
\mathcal{H}_\alpha^{-1}(B)_{ij} \;=\; m_\alpha(p_i, p_j)\,B_{ij}.
}\end{equation}

\end{corollary}

\begin{proof}
By (\ref{eq-hessian-explicit}),
\(\langle A, \mathcal{H}_\alpha A\rangle = \sum_{ij} h_\alpha(p_i,p_j)|A_{ij}|^2\)
on \(T_\sigma\mathcal{S}_n\), and \(h_\alpha > 0\) on \((0,\infty)^2\)
for \(\alpha \geq 1/2\) (for \(\alpha < 1\) both the prefactor and the
divided difference in (\ref{eq-fisher-kernel}) are negative); this is
positivity. With \(K_\alpha\) the Schur multiplier by
\(h_\alpha(p_i,p_j)\) and \(P_T(X) = X|_T\),
\(\mathcal{H}_\alpha = P_T K_\alpha P_T\) on \(T_\sigma\mathcal{S}_n\)
(the multiplier itself does not preserve tracelessness). Put
\(X := m_\alpha \circ B\) (entrywise product): since
\(m_\alpha(a,a) = a/\alpha\),
\(\operatorname{tr} X = \frac{1}{\alpha}\operatorname{tr}(\sigma B) = 0\),
so \(X \in T_\sigma\mathcal{S}_n\) and \(P_TK_\alpha(X) = B|_T\),
i.e.~\(X = \mathcal{H}_\alpha^{-1}(B|_T) = \mathcal{H}_\alpha^{-1}(B)\)
under the extension convention of Section~\ref{sec-hessian-bridge} ---
which is why \(\hat L\) is \(\sigma\)-centered. \(\square\)
\end{proof}

\begin{refremark}[Monotonicity of the Hessian metric]
For each \(\alpha \in [1/2, \infty)\) the rescaled kernel
\(h_\alpha/\alpha\) is a Petz monotone metric on \(\mathcal{S}_n\), with
representing function \(\mathfrak{f}_\alpha(t) = M_{1/\alpha}(t,1)\)
\citep[Thm.~5.7]{petz1996, hiai-petz2014} in the power-difference family
of (\ref{eq-power-difference}); it is monotone \emph{if and only if}
\(\alpha \in (-\infty,-1] \cup [1/2,\infty)\) \citep[Prop.
4.2]{takahashi-fujiwara2017, hiai-kosaki1999}, so the DPI threshold
\(\alpha = 1/2\) of the divergence is exactly the monotonicity threshold
of its Hessian.

\label{rem-monotone-metric}

\end{refremark}

\begin{corollary}[DRO variance for
\(H_\alpha\)]\protect\hypertarget{cor-dro-variance-sandwiched}{}\label{cor-dro-variance-sandwiched}

For \(\alpha \in [1/2, \infty)\) the DRO variance \(s_\alpha^2\) of
Proposition~\ref{prp-square-root} for \(\mathcal{D} = H_\alpha\) is
\begin{equation}\phantomsection\label{eq-dro-var-alpha}{
s_\alpha^2(\hat L)
\;:=\; F''(0)
\;=\; \sum_{i,j} m_\alpha(p_i, p_j)\,|\hat L_{ij}|^2,
\qquad
m_\alpha(a,a) = a/\alpha.
}\end{equation} In the commutative case \([\sigma, L] = 0\),
\(s_\alpha^2(\hat L) = (1/\alpha)\operatorname{Var}_\sigma(L)\).

\end{corollary}

\begin{proof}
The hypotheses of Theorem~\ref{thm-hessian-bridge} hold for
\(\mathcal{D} = H_\alpha\):
\(Q_\alpha(\rho\|\sigma) = \operatorname{tr}(\sigma^{-\beta}\rho\sigma^{-\beta})^\alpha\)
is \(C^\infty\) near \(\sigma\), where the sandwiched argument is
positive definite (likewise \(\operatorname{tr}\rho\log\rho\) at
\(\alpha = 1\), with kernel the continuous extension \(h_1\)), and
positivity of \(\mathcal{H}_\alpha\) is
Corollary~\ref{cor-hessian-superop}. Since
\(\operatorname{tr}(\sigma\hat L) = 0\),
(\ref{eq-hessian-inverse-diagonal}) gives
\(\mathcal{H}_\alpha^{-1}(\hat L)_{ij} = m_\alpha(p_i,p_j)\hat L_{ij}\),
and the pairing
\(\operatorname{tr}(\hat L\,\mathcal{H}_\alpha^{-1}(\hat L))\) in the
eigenbasis of \(\sigma\) is (\ref{eq-dro-var-alpha}); the commutative
case reads off the diagonal \(m_\alpha(p_i,p_i) = p_i/\alpha\).
\(\square\)
\end{proof}

\begin{refremark}[Quantum Fisher information and Cramér--Rao]
\(s_\alpha^2(\hat L) = \langle \hat L, \mathcal{H}_\alpha^{-1}(\hat L)\rangle\)
is the squared \emph{co-metric} norm of the differential of
\(\rho \mapsto \operatorname{tr}(L\rho)\) in the \(\alpha\)-information
metric: for any \(C^1\) curve \(\rho_t\) through \(\rho_0 = \sigma\),
Cauchy--Schwarz gives \[
\Bigl|\tfrac{d}{dt}\operatorname{tr}(L\rho_t)\big|_{t=0}\Bigr|^2
\;\leq\; \langle\dot\rho_0, \mathcal{H}_\alpha(\dot\rho_0)\rangle\;s_\alpha^2(\hat L),
\] with equality iff
\(\dot\rho_0 \propto \mathcal{H}_\alpha^{-1}(\hat L)\), the worst-case
direction (\ref{eq-rho-dot}); the square-root law
(\ref{eq-sqrt-law-hessian}) is this Cramér--Rao inequality integrated
over the \(\alpha\)-information ball. At \(\alpha = 1/2\) it is the SLD
quantum Cramér--Rao bound \citep{helstrom1976, braunstein-caves1994}, at
\(\alpha = 1\) the bound for the Kubo--Mori (Bogoliubov) metric
\citep{petz1993}; for a general monotone metric and its covariance the
inequality is in \citep{petz2002covariance}.

\label{rem-cramer-rao}

\end{refremark}

\begin{corollary}[Variance of the spectral
measure]\protect\hypertarget{cor-spectral-variance}{}\label{cor-spectral-variance}

For \(\alpha \in [1/2, \infty)\) let \(\omega_\alpha\) be the spectral
measure of Theorem~\ref{thm-spectral-rep}. Then
\begin{equation}\phantomsection\label{eq-spectral-variance}{
\mathbb{E}_{\omega_\alpha}[\ell] = \operatorname{tr}(L\sigma),
\qquad
\operatorname{Var}_{\omega_\alpha}(\ell)
\;=\; \alpha\,s_\alpha^2(\hat L)
\;=\; \alpha\sum_{i,j} m_\alpha(p_i, p_j)\,|\hat L_{ij}|^2.
}\end{equation}

\end{corollary}

\begin{proof}
Both sides of the variance identity are the coefficient of \(\sqrt{2q}\)
in \(V_\alpha(q) = V_\alpha^{\mathrm{cl}}(q;\omega_\alpha)\)
(\ref{eq-spectral-dro}); on the quantum side
\(V_\alpha(q) = \operatorname{tr}(L\sigma) + s_\alpha(\hat L)\sqrt{2q} + o(\sqrt q)\)
by Theorem~\ref{thm-hessian-bridge} and
Corollary~\ref{cor-dro-variance-sandwiched}, while on the classical side
\(H_\alpha^{\mathrm{cl}}\) is the \(f\)-divergence with
\(f(t) = (t^\alpha-1)/(\alpha-1)\), \(f''(1) = \alpha\), for which
\(V^{\mathrm{cl}}_\alpha(q;\nu) = \mathbb{E}_\nu[\ell] + \sqrt{2q\operatorname{Var}_\nu(\ell)/\alpha} + o(\sqrt q)\)
\citetext{\citealp[Thm.
3.1]{lam2016robust}; \citealp{duchi-namkoong2019}; \citealp{duchi-glynn-namkoong2021}}.
The mean is the identity of Theorem~\ref{thm-spectral-rep}. \(\square\)
\end{proof}

\subsection{Examples}\label{sec-examples}

The function \(m_\alpha\) at the three distinguished values
\(\alpha \in \{1/2, 1, 2\}\) recovers three classical means of
\(a, b > 0\), identified at the level of rescaled kernels
\(h_\alpha/\alpha\) with normalized Petz monotone metrics on
\(\mathcal{S}_n\).

\textbf{\(\alpha = 1\) (Kubo--Mori).} \(m_1\) is the logarithmic mean,
\(s_1^2(\hat L) = \int_0^1 \operatorname{tr}(\hat L\sigma^s\hat L\sigma^{1-s})\,ds\)
is the Kubo--Mori--Bogoliubov variance, and \(h_1\) generates the
Kubo--Mori--Bogoliubov metric
\citep{petz1996, hasegawa1993, lesniewski-ruskai1999}.

\textbf{\(\alpha = 1/2\) (Bures).} \(h_{1/2}(a,b) = 1/(a+b)\), so
\(s_{1/2}^2(\hat L) = 2\operatorname{Var}_\sigma(L)\), twice the Born
variance: \(H_{1/2}\) is the squared Bures distance
\citep{bures1969, uhlmann1976, fuchs1999}, its Hessian twice the Bures
metric, and the rescaled kernel \(2/(a+b)\) is the SLD/Bures metric
\citep{helstrom1976, braunstein-caves1994}.

\textbf{\(\alpha = 2\) (geometric mean).} At \(\alpha = 2\) both
\(1/\alpha\) and \((\alpha-1)/\alpha\) equal \(1/2\), the
divided-difference factor in (\ref{eq-fisher-kernel}) is identically
\(1\), and \[
h_2(a,b) \;=\; \frac{2}{\sqrt{ab}},
\qquad
m_2(a,b) \;=\; \tfrac{1}{2}\sqrt{ab},
\qquad m_2(a,a) = a/2.
\] Hence
\(s_2^2(\hat L) = \tfrac{1}{2}\operatorname{tr}[\sigma^{1/2}\hat L\,\sigma^{1/2}\hat L]\),
the quantity \(s_2^2\) of (\ref{eq-s2-def}), so the square-root law at
\(\alpha = 2\) is the small-\(q\) face of the exact formula
(\ref{eq-alpha-two-exact}); the same trace is the second moment of
\(\omega_2\) (Corollary~\ref{cor-spectral-variance}), whose deficit
against the Born variance is the Wigner--Yanase skew information
\citep{wigner-yanase1963},
\begin{equation}\phantomsection\label{eq-alpha-two-WY}{
\operatorname{Var}_\sigma(L) - \operatorname{Var}_{\omega_2}(\ell)
= \operatorname{tr}(\sigma L^2) -
\operatorname{tr}(\sigma^{1/2}L\sigma^{1/2}L)
= -\tfrac{1}{2}\operatorname{tr}[\sigma^{1/2}, L]^2
=: I_{\mathrm{WY}}(\sigma, L),
}\end{equation} non-negative and zero iff \([L, \sigma] = 0\). The
rescaled kernel \(h_2/\alpha = 1/\sqrt{ab}\) has representing function
\(\mathfrak{f}_2(t) = \sqrt{t}\), the \textbf{geometric-mean Petz
metric} (the Petz metric associated with the operator-monotone function
\(f(t)=\sqrt t\)) \citep{petz1996, hiai-kosaki1999}. This metric should
not be confused with the standard Wigner--Yanase monotone metric; the
Wigner--Yanase quantity that appears here is instead the variance
deficit in (\ref{eq-alpha-two-WY}).

\textbf{\(\alpha \to \infty\).} Since \(m_\alpha(a,b) \to 0\) pointwise,
\(s_\alpha^2(\hat L) \to 0\) and the square-root law degenerates:
\(H_\infty(\rho\|\sigma) = \|\sigma^{-1/2}\rho\sigma^{-1/2}\|_\infty - 1\)
is a cone at \(\rho = \sigma\),
\(H_\infty(\sigma+\epsilon A\|\sigma) = \epsilon\,\lambda_{\max}(\sigma^{-1/2}A\sigma^{-1/2})\),
not a paraboloid, and \(V_\infty(q)\) grows linearly in \(q\). The
spectral measure does not degenerate ---
\(\operatorname{Var}_{\omega_\alpha}(\ell) \to \operatorname{Var}_{\omega_\infty}(\ell) > 0\)
(Theorem~\ref{thm-variance-ordering}) --- only the budget-normalized
coefficient \(\operatorname{Var}_{\omega_\alpha}(\ell)/\alpha\)
vanishes, and the exact CVaR formula (\ref{eq-cvar-classical}) takes
over.

\begin{refremark}[Higher-order terms]
The expansion can be continued. By classicalization, every further term
is that of the classical small-radius expansion of the \(f\)-divergence
DRO of \(\omega_\alpha\) --- \citep{he-lam2025} for general \(f\),
\citep{lam2016robust} for the relative entropy,
\citep{gotoh-kim-lim2018} in penalty form --- evaluated on the moments
of \(\omega_\alpha\): the coefficient of \(q\) is a skewness correction,
proportional to \((2-\alpha)\) and to the third central moment of
\(\omega_\alpha\), which vanishes at \(\alpha = 2\) in agreement with
Corollary~\ref{cor-alpha-two-exact}; the same coefficient can be reached
intrinsically from the third derivative of \(H_\alpha(\cdot\|\sigma)\)
at \(\sigma\).

\label{rem-higher-order}

\end{refremark}

\subsection{Variance ordering}\label{sec-variance-ordering}

Comparing worst-case fluctuations across \(\alpha\) requires a
normalization: \(H_\alpha\) has classical curvature
\(f_\alpha''(1) = \alpha\), so a budget-\(q\) ball has a different local
size at each \(\alpha\), and comparing \(s_\alpha^2\) at equal \(q\)
conflates this scaling with the effect of interest. The clean comparison
rescales to unit curvature --- the Petz normalization
\(H_\alpha/\alpha\), under which every member of the family restricts to
the \emph{same} classical Fisher information on commuting directions
\citep{takahashi-fujiwara2017} --- where the DRO variance is the
spectral variance
\(\operatorname{Var}_{\omega_\alpha}(\ell) = \alpha\,s_\alpha^2(\hat L)\)
of Corollary~\ref{cor-spectral-variance}, equal to
\(\operatorname{Var}_\sigma(L)\) for every \(\alpha\) in the commuting
case; any ordering across \(\alpha\) is then a pure noncommutativity
effect. The normalized mean function is the power-difference mean
\(\alpha\,m_\alpha = M_{1/\alpha}\) of (\ref{eq-power-difference}) ---
the ladder \(M_2\) (arithmetic), \(M_1\) (logarithmic), \(M_{1/2}\)
(geometric), \(M_0 = ab\log(a/b)/(a-b)\) of Table~\ref{tbl-metrics} read
in the variable \(1/\alpha\) --- for which Hiai--Kosaki prove the
pointwise order \(M_r \le M_{r'}\), \(r < r'\) \citep[Thm.
2.1]{hiai-kosaki1999}. The ordering on \([1/2,\infty)\) is reproved in
Lemma~\ref{lem-hyperbolic-mean} by a log-convexity argument in the
hyperbolic parametrization of \citep[proof of Thm.
2.1]{hiai-kosaki1999}, because it also gives strictness off the diagonal
and the equality cases, which we did not find stated.

\begin{lemma}[Hyperbolic representation of
\(m_\alpha\)]\protect\hypertarget{lem-hyperbolic-mean}{}\label{lem-hyperbolic-mean}

For \(a, b > 0\) put \(u := \tfrac{1}{2}\log(a/b)\) and \[
\Psi(x) \;:=\; \frac{\sinh(xu)}{x} \;=\; \frac{1}{2}\int_{-u}^{u} e^{xt}\,dt,
\qquad \Psi(0) := u.
\] Then for all \(\alpha \in [1/2, \infty)\), with
\(\xi := 1/\alpha \in (0,2]\),
\begin{equation}\phantomsection\label{eq-hyperbolic-rep}{
\alpha\,m_\alpha(a,b) \;=\; \sqrt{ab}\;\frac{\Psi(\xi)}{\Psi(1-\xi)},
}\end{equation} the value at \(a = b\) understood by continuity.
Consequently, for \(a \neq b\),
\begin{equation}\phantomsection\label{eq-am-bound}{
\frac{ab\,\log(a/b)}{a-b} \;<\; \alpha\,m_\alpha(a,b) \;\leq\; \frac{a+b}{2},
}\end{equation} with equality on the right iff \(\alpha = 1/2\), the
left value attained as \(\alpha \to \infty\); and
\(\alpha \mapsto \alpha\,m_\alpha(a,b)\) is strictly decreasing on
\([1/2,\infty)\).

\end{lemma}

The proof is in \hyperref[sec-app-moment-tools]{Appendix A}.

Since \(\Psi\) is even, (\ref{eq-hyperbolic-rep}) gives
\(\alpha' m_{\alpha'}(a,b) = \bigl(\alpha\,m_\alpha(a^{-1},b^{-1})\bigr)^{-1}\)
for the Hölder conjugate \(\alpha'\) --- the duality
\(M_{1-r} = M_r^{(-)}\) of \citep[p.~907]{hiai-kosaki1999} --- which
pairs Bures (\(\alpha = 1/2\)) with the RLD kernel (\(\alpha' = -1\)),
Kubo--Mori with the CVaR kernel (\(\alpha' = \infty\)), and the
geometric mean with itself; hence the \(\alpha \to \infty\) row of
Table~\ref{tbl-metrics}.

\begin{table}[htbp]

\caption{\label{tbl-metrics}DRO mean functions \(m_\alpha = 1/h_\alpha\)
for the sandwiched Rényi family, and the rescaled kernels
\(h_\alpha/\alpha\) identifying the corresponding normalized Petz
monotone metric.}

\centering{

\begin{tabular}{@{}ccc@{}}
\toprule
$\alpha$ & $m_\alpha(a,b)$ & Rescaled kernel $h_\alpha/\alpha$ \\
\midrule
$1/2$ & $a+b$ & SLD/Bures: $2/(a+b)$ \\
$1$ & $(a-b)/(\log a - \log b)$ & Kubo--Mori: $(\log a-\log b)/(a-b)$ \\
$2$ & $\sqrt{ab}/2$ & Geometric: $1/\sqrt{ab}$ \\
$\to\infty$ & $\alpha m_\alpha \downarrow ab\log(a/b)/(a-b)$ & harmonic-log: $m_1(a^{-1},b^{-1})$ \\
\bottomrule
\end{tabular}

}

\end{table}%

\begin{theorem}[Variance
ordering]\protect\hypertarget{thm-variance-ordering}{}\label{thm-variance-ordering}

Let \(\hat L \neq 0\). The map
\(\alpha \mapsto \operatorname{Var}_{\omega_\alpha}(\ell)\) is
non-increasing on \([1/2, \infty]\), constant if \([\sigma, L] = 0\) and
strictly decreasing otherwise, with
\begin{equation}\phantomsection\label{eq-var-omega-infty}{
\operatorname{Var}_\sigma(L) \;=\; \operatorname{Var}_{\omega_{1/2}}(\ell)
\;\geq\; \operatorname{Var}_{\omega_\alpha}(\ell)
\;\geq\; \operatorname{Var}_{\omega_\infty}(\ell)
\;=\; \sum_{i,j}\frac{p_i p_j\,\log(p_i/p_j)}{p_i - p_j}\,|\hat L_{ij}|^2 \;>\; 0
}\end{equation} (kernel value \(p_i\) at \(p_i = p_j\)).

\end{theorem}

Here \(\omega_{1/2} = \nu_L\) is the Born measure (\ref{eq-omega-half}),
so the left end is
\(\operatorname{Var}_{\nu_L}(\ell) = \operatorname{tr}(\sigma\hat L^2)\),
and \(\omega_\infty\) is the CVaR measure of Section~\ref{sec-stoploss};
in the commuting case all variances equal
\(\operatorname{Var}_\sigma(L)\). That a variance built from a monotone
metric lies below the Born variance, with a deficit of skew-information
type, is known
\citep{petz2002covariance, gibilisco-hiai-petz2009, hansen2008skew, luo2003},
and the monotonicity in \(\alpha\) is \citep[Thm. 2.1]{hiai-kosaki1999}
for the kernels and \citep[Thm. 30]{wilde2025fisher} for the sandwiched
information matrices; the theorem adds strictness with the equality
cases, and the fact that these are the variances of the laws
\(\omega_\alpha\), with \(\omega_\infty\) the CVaR measure at the right
end.

\begin{proof}
By Corollary~\ref{cor-spectral-variance},
\(\operatorname{Var}_{\omega_\alpha}(\ell) = \sum_{ij} \alpha\,m_\alpha(p_i,p_j)\,|\hat L_{ij}|^2\).
By Lemma~\ref{lem-hyperbolic-mean} each kernel entry
\(\alpha\,m_\alpha(p_i,p_j)\) is strictly decreasing in \(\alpha\) when
\(p_i \neq p_j\) and constant (\(= p_i\)) when \(p_i = p_j\); hence the
sum is strictly decreasing iff \(\hat L_{ij} \neq 0\) for some pair with
\(p_i \neq p_j\), i.e.~iff \([\sigma, L] \neq 0\), and in the commuting
case a joint eigenbasis gives
\(\operatorname{Var}_{\omega_\alpha} = \operatorname{Var}_\sigma(L)\)
for all \(\alpha\) (with \(\omega_\alpha = \sum_i p_i \delta_{l_i}\)
independent of \(\alpha\)).

\emph{Left endpoint.} At \(\alpha = 1/2\) the upper bound of
(\ref{eq-am-bound}) is an equality, \(\alpha\,m_\alpha(a,b) = (a+b)/2\),
so the sum is
\(\sum_{ij}\tfrac{p_i+p_j}{2}|\hat L_{ij}|^2 = \operatorname{tr}(\sigma \hat L^2) = \operatorname{Var}_\sigma(\hat L)\),
agreeing with the direct evaluation
\(\operatorname{Var}_{\nu_L}(\ell) = \operatorname{tr}(\sigma\hat L^2)\)
against \(\omega_{1/2} = \nu_L\).

\emph{Right endpoint.} By Lemma~\ref{lem-hyperbolic-mean},
\(\alpha\,m_\alpha(a,b) \downarrow ab\log(a/b)/(a-b)\) as
\(\alpha \to \infty\) (value \(a\) at \(a = b\)), so the finite sum
\(\sum_{ij}\alpha\,m_\alpha(p_i,p_j)|\hat L_{ij}|^2\) converges to the
right side of (\ref{eq-var-omega-infty}), which is strictly positive for
\(\hat L \neq 0\) since every kernel entry is. This limit is
\(\operatorname{Var}_{\omega_\infty}(\ell)\) because
\(\omega_\alpha \to \omega_\infty\) weakly as \(\alpha \to \infty\): by
(\ref{eq-classicalization-alpha-big}) and (\ref{eq-cvar-classical}),
\(\int(\ell-c)_+^{\alpha'}d\omega_\alpha = \operatorname{tr}[\sigma^\beta(L-cI)\sigma^\beta]_+^{\alpha'} \to \operatorname{tr}[\sigma^{1/2}(L-cI)\sigma^{1/2}]_+ = \int(\ell-c)_+\,d\omega_\infty\)
for every \(c\) (continuity of eigenvalues as \(\beta \to \tfrac12\),
\(\alpha' \to 1\)), while \(x^{\alpha'} \to x\) uniformly on
\([0, l_n - l_1]\) and the functions \(1\), \((\cdot - c)_+\) span a
dense subspace of \(C([l_1,l_n])\), which carries every
\(\omega_\alpha\); the second moment is continuous under weak
convergence. \(\square\)
\end{proof}

\section{Concentration and distinguishability: the quantum
deflation}\label{sec-concentration}

The quantum DRO value lies \emph{below} its measured shadow. Let
\(\nu_L =
\sum_i \langle i|\sigma|i\rangle\,\delta_{l_i}\) be the Born measure of
\(L\) in the state \(\sigma\) (\ref{eq-omega-half}) and \(V^{\rm
meas}_\alpha(q) := V^{\rm cl}_\alpha(q;\nu_L)\) its classical DRO, the
\textbf{measured DRO}: then \(V_\alpha(q) \leq V^{\rm meas}_\alpha(q)\)
for every budget, with equality for all \(q\) only if \([\sigma,L] = 0\)
or \(\alpha = \tfrac12\). The deficit is noncommutativity, and this
section shows that it is seen at three levels: pointwise in the dual
variable, where, as in \citep{berta2017}, Golden--Thompson and
Araki--Lieb--Thirring give the comparison with the measured problem
(Proposition~\ref{prp-born-domination}); in the variance, where
\(\operatorname{Var}_{\omega_\alpha}(\ell) < \operatorname{Var}_\sigma(L)\)
strictly for \(\alpha > \tfrac12\) when \([\sigma,L] \neq 0\)
(Theorem~\ref{thm-variance-ordering}); and, non-asymptotically, in every
concentration bound that reads the second moment --- the spectral
measure \(\omega_\alpha\) of Theorem~\ref{thm-spectral-rep} is a
\emph{deflated} Born measure, and a Bernstein bound built on it is never
larger than the same bound built on \(\nu_L\), and strictly smaller
below saturation when \([\sigma,L]\ne0\) and \(\alpha\in(\tfrac12,2]\)
(Proposition~\ref{prp-deflation}). The classical layer is standard ---
the \(\varphi\)-divergence DRO duals
\citep{ben-tal-etal2013, breuer-csiszar2016}, the concentration
inequalities of Hoeffding, Bernstein and Bennett
\citep{hoeffding1963, bennett1962, boucheron-lugosi-massart2013}, and
the Markov--Krein theory of Chebyshev systems
\citep{karlin-studden1966, krein-nudelman1977} and the stochastic orders
\citep{shaked-shanthikumar2007}, whose proofs are collected in
\hyperref[sec-app-moment-tools]{Appendix A} --- and the quantum input is
the transform identities of Section~\ref{sec-spectral-measure} together
with the deflated variance. Throughout,
\(\mu := \operatorname{tr}(L\sigma)\) is the common mean of
\(\omega_\alpha\) and \(\nu_L\), \(\hat L = L - \mu I\), \(a
:= l_n - \mu\) and \(b := \mu - l_1\) (both positive when \(L \notin
\mathbb{R}I\)), and every upper bound below is two-sided once applied to
\(-L\) as well (the spectral measure of \(-L\) is the reflection of
\(\omega_\alpha\); (\ref{eq-two-sided})).

One monotonicity is used throughout. Write, for a probability measure
\(\nu\) on \([l_1,l_n]\), \(\Pi(c;\nu) := \int(\ell-c)_+^{\alpha'}
d\nu\) (\(\alpha>1\)), \(M(t;\nu) := \int e^{t\ell}d\nu\) (\(\alpha=1\))
and \(S_p(c;\nu) := \int(c-\ell)^{-p}d\nu\) (\(\alpha<1\)).

\begin{lemma}[Transform
monotonicity]\protect\hypertarget{lem-transform-monotone}{}\label{lem-transform-monotone}

If the transform of \(\nu\) is pointwise dominated by that of \(\nu'\)
--- \(\Pi(c;\nu) \leq \Pi(c;\nu')\) for all \(c \in \mathbb{R}\)
(\(1 < \alpha
\leq \infty\), with \(\alpha' = 1\) at \(\alpha = \infty\)),
\(M(t;\nu) \leq
M(t;\nu')\) for all \(t \geq 0\) (\(\alpha = 1\)), or \(S_p(c;\nu) \leq
S_p(c;\nu')\) for all \(c > l_n\) (\(\alpha < 1\)) --- then
\(V_\alpha^{\rm cl}(q;\nu) \leq V_\alpha^{\rm cl}(q;\nu')\) for every
\(q \geq 0\).

\end{lemma}

\begin{proof}
By Lemma~\ref{lem-classical-dual} the value is an infimum over the dual
variable of an objective that increases with the transform at that
point: \(c +
\kappa\,\Pi(c;\nu)^{1/\alpha'}\) with
\(\kappa = (1+(\alpha-1)q)^{1/\alpha}\) (and \(c + (1+q)\Pi(c;\nu)\) at
\(\alpha = \infty\)), \((q + \log
M(t;\nu))/t\), and \(c - \kappa\,S_p(c;\nu)^{-1/p}\). Pointwise
domination of the objectives passes to the infima. \(\square\)
\end{proof}

\subsection{Born domination}\label{sec-stochastic-order}

\begin{proposition}[Born
domination]\protect\hypertarget{prp-born-domination}{}\label{prp-born-domination}

For every \(\alpha \in [\tfrac12, \infty]\) the trace functional of the
quantum dual (Theorem~\ref{thm-main}, Theorem~\ref{thm-DRO-1},
Theorem~\ref{thm-quantum-cvar}) is pointwise dominated by the
corresponding transform of the Born measure \(\nu_L\):

\begin{enumerate}
\def\labelenumi{(\roman{enumi})}
\item
  for \(\alpha \in (1, \infty)\) and every \(c \in \mathbb{R}\),
  \begin{equation}\phantomsection\label{eq-alt-domination}{
  \operatorname{tr}\bigl[\sigma^\beta(L - cI)\sigma^\beta\bigr]_+^{\alpha'}
  \;\leq\;
  \operatorname{tr}\bigl[\sigma\,\bigl((L - cI)_+\bigr)^{\alpha'}\bigr];
  }\end{equation}
\item
  at \(\alpha = 1\), for every \(t \geq 0\),
  \(\operatorname{tr}\,e^{\log\sigma + tL} \leq
  \operatorname{tr}\bigl[\sigma\,e^{tL}\bigr]\);
\item
  for \(\alpha \in [\tfrac12, 1)\) and every \(c > l_n\),
  \(\operatorname{tr}\bigl[\sigma^\beta(cI-L)\sigma^\beta\bigr]^{-p}
  \leq \operatorname{tr}\bigl[\sigma\,(cI-L)^{-p}\bigr]\);
\item
  at \(\alpha = \infty\), for every \(c \in \mathbb{R}\),
  \(\operatorname{tr}\bigl[\sigma^{1/2}(L - cI)\sigma^{1/2}\bigr]_+
  \leq \operatorname{tr}\bigl[\sigma\,(L - cI)_+\bigr]\), i.e.
  \begin{equation}\phantomsection\label{eq-cx-endpoint}{
  \omega_\infty \;\preceq_{\rm cx}\; \nu_L
  }\end{equation} in the convex order.
\end{enumerate}

Consequently \begin{equation}\phantomsection\label{eq-dpi-value}{
V_\alpha(q) \;\leq\; V^{\rm meas}_\alpha(q)
\qquad \text{for all } q \geq 0,
}\end{equation} with equality for all \(q \geq 0\) if and only if
\(\alpha = \tfrac12\) or \([\sigma, L] = 0\).

\end{proposition}

\begin{proof}
Write \(X = L - cI\). (i) \(\sigma^\beta X\sigma^\beta \leq \sigma^\beta
X_+\sigma^\beta\), so
\(\operatorname{tr}[(\sigma^\beta X\sigma^\beta)_+^{\alpha'}]
\leq \operatorname{tr}[(\sigma^\beta X_+\sigma^\beta)^{\alpha'}] \leq
\operatorname{tr}[\sigma^{2\beta\alpha'}X_+^{\alpha'}]\), by Weyl
monotonicity and the Araki--Lieb--Thirring inequality
\citep{lieb-thirring1976, araki1990}
(\(\operatorname{tr}(B^{1/2}AB^{1/2})^r
\leq \operatorname{tr}(B^{r/2}A^rB^{r/2})\) for \(A,B \geq 0\),
\(r \geq 1\)) with \(r = \alpha'\); and \(2\beta\alpha' = 1\). (ii) is
the Golden--Thompson inequality \citep{golden1965, thompson1965}. (iii)
is Araki--Lieb--Thirring with \(A = (cI-L)^{-1}\), \(B =
\sigma^{-2\beta}\), \(r = p\), since \(-2\beta p = 1\). (iv) is the Weyl
step of (i) with \(\alpha' = 1\); its two sides are the order-one
stop-loss transforms of \(\omega_\infty\) and \(\nu_L\), two laws with
equal means, so it is the convex order. The dual objectives of
Theorem~\ref{thm-main}, Theorem~\ref{thm-DRO-1} and
Theorem~\ref{thm-quantum-cvar} are increasing functions of the left
sides, and those of Lemma~\ref{lem-classical-dual} for \(\nu_L\) the
same functions of the right sides, so (\ref{eq-dpi-value}) follows by
taking infima. If \(\alpha = \tfrac12\) or \([\sigma,L] = 0\) then
\(\omega_\alpha = \nu_L\) (Theorem~\ref{thm-spectral-rep}) and equality
holds. Conversely, equality in (\ref{eq-dpi-value}) for all small \(q\)
forces equal square-root coefficients
(Corollary~\ref{cor-spectral-variance}), i.e.
\(\operatorname{Var}_{\omega_\alpha}(\ell) = \operatorname{Var}_{\nu_L}(\ell)\),
and at \(\alpha = \infty\) equality of the two CVaRs at every level
forces \(\omega_\infty = \nu_L\); either contradicts
Theorem~\ref{thm-variance-ordering} when \(\alpha > \tfrac12\) and
\([\sigma,L] \neq
0\). \(\square\)
\end{proof}

The proposition is the dual-variable form of the comparison of the
measured and the sandwiched divergences by Berta, Fawzi and Tomamichel
\citep[Prop. 5 and Thm. 6]{berta2017}, through the same two trace
inequalities, Golden--Thompson at \(\alpha = 1\) and
Araki--Lieb--Thirring otherwise, with equality only for commuting pairs;
(\ref{eq-dpi-value}) is the data-processing inequality
\citep{frank-lieb2013, beigi2013} for the measurement of \(L\), and at
\(\alpha = 1\) item (ii) with (\ref{eq-dpi-value}) is the robustness
bound of Yamamoto and Bouten \citep{yamamoto-bouten2009} and of
Vladimirov, Petersen and James \citep[ eq.
(53)]{vladimirov-petersen-james2018}. What we add is the convex-order
endpoint (iv) and the equality case for the value.

\subsection{Moment certificates}\label{sec-moment-reduction}

Fix \(\alpha\) and write \(\omega = \omega_\alpha\). Let
\(\mathcal{M}_k\) be the probability measures on \([l_1, l_n]\) whose
first \(k\) moments agree with those of \(\omega\); since
\(\omega \in \mathcal{M}_k\),
\begin{equation}\phantomsection\label{eq-moment-sandwich}{
\inf_{\nu \in \mathcal{M}_k}
V_\alpha^{\rm cl}(q;\,\nu)
\;\leq\; V_\alpha(q) \;\leq\;
\sup_{\nu \in \mathcal{M}_k}
V_\alpha^{\rm cl}(q;\,\nu).
}\end{equation} What makes (\ref{eq-moment-sandwich}) sharp and
computable is the theory of Chebyshev systems
\citep{karlin-studden1966, krein-nudelman1977}. A function \(f\) is
\textbf{\((k{+}1)\)-convex} if its divided differences of order \(k+1\)
are nonnegative (\(f^{(k+1)} \geq 0\) for \(f \in C^{k+1}\)). For an
interior moment point, \(\mathcal{M}_k\) contains exactly two
representing measures of minimal index (interior atoms counting \(1\),
endpoint atoms \(\tfrac12\)), the \textbf{lower and upper principal
representations} \(\nu_-^{(k)}\) and \(\nu_+^{(k)}\), distinguished by
\(l_n \in \operatorname{supp}\nu_+^{(k)}\); they are the Gauss,
Gauss--Radau and Gauss--Lobatto quadratures of the moment data. The
first two levels are explicit: \(\nu_-^{(1)} = \delta_\mu\) and
\(\nu_+^{(1)}\) is the two-point measure on \(\{l_1,l_n\}\) with mean
\(\mu\), the \emph{Hoeffding measure}; at \(k = 2\), with \(v :=
\operatorname{Var}_\omega(\ell)\),
\begin{equation}\phantomsection\label{eq-bernstein-measure}{
\nu_+^{(2)}
= \Bigl(1 - \tfrac{v}{v + a^2}\Bigr)\,\delta_{x_+}
+ \tfrac{v}{v + a^2}\,\delta_{l_n},
\qquad x_+ = \mu - \frac{v}{a},
}\end{equation} which we call the \emph{Bernstein measure}; it is the
extremal law of Bennett's lemma \citep{bennett1962, hoeffding1963}, and
its moment generating function is the Bennett bound of
\citep{gourgoulias-model-bias2020}. Symmetrically \(\nu_-^{(2)}\) is
carried by \(\{l_1, x_-\}\), \(x_- = \mu + v/b\). The Markov--Krein
theorem \citep[Ch. III--IV]{karlin-studden1966} states that for every
\((k{+}1)\)-convex \(f\) and \(\nu \in \mathcal{M}_k\),
\begin{equation}\phantomsection\label{eq-markov-krein}{
\int f\,d\nu_-^{(k)}
\;\leq\; \int f\,d\nu
\;\leq\; \int f\,d\nu_+^{(k)},
}\end{equation} both bounds attained. Whether the dual integrands of
Lemma~\ref{lem-classical-dual} --- \((\ell-c)_+^{\alpha'}\),
\(e^{t\ell}\), \((c-\ell)^{-p}\) --- lie in this cone decides where the
certificate is sharp:

\begin{proposition}[Convexity-cone
stratification]\protect\hypertarget{prp-cone-stratification}{}\label{prp-cone-stratification}

Fix an integer \(k \geq 0\) and work on \([l_1, l_n]\). (i) For every
\(p > 0\) and \(c > l_n\), \(\ell \mapsto (c-\ell)^{-p}\) is
\((k{+}1)\)-convex, and so is \(\ell \mapsto e^{t\ell}\) for every
\(t \geq
0\). (ii) For \(\alpha' > 0\), \(\ell \mapsto (\ell-c)_+^{\alpha'}\) is
\((k{+}1)\)-convex for every \(c \in \mathbb{R}\) if and only if
\(k \leq
\alpha'\); for \(k > \alpha'\) it fails at every interior \(c\).
Consequently, whenever \(k \leq \alpha'\) --- and, by (i), for every
\(k\) when \(\alpha \leq 1\) ---
\begin{equation}\phantomsection\label{eq-mk-certificate}{
V_\alpha^{\rm cl}\bigl(q;\,\nu_-^{(k)}\bigr)
\;\leq\; V_\alpha(q) \;\leq\;
V_\alpha^{\rm cl}\bigl(q;\,\nu_+^{(k)}\bigr),
}\end{equation} and both bounds are attained on \(\mathcal{M}_k\).

\end{proposition}

Parts (i) and (ii) are classical
\citep{karlin-studden1966, denuit-lefevre-shaked1998}; the proof is in
\hyperref[sec-app-moment-tools]{Appendix A}. The certificate follows
from (i)--(ii) because (\ref{eq-markov-krein}) then brackets the
transform of \(\omega\) uniformly in the dual variable, and
Lemma~\ref{lem-transform-monotone} turns the bracket into
(\ref{eq-mk-certificate}).

\begin{figure}[htbp]

\centering{

\includegraphics[width=1\linewidth,height=\textheight,keepaspectratio]{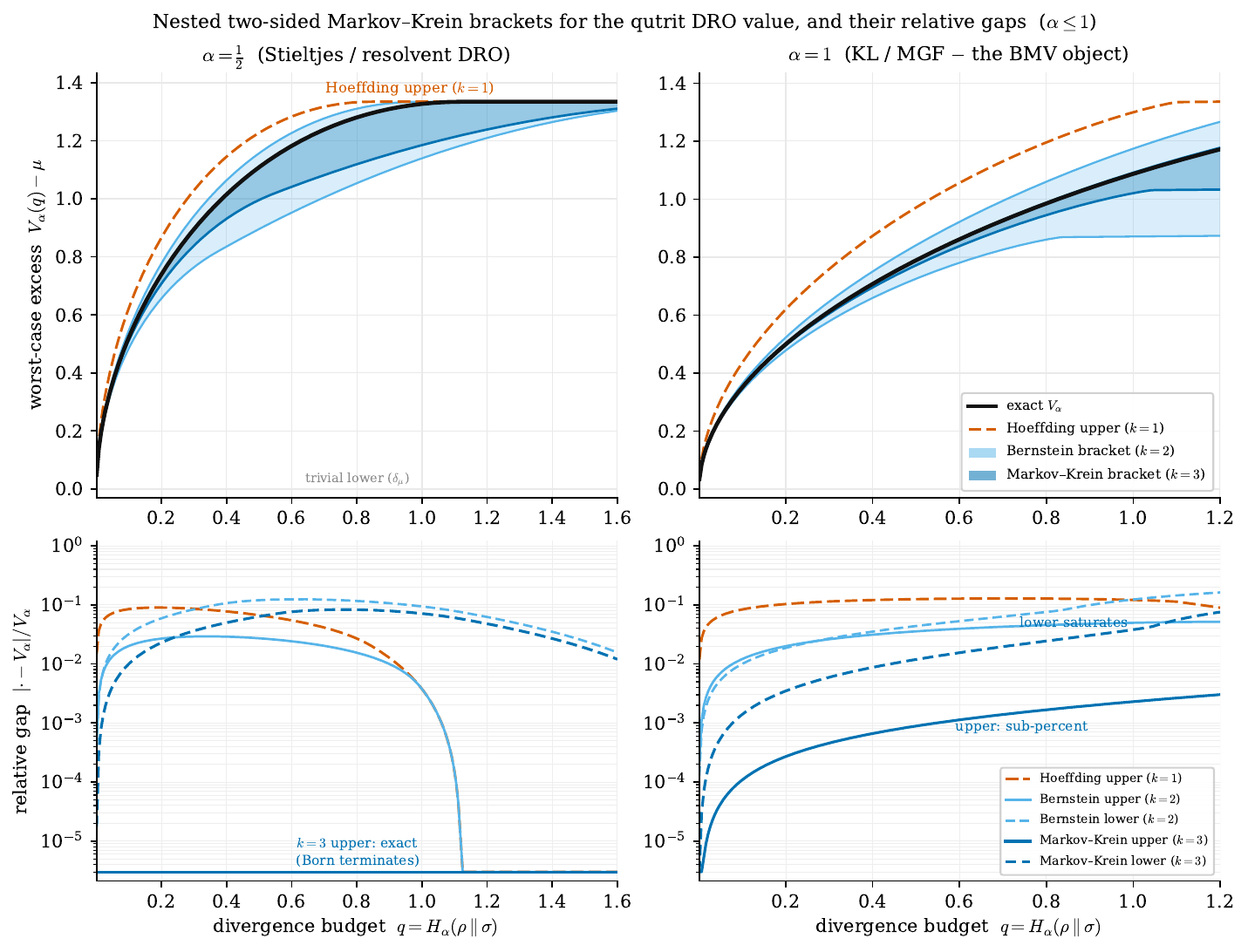}

}

\caption{\label{fig-dro-bracket}Nested Markov--Krein certificates for a
qutrit (\(L =
\operatorname{diag}(0,1,2)\), \(\sigma\) tilted so that
\([\sigma,L] \neq
0\)) at \(\alpha = \tfrac12\) and \(\alpha = 1\): the exact excess
\(V_\alpha(q) - \mu\) inside the Hoeffding cap (\(k = 1\)), the
Bernstein bracket (\(k = 2\)) and the \(k = 3\) bracket, with relative
gaps below. At \(\alpha = \tfrac12\) three moments pin the three-atom
Born measure and the \(k = 3\) upper gap vanishes; at \(\alpha = 1\) the
density of \(\omega_1\) keeps every bound two-sided, the \(k = 3\) upper
gap staying below one percent.}

\end{figure}

\begin{refremark}[The ladder for \(\alpha \le 1\): Gaussian quadrature
and Padé]
For \(\alpha \leq 1\) the ladder is unlimited, and it is a classical
object under another name. The extremal measures \(\nu_\pm^{(k)}\) are
the principal (Gauss, Radau, Lobatto) quadrature measures of the
truncated Hausdorff moment problem, their nodes and weights governed by
the orthogonal polynomials of \(\omega_\alpha\)
\citep{golub-meurant2010}; their Cauchy transforms --- nodes the poles,
weights the residues --- are the Padé approximants to the Stieltjes
transform of \(\omega_\alpha\) \citep{baker-gravesmorris1996}; the same
principal measures give the sharp brackets (\ref{eq-mk-certificate})
through (\ref{eq-markov-krein}).

At \(\alpha = 1\) this is the mechanism of Bessis--Moussa--Villani
\citep{bessis1975}, who conjectured the positivity of the representing
measure of \(\operatorname{tr}\,e^{\log\sigma + tL}\) precisely to
license monotone two-sided Padé bounds on the partition function, with
Gibbs--Bogoliubov as the \(k = 1\) rung; Stahl's theorem
\citep{stahl2013} licenses them, and classicalization makes them
certificates for the DRO value.

The bounds are monotone in \(k\) --- \(\mathcal{M}_{k+1} \subseteq
\mathcal{M}_k\), so (\ref{eq-markov-krein}) at level \(k\) orders the
transforms of \(\nu_\pm^{(k+1)}\) and \(\nu_\pm^{(k)}\), hence the
values by Lemma~\ref{lem-transform-monotone} --- and the upper
certificates converge to \(V_\alpha(q)\): by Hausdorff determinacy
\(\nu_\pm^{(k)} \to
\omega_\alpha\) weakly, so the objective at \(\nu_+^{(k)}\) converges to
that at \(\omega_\alpha\) for each fixed dual variable, which bounds
\(\limsup_k V_\alpha^{\rm cl}(q;\nu_+^{(k)})\) by \(V_\alpha(q)\) from
above. The moments are computable: at \(\alpha = 1\), \(m_k =
\tfrac{d^k}{dt^k}\operatorname{tr}\,e^{\log\sigma + tL}\big|_{t=0}\),
and for \(\alpha < 1\) they are the coefficients of
\(\operatorname{tr}[\sigma^\beta(cI-L)\sigma^\beta]^{-p}\) in powers of
\(1/c\) (Lemma~\ref{lem-transform-unique}). Figure~\ref{fig-dro-bracket}
shows the ladder closing on a qutrit.

\label{rem-pade}

\end{refremark}

\subsection{The spectral Bernstein bound and the
deflation}\label{sec-bernstein-bound}

The \textbf{spectral variance}
\begin{equation}\phantomsection\label{eq-v-def}{
v \;:=\; \operatorname{Var}_{\omega_\alpha}(\ell)
\;=\; \alpha\,s_\alpha^2(\hat L)
\;=\; \alpha\sum_{i,j} m_\alpha(p_i,p_j)\,|\hat L_{ij}|^2
}\end{equation} is computable in closed form for every \(\alpha\) by
Corollary~\ref{cor-spectral-variance} (at \(\alpha = \infty\) by
(\ref{eq-var-omega-infty})), and provides the \(k = 2\) level of the
hierarchy.\footnote{The names \emph{Hoeffding} and \emph{Bernstein}
  refer, as classically \citep{hoeffding1963, bennett1962}, to the
  moment data consumed --- support, resp.~support and variance --- and
  not to the quantum Hoeffding bound of hypothesis testing
  \citep{hayashi2007, audenaert-nussbaum2008} or the matrix Bernstein
  inequalities \citep{ahlswede-winter2002, tropp2015}. We use
  \emph{Bernstein} for the whole mean--variance--support tier;
  classically, the sharp two-point bound of that tier is Bennett's and
  its explicit form \(\sqrt{2vq} + aq/3\) is Bernstein's
  \citep{bennett1962, boucheron-lugosi-massart2013}.} Let
\(\mathcal{M}(\mu,v)\) be the probability measures on \([l_1,l_n]\) with
mean \(\mu\) and variance \(v\); note \(v \leq ab\) (the Bhatia--Davis
inequality \citep{bhatia-davis2000}).

\begin{theorem}[Spectral Bernstein
bound]\protect\hypertarget{thm-spectral-bernstein}{}\label{thm-spectral-bernstein}

Let \(\alpha \in [\tfrac12, \infty]\), \(q \geq 0\), and \(\rho\) with
\(H_\alpha(\rho\|\sigma) \leq q\).

\begin{enumerate}
\def\labelenumi{(\roman{enumi})}
\item
  For \(\alpha \in [\tfrac12, 2]\),
  \(V_\alpha^{\rm cl}(q;\nu_-^{(2)}) \leq V_\alpha(q) \leq
  V_\alpha^{\rm cl}(q;\nu_+^{(2)})\), both bounds attained on
  \(\mathcal{M}(\mu,v)\); in particular
  \begin{equation}\phantomsection\label{eq-bernstein-bound}{
  \operatorname{tr}\bigl(L(\rho - \sigma)\bigr)
  \;\leq\; \mathcal{B}_\alpha(q;v)
  := V_\alpha^{\rm cl}\bigl(q;\,\nu_+^{(2)}\bigr) - \mu .
  }\end{equation}
\item
  For \(\alpha = 2\), with \(\operatorname{Var}_{\omega_2}(\ell)\) as in
  (\ref{eq-alpha-two-WY}),
  \begin{equation}\phantomsection\label{eq-bernstein-chi2}{
  \boxed{\;
  \operatorname{tr}\bigl(L(\rho-\sigma)\bigr)
  \;\leq\;
  \min\Bigl(\sqrt{q\,\operatorname{Var}_{\omega_2}(\ell)}\,,\;
  a\Bigr).
  \;}
  }\end{equation}
\item
  For \(\alpha = \infty\), with \(u = (1+q)^{-1}\),
  \begin{equation}\phantomsection\label{eq-cvar-cantelli}{
  \operatorname{tr}\bigl(L(\rho - \sigma)\bigr)
  \;\leq\;
  \sup_{\nu \in \mathcal{M}(\mu,v)} \mathrm{CVaR}_u(\nu) - \mu
  \;=\; \min\bigl(\,q\,b,\;\sqrt{q\,v},\;a\,\bigr).
  }\end{equation}
\end{enumerate}

\end{theorem}

\begin{proof}
(i) For \(\alpha \in (1,2]\) one has \(\alpha' \geq 2\), and for
\(\alpha
\leq 1\) the cone is unlimited
(Proposition~\ref{prp-cone-stratification}), so
(\ref{eq-mk-certificate}) holds at \(k = 2\) with \(\mathcal{M}_2 =
\mathcal{M}(\mu,v)\). (ii) \(\nu_+^{(2)}\) is a two-point law; the
classical \(\chi^2\)-DRO of a law \(\nu\) with mean \(\mu\) and variance
\(v\) equals \(\mu + \sqrt{qv}\) as long as the linear tilt \(1 +
\sqrt{q/v}\,(\ell-\mu)\) is nonnegative on the support
\citep[§2.1]{duchi-namkoong2019} --- for \(\nu_+^{(2)}\), until
\(\sqrt{qv} =
a\) at the atom \(x_+\) --- and \(a\) beyond; the variance identity is
(\ref{eq-alpha-two-WY}). (iii) is Hürlimann's theorem \citep[Thm.
3.1]{hurlimann2002}. \(\square\)
\end{proof}

In (iii), Hürlimann's formula for the maximal CVaR at given range, mean
and variance, the three terms are attained on \(q \leq v/b^2\), on
\(v/b^2 \leq q \leq a^2/v\) and on \(q \geq a^2/v\), each by a law with
at most three atoms. At \(\alpha = 1\) the upper certificate of (i) is
the two-point bound of Bennett and Hoeffding
\citep{bennett1962, hoeffding1963}, in the form of
\citep{gourgoulias-model-bias2020}, where its sharpness over the moment
class is also proved. The quantum content of the theorem is its input
\(v\): \(\omega_\alpha\) and \(\nu_L\) share \((\mu, l_1, l_n)\), so the
deflation of \(v\) is the whole difference between the spectral bound
and its measured counterpart.

\begin{proposition}[Quantum deflation of the Bernstein
bound]\protect\hypertarget{prp-deflation}{}\label{prp-deflation}

Let \(\alpha \in [\tfrac12, 2]\) and \(q \geq 0\). The certificate
\(v \mapsto \mathcal{B}_\alpha(q;v)\) of (\ref{eq-bernstein-bound}) is
nondecreasing on \((0, ab]\), and, for \(q > 0\), strictly increasing
below saturation: \(\mathcal{B}_\alpha(q;v) < \mathcal{B}_\alpha(q;v')\)
whenever \(v < v'\) and \(\mathcal{B}_\alpha(q;v') < a\). Consequently,
for every \(\rho\) with \(H_\alpha(\rho\|\sigma) \leq q\),
\begin{equation}\phantomsection\label{eq-deflation}{
\operatorname{tr}\bigl(L(\rho-\sigma)\bigr)
\;\leq\; \mathcal{B}_\alpha\bigl(q;\operatorname{Var}_{\omega_\alpha}(\ell)\bigr)
\;\leq\; \mathcal{B}_\alpha\bigl(q;\operatorname{Var}_\sigma(L)\bigr),
}\end{equation} and the second inequality is strict whenever
\([\sigma,L] \neq 0\), \(\alpha >
\tfrac12\), \(q > 0\), and the measured bound is below saturation.

\end{proposition}

At \(\alpha = \infty\) the chain (\ref{eq-deflation}) holds for
(\ref{eq-cvar-cantelli}), which is nondecreasing in \(v\); its second
inequality is strict exactly when
\(\sqrt{q\operatorname{Var}_{\omega_\infty}(\ell)} < \min(qb,\,a)\),
i.e.~when the \(\sqrt{qv}\) term is the active one at the deflated
variance, and at small budgets, where the support term \(qb\) is active
for both measures, it is an equality.

\begin{proof}
Let \(v < v'\) and write \(\nu_+ := \nu_+^{(2)}(v)\), \(\nu'_+ :=
\nu_+^{(2)}(v')\). Since \(x_+(v) = \mu - v/a > \mu - v'/a = x_+(v')\),
the law \(\nu_+\) is supported in \([x_+(v'), l_n]\) and has mean
\(\mu\), while \(\nu'_+\) is the two-point law on the endpoints of that
interval with mean \(\mu\); hence
\(\int f\,d\nu_+ \leq \int f\,d\nu'_+\) for every convex \(f\) (Jensen
along the chord), i.e.~\(\nu_+ \preceq_{\rm cx}
\nu'_+\). The dual integrands are convex, so
Lemma~\ref{lem-transform-monotone} gives \(\mathcal{B}_\alpha(q;v) \leq
\mathcal{B}_\alpha(q;v')\). For strictness, let \(c'\) be a minimizer of
the \(\nu'_+\)-objective, which exists in \((-\infty, l_n)\)
(resp.~\((l_n,\infty)\) for \(\alpha < 1\), \(t' \in (0,\infty)\) for
\(\alpha = 1\)) when \(q > 0\) and \(\mathcal{B}_\alpha(q;v') < a\), the
objective being continuous and coercive with value \(\geq a\) at
\(c \geq l_n\); there the integrand --- \((\ell-c')_+^{\alpha'}\) with
\(c' < l_n\), \(e^{t'\ell}\), \((c'-\ell)^{-p}\) --- is convex and not
affine on \([x_+(v'), l_n]\), and \(\nu_+\) charges the interior point
\(x_+(v)\), so the chord inequality is strict and
\(\mathcal{B}_\alpha(q;v) \leq \text{objective}_{\nu_+}(c') <
\text{objective}_{\nu'_+}(c') = \mathcal{B}_\alpha(q;v')\). The chain
(\ref{eq-deflation}) is (\ref{eq-bernstein-bound}) followed by
monotonicity with \(v = \operatorname{Var}_{\omega_\alpha}(\ell) \leq
\operatorname{Var}_{\nu_L}(\ell) = \operatorname{Var}_\sigma(L)\),
strict for \(\alpha > \tfrac12\) and \([\sigma,L] \neq 0\) by
Theorem~\ref{thm-variance-ordering}; and \(\nu_L \in \mathcal{M}(\mu,
\operatorname{Var}_\sigma(L))\) makes the right side the measured
certificate. \(\square\)
\end{proof}

\begin{figure}[htbp]

\centering{

\includegraphics[width=1\linewidth,height=\textheight,keepaspectratio]{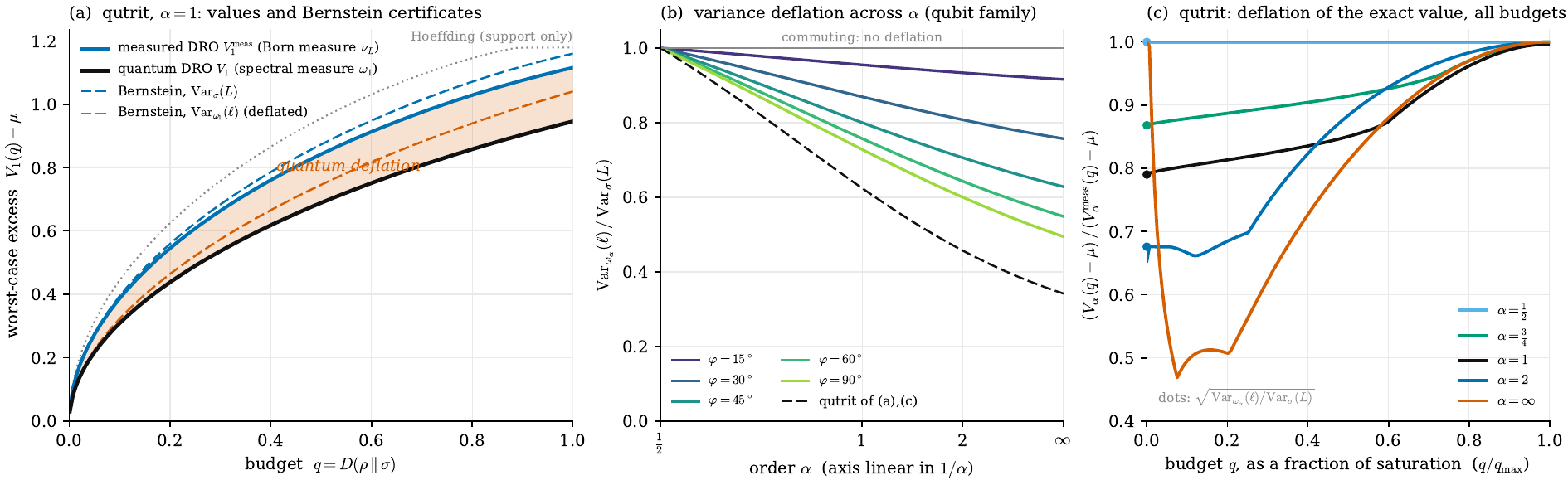}

}

\caption{\label{fig-deflation}The quantum deflation. \emph{(a)} A qutrit
(\(L = \operatorname{diag}(0,1,2)\), \(\sigma\) with eigenvalues
\((0.90, 0.07, 0.03)\) rotated so that \([\sigma,L] \neq 0\); Born
weights \((0.47, 0.24, 0.29)\)) at \(\alpha =
1\): worst-case excess against the budget \(q = D(\rho\|\sigma)\) for
the measured DRO (Born measure \(\nu_L\)) and the quantum DRO (spectral
measure \(\omega_1\)), with the two \(k = 2\) Bernstein certificates of
Theorem~\ref{thm-spectral-bernstein} built on
\(\operatorname{Var}_\sigma(L)\) and on the deflated
\(\operatorname{Var}_{\omega_1}(\ell) = 0.62\,
\operatorname{Var}_\sigma(L)\), and the Hoeffding bound above them. The
shaded band is the deflation; the deflated certificate lies below the
exact \emph{measured} value. \emph{(b)} The ratio
\(\operatorname{Var}_{\omega_\alpha}(\ell)/\operatorname{Var}_\sigma(L)\)
against \(\alpha\) (axis linear in \(1/\alpha\)) for the qubit
\(\sigma =
\operatorname{diag}(0.9, 0.1)\) and \(L_\varphi = \cos\varphi\,Z +
\sin\varphi\,X\), \(\varphi\) from \(15^\circ\) to \(90^\circ\), and
(dashed) for the qutrit of (a): equal to \(1\) at \(\alpha = \tfrac12\)
and for commuting pairs, strictly decreasing in \(\alpha\) otherwise
(Theorem~\ref{thm-variance-ordering}), down to the harmonic-logarithmic
variance at \(\alpha = \infty\). \emph{(c)} For the qutrit of (a), the
ratio of the exact quantum to the exact measured excess against the
budget, as a fraction of the saturation budget, at
\(\alpha = \tfrac12, \tfrac34, 1, 2, \infty\); the dots at \(q = 0\) are
\(\sqrt{\operatorname{Var}_{\omega_\alpha}(\ell)/\operatorname{Var}_\sigma(L)}\)
(Corollary~\ref{cor-spectral-variance}). No deflation at
\(\alpha = \tfrac12\); at \(\alpha = \infty\) the small-budget regime is
linear for both measures, so the ratio starts at \(1\) and the deflation
is strongest at intermediate budgets.}

\end{figure}

Since \(\alpha \mapsto \operatorname{Var}_{\omega_\alpha}(\ell)\) is
strictly decreasing for \([\sigma,L] \neq 0\)
(Theorem~\ref{thm-variance-ordering}), the deflation grows with
\(\alpha\): from none at the fidelity endpoint, where
\(\omega_{1/2} = \nu_L\), to the Wigner--Yanase skew information at
\(\alpha = 2\) --- (\ref{eq-bernstein-chi2}) reads
\(\sqrt{q(\operatorname{Var}_\sigma(L) - I_{\rm WY}(\sigma,L))}\)
against the measured \(\sqrt{q\operatorname{Var}_\sigma(L)}\)
(\ref{eq-alpha-two-WY}) --- and to the harmonic-logarithmic variance
(\ref{eq-var-omega-infty}) at \(\alpha = \infty\).
Figure~\ref{fig-deflation} shows it on a qutrit, across \(\alpha\) and
across budgets.

The certificates of Theorem~\ref{thm-spectral-bernstein} are sharp but
quadrature-shaped. Explicit formulas exist at three \emph{anchors} ---
\(\alpha = \tfrac12\), where the two-point certificate is solved by the
Bures angle, \(\alpha = 1\), where Bernstein's inequality applies, and
\(\alpha = 2\), where the value is a square root --- and the
monotonicity of \(H^{\rm cl}_\alpha\) in \(\alpha\)
(Remark~\ref{rem-alpha-monotone}) transports each anchor to the orders
above it, at the price of a constant. Write, for \(w \in [0,1]\) and
\(\mathrm{Fid} := 1 - q/2\),
\(\Delta^*_q(w) := \sin(2\arcsin\sqrt w + \arccos\mathrm{Fid})\,\sqrt{1-\mathrm{Fid}^2}\)
for \(w \leq \mathrm{Fid}^2\) and \(\Delta^*_q(w) := 1 - w\) otherwise,
the fidelity profile of Theorem~\ref{thm-fuchs-vdg}.

\begin{proposition}[Explicit Bernstein
bounds]\protect\hypertarget{prp-explicit-bernstein}{}\label{prp-explicit-bernstein}

Let \(q \geq 0\), \(\rho\) with \(H_\alpha(\rho\|\sigma) \leq q\), and
\(v = \operatorname{Var}_{\omega_\alpha}(\ell)\).

\begin{enumerate}
\def\labelenumi{(\roman{enumi})}
\item
  For \(\alpha \in [\tfrac12, 1)\) and \(q \leq 2\),
  \begin{equation}\phantomsection\label{eq-fidelity-anchor}{
  \operatorname{tr}\bigl(L(\rho-\sigma)\bigr)
  \;\leq\; \frac{a^2 + v}{a}\;\Delta^*_q\Bigl(\frac{v}{v + a^2}\Bigr).
  }\end{equation}
\item
  For \(\alpha \in [1, 2)\),
  \begin{equation}\phantomsection\label{eq-bennett-bound}{
  \operatorname{tr}\bigl(L(\rho-\sigma)\bigr)
  \;\leq\;
  \min\Bigl(\sqrt{2vq} + \frac{aq}{3}\,,\; a\Bigr).
  }\end{equation}
\item
  For \(\alpha \in [2,\infty)\),
  \(\operatorname{tr}(L(\rho - \sigma)) \leq \min\bigl(\sqrt{vq},\ \sqrt{v(\kappa_\alpha(q)^2 - 1)},\ a\bigr)\),
  \(\kappa_\alpha(q) := (1+(\alpha-1)q)^{1/\alpha}\).
\end{enumerate}

\end{proposition}

\begin{proof}
For (i)--(ii),
\(\operatorname{tr}(L(\rho-\sigma)) \leq \mathcal{B}_\alpha(q;v) = V^{\rm cl}_\alpha(q;\nu_+^{(2)}) - \mu\)
by Theorem~\ref{thm-spectral-bernstein}(i), with
\(\nu_+^{(2)} = \nu_+^{(2)}(v)\) the two-point law
(\ref{eq-bernstein-measure}); this is where the deflated variance
enters. For an anchor \(\alpha_0\le\alpha\) within the range where the
\(k=2\) certificate is valid, monotonicity gives
\(V^{\rm cl}_\alpha(q;\nu_+^{(2)}) \leq V^{\rm cl}_{\alpha_0}(q;\nu_+^{(2)})\).
Thus (i) \(\alpha_0 = \tfrac12\): \(\nu_+^{(2)}\) is the
Bernoulli\((w)\) law on \(\{x_+, l_n\}\), \(w = v/(v+a^2)\), of mean
\(\mu\), and by the computation in the proof of
Theorem~\ref{thm-fuchs-vdg} its fidelity DRO excess is
\((l_n - x_+)\,\Delta^*_q(w)\) with \(l_n - x_+ = (a^2+v)/a\); at
\(\alpha = \tfrac12\) the inclusion is an equality. (ii)
\(\alpha_0 = 1\): a law with mean \(\mu\), variance \(v\) and support in
\((-\infty, l_n]\) obeys Bernstein's majorant
\(\log\int e^{t\hat\ell}\,d\nu \leq vt^2/(2(1-ta/3))\), \(0 < t < 3/a\),
and optimizing the \(\alpha = 1\) dual gives \(\sqrt{2vq} + aq/3\)
(\hyperref[sec-app-moment-tools]{Appendix A}). (iii) For \(\alpha\ge2\)
we do \textbf{not} invoke the \(k=2\) Markov--Krein certificate, whose
sharp validity stops at \(\alpha=2\). Instead classicalization gives a
probability \(P\) with \(H_\alpha^{\rm cl}(P\|\omega_\alpha)\le q\).
Since \(H_2^{\rm cl}\le H_\alpha^{\rm cl}\), Cauchy--Schwarz
(Hammersley--Chapman--Robbins) directly yields
\(\int\hat\ell\,dP\le\sqrt{vq}\). The second estimate follows from the
exact \(\alpha\)-dual at the shifted level \(\tilde c = c - \mu\) and
Lyapunov's inequality
\(\|(\hat\ell-\tilde c)_+\|_{\alpha'}\le\|\hat\ell-\tilde c\|_2=\sqrt{v+\tilde c^{\,2}}\);
minimizing \(\tilde c + \kappa_\alpha(q)\sqrt{v+\tilde c^{\,2}}\) over
\(\tilde c\) gives \(\sqrt{v(\kappa_\alpha(q)^2-1)}\)
(\hyperref[sec-app-moment-tools]{Appendix A}). Finally \(a\) is the
support bound. \(\square\)
\end{proof}

The anchors are \(\alpha = \tfrac12\) (fidelity) in (i), \(\alpha = 1\)
(Bernstein) in (ii) and \(\alpha = 2\) (Cauchy--Schwarz) in (iii), whose
second term is the Lyapunov bound. In (i) equality holds at \(\alpha =
\tfrac12\) for the certificate; in (ii) at \(\alpha = 1\),
\(q = D(\rho\|\sigma)\) and \(v\) is the Kubo--Mori variance. At
\(\alpha = \tfrac12\), where \(v = \operatorname{Var}_\sigma(L)\), (i)
is, after a reflection of the observable, the mean--variance bound of
Weber et al. \citep[Thm. 2]{weber-etal2022}, with the same validity
condition, here the \(k = 2\) certificate, which for \(\alpha \in
(\tfrac12, 1)\) holds with the smaller variance
\(\operatorname{Var}_{\omega_\alpha}(\ell)\); for a classical pair (ii)
is Bernstein's inequality \citep{boucheron-lugosi-massart2013}. As
\(q \to 0\) the three bounds behave as \(2\sqrt{qv}\), \(\sqrt{2qv}\)
and \(\sqrt{qv}\) against the sharp \(\sqrt{2qv/\alpha}\) of
Corollary~\ref{cor-spectral-variance}: loss factors \(\sqrt{2\alpha}\),
\(\sqrt{\alpha}\) and \(\sqrt{\alpha/2}\), each equal to \(1\) at its
anchor. The Lyapunov term in (iii), of order
\(\sqrt{2qv(\alpha-1)/\alpha}\), is worse than \(\sqrt{qv}\) at small
\(q\) but sublinear in \(q\), hence better at large budgets; all bounds
are monotone in \(v\) and inherit the deflation.

\begin{refremark}[The \(\alpha = 2\) anchor is
Hammersley--Chapman--Robbins]
The Cauchy--Schwarz step of (iii) is the classical
Hammersley--Chapman--Robbins inequality
\((\mathbb{E}_Pg - \mathbb{E}_Qg)^2
\leq \operatorname{Var}_Q(g)\,\chi^2(P\|Q)\)
\citep{hammersley1950, chapman-robbins1951}, and its quantum form at
\(\alpha = 2\), \begin{equation}\phantomsection\label{eq-hcr}{
\bigl|\operatorname{tr}(L(\rho-\sigma))\bigr|^2
\;\le\; \operatorname{Var}_{\omega_2}(\ell)\; H_2(\rho\|\sigma),
}\end{equation} follows directly by splitting
\(\operatorname{tr}(\hat L X) =
\operatorname{tr}[(\sigma^{1/4}\hat L\sigma^{1/4})(\sigma^{-1/4}X\sigma^{-1/4})]\),
\(X = \rho - \sigma\): here
\(H_2 = \operatorname{tr}(\sigma^{-1/2}X\sigma^{-1/2}X)\) is the
geometric-mean member of the quantum \(\chi^2\) family of
\citep{temme-etal2010}, and the variance factor is the spectral variance
(\ref{eq-alpha-two-WY}), deflated by the skew information. It is exact:
equality holds at \(X \propto \sigma^{1/2}\hat L\sigma^{1/2}\) as long
as \(\rho\) stays positive, i.e.~for \(H_2 \leq q_0\), where
(\ref{eq-hcr}) is the DRO value itself
(Corollary~\ref{cor-alpha-two-exact}); and for \(\rho = \sigma +
\varepsilon A\), \(H_2 = \tfrac12\varepsilon^2\langle A, \mathcal H_2
A\rangle\) exactly, so (\ref{eq-hcr}) is the quantum Cramér--Rao bound
of Remark~\ref{rem-cramer-rao}.

\label{rem-hcr}

\end{refremark}

\subsection{The Hoeffding bound}\label{sec-hoeffding-bound}

The \(k = 1\) certificate uses only \((\mu, l_1, l_n)\), which
\(\omega_\alpha\) shares with \(\nu_L\); it therefore needs neither the
spectral measure nor the classicalization, only Born domination. Write
\(\Delta = a + b\) and \(\theta = b/\Delta\), the weight of
\(\nu_+^{(1)}\) at \(l_n\).

\begin{theorem}[Spectral Hoeffding
bound]\protect\hypertarget{thm-spectral-hoeffding}{}\label{thm-spectral-hoeffding}

For every \(\alpha \in [\tfrac12, \infty]\), \(q \geq 0\), and every
\(\rho\) with \(H_\alpha(\rho\|\sigma) \leq q\):
\begin{equation}\phantomsection\label{eq-hoeffding-bound}{
-\Delta\cdot\eta_\alpha(q,\,1-\theta)
\;\leq\;
\operatorname{tr}\bigl(L(\rho - \sigma)\bigr)
\;\leq\;
\Delta\cdot\eta_\alpha(q,\,\theta),
}\end{equation} where \(\eta_\alpha(q,\theta)\) is the \textbf{Hoeffding
profile}, the excess DRO value of a Bernoulli\((\theta)\) on
\(\{0,1\}\),
\begin{equation}\phantomsection\label{eq-hoeffding-profile}{
\eta_\alpha(q,\theta)
:= V_\alpha^{\rm cl}\bigl(q;\,
\theta\delta_1 + (1-\theta)\delta_0\bigr) - \theta.
}\end{equation}

\end{theorem}

\begin{proof}
By (\ref{eq-dpi-value}),
\(V_\alpha(q) \leq V_\alpha^{\rm cl}(q;\nu_L)\). The dual integrands are
convex (\(k = 1 \leq \alpha'\) always, and
Proposition~\ref{prp-cone-stratification}(i)), and \(\nu_L\) has mean
\(\mu\) on \([l_1,l_n]\), so Jensen along the chord bounds each
transform of \(\nu_L\) by that of the Hoeffding measure \(\nu_+^{(1)}\);
Lemma~\ref{lem-transform-monotone} gives
\(V_\alpha^{\rm cl}(q;\nu_L) \leq
V_\alpha^{\rm cl}(q;\nu_+^{(1)}) = \mu + \Delta\,\eta_\alpha(q,\theta)\),
the last step by the affine rescaling of \([l_1,l_n]\) onto \(\{0,1\}\).
The lower bound is the upper bound for \(-L\), whose Hoeffding parameter
is \(1-\theta\). \(\square\)
\end{proof}

The classical statement is Hoeffding's \citep{hoeffding1963}, at
\(\alpha = 1\) in the form of \citep{gourgoulias-model-bias2020}; the
quantum step is data processing. At \(\alpha = \tfrac12\) the bound,
with the closed form (\ref{eq-fvdg-excess}) of its profile, is the
robustness interval of Weber et al. \citep[Thm. 1]{weber-etal2022}; at
\(\alpha = 1\) it follows from the measured bound discussed after
Proposition~\ref{prp-born-domination}; what the theorem adds is the
statement for all \(\alpha\) at once. The bound is sharp given only the
data \((\mu, l_1, l_n)\). For \(\alpha \in (\tfrac12, \infty)\) it is
attained in the quantum model at every budget precisely when
\([\sigma,L] = 0\) and \(L\) has at most two distinct eigenvalues: only
then does \(\omega_\alpha\) collapse onto \(\nu_L\)
(Theorem~\ref{thm-spectral-rep}(i)), discrete on \(\{l_1,l_n\}\), hence
equal to \(\nu_+^{(1)}\); at \(\alpha = \tfrac12\), where
\(\omega_{1/2} = \nu_L\), it is attained for every \(L\) with at most
two distinct eigenvalues, commuting with \(\sigma\) or not. For
\(\alpha \in (\tfrac12, \infty)\) and a noncommuting \(L\) with two
eigenvalues, \(\omega_\alpha\) carries a density
(Theorem~\ref{thm-spectral-rep}(ii)) and the bound is strictly loose at
every budget below saturation, and for \(L\) with three or more
eigenvalues it is generically only approached
(Figure~\ref{fig-dro-bracket}). At \(\alpha = \infty\) the integrand
\((\ell-c)_+\) is affine above \(c\) and the bound is attained at small
budgets for every \(L\) with simple spectrum, commuting or not: for
\(q \leq
w_1/(1-w_1)\),
\(w_1 := \omega_\infty(\{l_1\}) = 1/\langle 1|\sigma^{-1}|1\rangle\),
the bottom \((1-u)\)-mass of \(\omega_\infty\) sits at \(l_1\) and
\(V_\infty(q) - \mu = qb = \Delta\,\eta_\infty(q,\theta)\). At
\(\alpha = \infty\) the profile is piecewise linear,
\(\eta_\infty(q,\theta) = \min(\theta q,\,1-\theta)\), the CVaR of a
Bernoulli\((\theta)\).

\subsection{Trace distance and fidelity}\label{sec-state-discrimination}

Applied to the Helstrom observable, the bounds of this section become
inequalities between a divergence and the trace distance ---
Pinsker-type inequalities
\citep{pinsker1964, csiszar1967, hiai-ohya-tsukada1981, fuchs1999}.
Throughout, \(\sigma > 0\),
\(T(\rho,\sigma) := \tfrac12\|\rho-\sigma\|_1
= \max_{0 \le L \le I}\operatorname{tr}(L(\rho-\sigma))\), attained at
the Helstrom projector \(L_0 := \mathbb 1_{\{\rho > \sigma\}}\)
\citep{helstrom1976}, of some rank \(k \in \{1,\dots,n-1\}\) when
\(T > 0\); \(\theta_0 := \operatorname{tr}(\sigma L_0) \in (0,1)\); and
\(L_H := 2L_0
- I\), so that \(L_H^2 = I\), \(\operatorname{tr}(L_H(\rho-\sigma)) =
\|\rho-\sigma\|_1\), \(\operatorname{tr}(\sigma L_H) = 2\theta_0 - 1\)
and \(\operatorname{Var}_\sigma(L_H) = 4\theta_0(1-\theta_0)\).

\begin{corollary}[Pinsker-type inequalities at
\(\alpha = 1\)]\protect\hypertarget{cor-pinsker}{}\label{cor-pinsker}

Let \(r := D(\rho\|\sigma)\) and \(s_1^2 := s_1^2(\hat L_H) =
\operatorname{Var}_{\omega_1}(\ell)\), the Kubo--Mori variance of
\(L_H\) (Corollary~\ref{cor-spectral-variance}).

\begin{enumerate}
\def\labelenumi{(\roman{enumi})}
\item
  \(T(\rho,\sigma) \le \eta_1(r,\theta_0) \le \sup_{\theta \in (0,1)}
  \eta_1(r,\theta) \le \sqrt{r/2}\).
\item
  \begin{equation}\phantomsection\label{eq-bernstein-pinsker}{
  \|\rho-\sigma\|_1 \;\le\; \sqrt{2 s_1^2\, r} + \tfrac23(1-\theta_0)\,r,
  \qquad
  s_1^2 \;\le\; \operatorname{Var}_\sigma(L_H) = 4\theta_0(1-\theta_0) \;\le\; 1,
  }\end{equation} with \(s_1^2 < 4\theta_0(1-\theta_0)\) whenever
  \([\sigma, L_H] \neq 0\).
\end{enumerate}

\end{corollary}

\begin{proof}
(i) Theorem~\ref{thm-spectral-hoeffding} at \(\alpha = 1\) for
\(L = L_0\), which has \(l_1 = 0\), \(l_n = 1\), \(\Delta = 1\) and
\(\theta = \theta_0\), gives \(T =
\operatorname{tr}(L_0(\rho-\sigma)) \le \eta_1(r,\theta_0)\). For a
binary pair the mean shift is the total variation, so
\(\eta_1(r,\theta)\) is the largest total variation of a binary pair
\((P,Q)\) with \(Q = \mathrm{Ber}(\theta)\) and \(D(P\|Q) \le r\); its
supremum over \(\theta\) is the optimal classical curve, and
\(\eta_1(r,\theta) \le \sqrt{r/2}\) is Pinsker's inequality for binary
pairs. (ii) Proposition~\ref{prp-explicit-bernstein}(ii) for
\(L = L_H\), with \(q = r\),
\(\operatorname{tr}(L_H(\rho-\sigma)) = \|\rho-\sigma\|_1\) and \(a =
\lambda_{\max}(\hat L_H) = 1 - (2\theta_0-1) = 2(1-\theta_0)\). The
variance comparison is Theorem~\ref{thm-variance-ordering} at
\(\alpha = 1\), strict iff \([\sigma,L_H] \neq 0\), and
\(\operatorname{Var}_\sigma(L_H) = 1 -
(2\theta_0-1)^2\). The regime of improvement: \(\sqrt{2s_1^2 r} +
2(1-\theta_0) r/3 < \sqrt{2r}\) iff
\(\sqrt r < 3\sqrt2\,(1-s_1)/(2(1-\theta_0))\). \(\square\)
\end{proof}

Part (ii) improves on Pinsker's \(\|\rho-\sigma\|_1 \le \sqrt{2r}\)
exactly when \(r < 9(1-s_1)^2/(2(1-\theta_0)^2)\), in particular for all
small \(r > 0\) as soon as \(s_1^2 < 1\). In (i) the middle term is the
optimal classical trade-off between relative entropy and total variation
\citep{fedotov-harremoes-topsoe2003} and the last inequality is
Pinsker's; all of (i) is known: the reduction to a binary pair is the
argument of Hiai, Ohya and Tsukada \citep{hiai-ohya-tsukada1981}, and
the sharp bound is \citep[Thm. 4]{audenaert-eisert2005}. Replacing
\(s_1^2\) by \(4\theta_0(1-\theta_0)\) gives the \emph{measured}
Bernstein--Pinsker bound, which is classical
\citep{boucheron-lugosi-massart2013}; the strict inequality \(s_1^2 <
4\theta_0(1-\theta_0)\) is the quantum layer,
Proposition~\ref{prp-deflation} at \(\alpha =
1\) for the Helstrom observable, and the only place in this subsection
where the spectral measure, hence Stahl's theorem, is used. The
\(\alpha = 2\) member needs no inversion: (\ref{eq-hcr}) with
\(L = L_H\) reads \(\|\rho-\sigma\|_1^2
\leq \operatorname{Var}_{\omega_2}(\ell)\,H_2(\rho\|\sigma)\) with
\(\operatorname{Var}_{\omega_2}(\ell) = 4\theta_0(1-\theta_0) -
I_{\rm WY}(\sigma, L_H)\).

\begin{refremark}[No quantum penalty]
Let \(\Phi_\alpha(q) := \sup_{\theta \in (0,1)} \eta_\alpha(q,\theta)\),
the largest total variation of a binary pair with \(H^{\rm cl}_\alpha
\le q\) \citep{harremoes-vajda2011, sason-verdu2016}. The argument of
Corollary~\ref{cor-pinsker}(i), which is that of
\citep{hiai-ohya-tsukada1981} and uses only monotonicity under
measurements, runs verbatim at every \(\alpha
\in [\tfrac12,\infty]\):
\(T(\rho,\sigma) \le \eta_\alpha(q,\theta_0) \le \Phi_\alpha(q)\)
whenever \(H_\alpha(\rho\|\sigma) \le q\), while the commuting binary
pairs \(\sigma = \operatorname{diag}(1-\theta,\theta)\), \(\rho =
\operatorname{diag}(1-\theta-x, \theta+x)\) have \(H_\alpha = H^{\rm
cl}_\alpha\) and \(T = x\), and realize \(\eta_\alpha(q,\theta)\) (a
continuous objective over a compact feasible set). Hence
\begin{equation}\phantomsection\label{eq-optimality-transfer}{
\sup\bigl\{T(\rho,\sigma) : n \ge 1,\ \sigma > 0,\
H_\alpha(\rho\|\sigma) \le q\bigr\} \;=\; \Phi_\alpha(q):
}\end{equation} the optimal trace-distance versus \(H_\alpha\)
inequality is the classical one --- at \(\alpha = 1\) the theorem of
Hiai--Ohya--Tsukada \citep[Thm.
4]{hiai-ohya-tsukada1981, audenaert-eisert2005}, at \(\alpha =
\tfrac12\) the sharpness of Fuchs--van de Graaf
(Theorem~\ref{thm-fuchs-vdg}). This does not contradict the deflation:
deflation is an observable-by-observable statement, and the extremal
pairs of the universal curve are commuting, so there is nothing to
deflate.

\label{rem-no-penalty}

\end{refremark}

\textbf{The fidelity endpoint.} At \(\alpha = \tfrac12\) the spectral
measure is the Born measure and \(V_{1/2}(q) = V^{\rm meas}_{1/2}(q)\)
(\ref{eq-omega-half}); for the Helstrom projector
\(\nu_{L_0} = \mathrm{Ber}(\theta_0)\) and the fidelity dual
Corollary~\ref{cor-fidelity-dro} is a two-point computation, which is
Fuchs--van de Graaf. The argument is the original proof of Fuchs and van
de Graaf \citep{fuchs1999} in dual form: Alberti's formula
\(\mathrm{Fid}(\rho,\sigma)^2 \le
\operatorname{tr}(\rho X)\operatorname{tr}(\sigma X^{-1})\)
\citep{alberti1983} for \(X = cI - L_0\) is the data-processing
inequality for the fidelity followed by a classical Cauchy--Schwarz
step.

\begin{theorem}[Fuchs--van de Graaf via
DRO]\protect\hypertarget{thm-fuchs-vdg}{}\label{thm-fuchs-vdg}

Let \(\mathrm{Fid} := \mathrm{Fid}(\rho,\sigma)\)
(Section~\ref{sec-fidelity-boundary}),
\(\psi := \arccos \mathrm{Fid} \in [0,\tfrac\pi2)\) and, for
\(\theta \in
[0,1]\), \(\phi_\theta := \arcsin\sqrt\theta\). Then
\begin{equation}\phantomsection\label{eq-fvdg-excess}{
T(\rho,\sigma) \;\le\; \Delta^*(\theta_0) \;\le\; \sqrt{1-\mathrm{Fid}^2},
\qquad
\Delta^*(\theta) :=
\begin{cases}
\sin(2\phi_\theta + \psi)\,\sin\psi, & \theta \le \mathrm{Fid}^2,\\[2pt]
1 - \theta, & \theta \ge \mathrm{Fid}^2,
\end{cases}
}\end{equation} The second inequality is an equality iff
\(\theta_0 = \theta^* :=
\tfrac12(1 - \sqrt{1-\mathrm{Fid}^2})\), and for
\(\theta_0 \le \mathrm{Fid}^2\)
\begin{equation}\phantomsection\label{eq-fvdg-angle}{
T(\rho,\sigma) \;\le\; \sqrt{1-\mathrm{Fid}^2}\;
\cos\bigl(2(\phi_{\theta_0} - \phi_{\theta^*})\bigr).
}\end{equation}

\end{theorem}

\begin{proof}
With \(q = H_{1/2}(\rho\|\sigma) = 2(1-\mathrm{Fid})\) and \(L = L_0\),
Corollary~\ref{cor-fidelity-dro} reads \(V_{1/2}(q) = \inf_{c>1}\{c -
\mathrm{Fid}^2/[(1-\theta_0)/c + \theta_0/(c-1)]\}\), which by
Lemma~\ref{lem-classical-dual} is the classical fidelity DRO of
\(\mathrm{Ber}(\theta_0)\); so \(T = \operatorname{tr}(L_0(\rho-\sigma))
\le V_{1/2}(q) - \theta_0 = \sup\{x : \mathrm{Fid}_{\rm
cl}(\mathrm{Ber}(\theta_0+x), \mathrm{Ber}(\theta_0)) \ge \mathrm{Fid}\}\).
The classical root fidelity of binary laws is an inner product of unit
vectors, \(\sqrt{(1-\theta)(1-\theta-x)} + \sqrt{\theta(\theta+x)} =
\cos(\phi_{\theta+x} - \phi_\theta)\), so the constraint is
\(\phi_{\theta_0+x} \le \phi_{\theta_0} + \psi\): \(x \le
\sin^2(\phi_{\theta_0}+\psi) - \theta_0\) if
\(\phi_{\theta_0} + \psi \le
\tfrac\pi2\), i.e.~\(\theta_0 \le \mathrm{Fid}^2\), and
\(x \le 1-\theta_0\) otherwise. The identity
\(\sin^2(\phi+\psi) - \sin^2\phi =
\sin(2\phi+\psi)\sin\psi\) gives the first form of \(\Delta^*\), and
expanding \(\sin(\phi_\theta+\psi) = \sqrt\theta\,\mathrm{Fid} +
\sqrt{1-\theta}\sqrt{1-\mathrm{Fid}^2}\) the second. On \(\theta \le
\mathrm{Fid}^2\) the argument \(2\phi_\theta + \psi\) ranges over
\([\psi,
\pi-\psi]\), so \(\Delta^*\) has the unique maximizer \(2\phi_\theta +
\psi = \tfrac\pi2\), i.e.~\(\theta = \sin^2(\tfrac\pi4 - \tfrac\psi2) =
\tfrac12(1-\sin\psi) = \theta^*\), with value \(\sin\psi =
\sqrt{1-\mathrm{Fid}^2}\); on \(\theta \ge \mathrm{Fid}^2\),
\(1 - \theta \le
1 - \mathrm{Fid}^2 = \sin^2\psi < \sin\psi\). Finally
\(2\phi_{\theta^*} +
\psi = \tfrac\pi2\) turns \(\sin(2\phi_{\theta_0}+\psi)\sin\psi\) into
\(\sqrt{1-\mathrm{Fid}^2}\cos\bigl(2(\phi_{\theta_0}-\phi_{\theta^*})\bigr)\),
which is (\ref{eq-fvdg-angle}). \(\square\)
\end{proof}

Here \(\Delta^*(\theta) = \eta_{1/2}(2(1-\mathrm{Fid}),\theta)\) is the
Hoeffding profile at \(\alpha = \tfrac12\), which for \(\theta \le
\mathrm{Fid}^2\) reads
\(\Delta^*(\theta) = (1-\mathrm{Fid}^2)(1-2\theta) +
2\,\mathrm{Fid}\sqrt{1-\mathrm{Fid}^2}\sqrt{\theta(1-\theta)}\). The
first inequality in (\ref{eq-fvdg-excess}) is the bound of Weber et al.
\citep[Thm. 1]{weber-etal2022} for the Helstrom observable \(2L_0 - I\);
its angle form (\ref{eq-fvdg-angle}), and the fact that its maximum over
\(\theta_0\) is the Fuchs--van de Graaf bound, are what the theorem
adds. The outer inequality is the upper half of Fuchs--van de Graaf
\citep{fuchs1999}; the lower half \(1 - \mathrm{Fid} \le T\) is not a
DRO statement. Nothing in this argument deflates: at
\(\alpha = \tfrac12\) the \(k = 2\) certificate of
Theorem~\ref{thm-spectral-bernstein} uses the Born variance, and for the
Helstrom projector, a two-point Born measure of maximal variance
\(\theta_0(1-\theta_0)\), it sharpens nothing. The lever that remains is
the spectrum of \(\sigma\), which constrains \(\theta_0\).

\begin{proposition}[Spectrum-dependent Fuchs--van de
Graaf]\protect\hypertarget{prp-spectrum-fvdg}{}\label{prp-spectrum-fvdg}

Let \(p_1 \ge \dots \ge p_n\) be the eigenvalues of \(\sigma\) and
\(\mathcal K_k(\sigma) := [\sum_{i > n-k} p_i,\ \sum_{i \le k} p_i]\)
the \(k\)-th Ky Fan interval. Then
\begin{equation}\phantomsection\label{eq-spectrum-fvdg}{
T(\rho,\sigma) \;\le\; \max_{1 \le k \le n-1}\;
\max_{\theta \in \mathcal K_k(\sigma)} \Delta^*(\theta)
\;\le\; \sqrt{1-\mathrm{Fid}^2},
}\end{equation} and the second inequality is strict iff
\(\theta^* \notin \bigcup_k
\mathcal K_k(\sigma)\).

\end{proposition}

\begin{proof}
\(\theta_0 = \operatorname{tr}(\sigma L_0)\) with \(L_0\) a rank-\(k\)
projector, and by Ky Fan's maximum principle \citep[Ch. III]{bhatia1997}
\(\operatorname{tr}(\sigma P)\) ranges exactly over
\(\mathcal K_k(\sigma)\) as \(P\) ranges over rank-\(k\) projectors: the
endpoints at the spectral projectors onto the \(k\) smallest, resp.
largest, eigenvalues, the interior by continuity on the connected
Grassmannian. Hence \(T \le
\Delta^*(\theta_0) \le \max_{\mathcal K_k(\sigma)}\Delta^*\) by
Theorem~\ref{thm-fuchs-vdg}, and the maximum over \(k\) removes the
unknown rank. Since \(\theta^*\) is the unique maximizer of \(\Delta^*\)
on \([0,1]\), the constrained maximum is \(< \sqrt{1-\mathrm{Fid}^2}\)
iff no Ky Fan interval contains \(\theta^*\). \(\square\)
\end{proof}

The refinement is classical in content: it depends on the pair only
through \(\operatorname{spec}\sigma\) and the Born weight \(\theta_0\),
and noncommutativity works against it. For commuting pairs \(L_0\) is
diagonal and \(\theta_0\) is a sum of \(k\) eigenvalues of \(\sigma\),
so (\ref{eq-spectrum-fvdg}) holds with the finite set of \(k\)-fold
partial sums in place of \(\mathcal K_k(\sigma)\); the interval is the
relaxation forced by the connectedness of the Grassmannian.

\section{Discussion}\label{sec-discussion}

The quantum DRO problem for the sandwiched Rényi family has turned out
to be a classical problem in disguise, and the disguise is informative.
The dual (Theorem~\ref{thm-main}) is the optimized certainty equivalent
with a sandwiched Schatten norm in place of an \(L^{\alpha'}\) norm; the
classicalization (Theorem~\ref{thm-spectral-rep}) trades that norm for
the law \(\omega_\alpha\) of a single real random variable, whose
transform of the right kind --- stop-loss, Laplace or Stieltjes --- is a
sandwiched trace; the linearization (Theorem~\ref{thm-dro-variance})
identifies the variance of \(\omega_\alpha\) with a generalized variance
in the sense of Petz, dual to a quantum Fisher metric; and the
comparison with the Born measure (Proposition~\ref{prp-born-domination},
Proposition~\ref{prp-deflation}) says what noncommutativity does to the
worst case: it deflates it. Equality at all budgets characterizes the
commuting case (apart from \(\alpha=\tfrac12\)), while strictness at a
given budget depends on whether saturation or another
variance-insensitive branch is active. Locally the deflation is
quantified by a variance that decreases from the Born variance to the
harmonic-logarithmic one as \(\alpha\) runs from \(\tfrac12\) to
\(\infty\). Pinsker and Fuchs--van de Graaf are the Helstrom-observable
instances of the undeflated tier, which is why they carry no quantum
penalty; the deflated tier gives the Bernstein-type refinement of
Pinsker.

Several questions are left open. The sandwiched family is one of several
quantum Rényi families, and the DRO problem is equally natural for
Petz's quasi-entropies and for the \(\alpha\)-\(z\) divergences
\citep{petz1986, audenaert-datta2015}, which carry the operational
meaning in the error-exponent regime for \(\alpha < 1\)
\citep{hayashi2007, audenaert-nussbaum2008}; whether their duals close
and classicalize, and with which spectral measures, we do not know. The
measured Rényi divergences are a further case, of a different flavour:
their Donsker--Varadhan-type variational formulas \citep{berta2017} give
the DRO dual directly as an optimization over positive operators rather
than over a scalar, and since the measured divergence is dominated by
the sandwiched one by data processing and dominates the classical
divergence of any fixed measurement, the corresponding robust value lies
between \(V_\alpha\) and the measured shadow \(V^{\rm meas}_\alpha\) of
Section~\ref{sec-concentration} --- where in that window it falls, and
whether it too classicalizes, is open. A further family is that of the
quantum \(f\)-divergences defined through integral representations over
hockey-stick divergences
\citep{hirche-tomamichel2024, liu-hirche-cheng2025}, whose DRO problem
we have not studied. The size of the deflation is understood for a fixed
pair, through
\(\operatorname{Var}_{\omega_\alpha}(\ell)/\operatorname{Var}_\sigma(L)\)
(Theorem~\ref{thm-variance-ordering}, Figure~\ref{fig-deflation}), but
not as a function of the dimension: for a generic pair in high
dimension, or for local observables of a many-body Gibbs state, we
expect the ratio to be small, and quantifying this would turn the
deflation from a comparison into a resource. In the Padé bracket of
Remark~\ref{rem-pade} only the upper certificates are shown to converge;
the lower principal representations also converge weakly to
\(\omega_\alpha\), and we have not pursued the convergence of the lower
certificates. Everything here is finite-dimensional, as is Heinävaara's
tracial measure; an infinite-dimensional classicalization, for normal
states of a von Neumann algebra, would need a replacement for it.
Finally, the transport-type ambiguity sets of classical DRO ---
Wasserstein balls --- have quantum counterparts whose duals are of a
different nature, and we intend to take them up separately.

\section*{Acknowledgments}\label{acknowledgments}
\addcontentsline{toc}{section}{Acknowledgments}

This work was supported by the National Science Foundation under grants
DMS-2307115 and DMS-2606221. I thank the Département de Mathématiques of
the École Polytechnique Fédérale de Lausanne for its hospitality while
this work was completed. The author used Claude (Anthropic) extensively
in the preparation of this paper: for drafting and restructuring the
text, for cross-checking proofs and references, for numerical
verification of the inequalities and equality cases stated here, and for
the figures and their generating code. All mathematical claims were
verified by the author, who takes full responsibility for them.

\section*{Appendix A. Classical moment tools:
proofs}\label{sec-app-moment-tools}
\addcontentsline{toc}{section}{Appendix A. Classical moment tools:
proofs}

This appendix collects the proofs that are classical in content: the
classical duals (Lemma~\ref{lem-classical-dual}), transform uniqueness
(Lemma~\ref{lem-transform-unique}), the hyperbolic representation of the
mean functions (Lemma~\ref{lem-hyperbolic-mean}), the convexity-cone
stratification (Proposition~\ref{prp-cone-stratification}) and the
explicit Lyapunov and Bernstein bounds
(Proposition~\ref{prp-explicit-bernstein}). Notation is that of the
sections in which the statements appear.

\textbf{Proof of Lemma~\ref{lem-classical-dual}.} Let
\(q_{\max}^{\mathrm{cl}}(\nu) := H_\alpha^{\mathrm{cl}}
(\delta_{\ell_{\max}}\|\nu)\). For \(q < q_{\max}^{\mathrm{cl}}(\nu)\)
this is Proposition~\ref{prp-dro-duality} for the commutative model
\((\mathcal{P}(J),
\int\ell\,dP)\), i.e.~the classical divergence-DRO duality for a scalar
random variable: the Cressie--Read representation for \(\alpha
\neq 1\) (\citep[Lemma 1]{duchi-namkoong2021} for
\(\alpha \in (1,\infty)\),
\citep{ben-tal-etal2013, pichler-schlotter2020} for \(\alpha \in
[\tfrac12, 1)\), and
\citep{ben-tal-teboulle2007, ben-tal-etal2013, shapiro2017} for general
\(\varphi\)-divergences), the relative-entropy DRO at \(\alpha = 1\)
\citep{petersen-james-dupuis2000, chowdhary-dupuis2013, breuer-csiszar2016, glasserman-xu2014},
and the Rockafellar--Uryasev representation of the conditional
value-at-risk at \(\alpha = \infty\) \citep{rockafellar-uryasev2000}.
For \(q \geq q_{\max}^{\mathrm{cl}}(\nu)\) both sides equal
\(\ell_{\max}\): the value because \(\delta_{\ell_{\max}}\) is feasible,
and each infimum because, with \(w := \nu(\{\ell_{\max}\})\), bounding
the transform from below by the contribution of the atom shows that the
objective is \(\geq \ell_{\max}\) for every dual variable precisely when
\(q \geq q_{\max}^{\mathrm{cl}}(\nu)\), while it tends to
\(\ell_{\max}\) as \(c \to \ell_{\max}\) (resp.~\(t \to \infty\)). For
\(\alpha < 1\) and \(q \geq 1/(1-\alpha)\) the ball is all of
\(\mathcal{P}(J)\) and the positive part in the prefactor makes the
formula read \(\ell_{\max}\). \(\square\)

\textbf{Proof of Lemma~\ref{lem-transform-unique}.} Each transform
returns every moment of \(\nu\), which determine a measure on a compact
interval (polynomials are dense in \(C([l_1, l_n])\)): for
\(c < -\max_i|l_i|\) one has
\(\int(\ell - c)_+^{\alpha'}d\nu = (-c)^{\alpha'}\sum_k
\binom{\alpha'}{k}(-1/c)^k\int\ell^k d\nu\) with non-vanishing
coefficients when \(\alpha'\) is not an integer, while for integer
\(\alpha'\) differentiating \(\alpha'\) times gives
\((-1)^{\alpha'}\alpha'!\,\nu((c,\infty))\) for a.e.~\(c\); for
\(c > l_n\) large,
\(\int(c-\ell)^{-p}d\nu = c^{-p}\sum_k \binom{-p}{k}(-1/c)^k
\int\ell^k d\nu\) with all coefficients non-zero; and the Laplace
transform is entire with derivatives \(\int\ell^k d\nu\) at \(t = 0\).
\(\square\)

\textbf{Proof of Lemma~\ref{lem-hyperbolic-mean}.} Both sides of
(\ref{eq-hyperbolic-rep}) are homogeneous of degree one in \((a,b)\), so
we may normalize \(ab = 1\), i.e.~\((a,b) = (e^u, e^{-u})\), and by
symmetry take \(u > 0\). From (\ref{eq-fisher-kernel}), with
\(\eta := 1 - \xi = (\alpha-1)/\alpha\) and \((ab)^{\eta} = 1\), \[
\alpha\,m_\alpha(e^u, e^{-u})
\;=\; (\alpha-1)\,\frac{e^{\xi u} - e^{-\xi u}}{e^{\eta u} - e^{-\eta u}}
\;=\; \frac{\eta}{\xi}\cdot\frac{\sinh(\xi u)}{\sinh(\eta u)}
\;=\; \frac{\Psi(\xi)}{\Psi(\eta)},
\] including \(\alpha = 1\) (\(\eta = 0\)) by the convention
\(\Psi(0) = u\), which reproduces the logarithmic mean.

The integral formula exhibits \(\Psi\) as the two-sided Laplace
transform of the measure \(\tfrac{1}{2}\mathbf{1}_{[-u,u]}\,dt\): it is
even, and strictly log-convex by Hölder's inequality (equality would
force \(e^{xt} \propto e^{yt}\) a.e. on \([-u,u]\), i.e.~\(x = y\)).
Hence \(\Lambda := \log\Psi\) is even, \(\Lambda'\) is odd and strictly
increasing, and \(\Lambda' > 0\) on \((0,\infty)\). The function \[
\psi(\xi) := \Lambda(\xi) - \Lambda(1-\xi), \qquad \psi'(\xi) = \Lambda'(\xi) + \Lambda'(1-\xi),
\] is strictly increasing on \((0,2)\): for \(\xi \in (0,1]\) both terms
of \(\psi'\) are nonnegative and not both zero, while for
\(\xi \in (1,2)\), \(\psi'(\xi) = \Lambda'(\xi) - \Lambda'(\xi-1) > 0\).
Since \(\alpha\,m_\alpha = \sqrt{ab}\,e^{\psi(\xi)}\) with
\(\xi = 1/\alpha\) decreasing in \(\alpha\), \(\alpha\,m_\alpha\) is
strictly decreasing in \(\alpha\), with supremum
\(\psi(2) = \Lambda(2) - \Lambda(1) = \log\cosh u\) (as
\(\Lambda(-1) = \Lambda(1)\)), attained only at \(\xi = 2\), where
\(\sqrt{ab}\,\cosh u = (a+b)/2\), and infimum
\(\psi(0^+) = \Lambda(0) - \Lambda(1) = \log(u/\sinh u)\), where
\(\sqrt{ab}\,u/\sinh u = ab\log(a/b)/(a-b)\). \(\square\)

\textbf{Proof of Proposition~\ref{prp-cone-stratification}.} (i) Both
functions are absolutely monotone on \([l_1, l_n]\): for \(c >
l_n\),
\(\frac{d^j}{d\ell^j}(c-\ell)^{-p} = p(p+1)\cdots(p+j-1)\,(c-\ell)^{-p-j}
> 0\) and \((e^{t\ell})^{(j)} = t^j e^{t\ell} \geq 0\) for every
\(j \geq
1\); all derivatives being nonnegative, they are \((k{+}1)\)-convex for
every \(k\). (ii) Here \((k{+}1)\)-convexity is the statement that the
\((k{+}1)\)-st distributional derivative is a nonnegative measure. Let
\(\chi_c^\lambda := (\ell-c)_+^\lambda / \Gamma(\lambda+1)\), entire in
\(\lambda\), with \((\chi_c^\lambda)' = \chi_c^{\lambda-1}\) and
\(\chi_c^{-1-j} = \delta_c^{(j)}\). Then
\(\frac{d^{k+1}}{d\ell^{k+1}}\,(\ell-c)_+^{\alpha'} =
\Gamma(\alpha'+1)\,\chi_c^{\,\alpha'-k-1}\). For \(k < \alpha'\) the
right-hand side is the nonnegative locally integrable function
\(\Gamma(\alpha'+1)\Gamma(\alpha'-k)^{-1}(\ell-c)_+^{\alpha'-k-1}\), and
at \(k = \alpha' \in \mathbb{N}\) it is \(\alpha'!\,\delta_c\). For
\(k >
\alpha'\) it is \(\alpha'!\,\delta_c^{(k-\alpha')}\) when \(\alpha' \in
\mathbb{N}\) and a finite-part distribution otherwise; neither has order
zero, so neither is a measure. The obstruction sits at \(c\), so it is
seen on \([l_1, l_n]\) whenever \(c\) is interior (for \(c < l_1\) and
non-integer \(\alpha'\) the sign of \(\Gamma(\alpha'-k)\) decides, so
\((k{+}1)\)-convexity may fail there as well; for \(c \geq l_n\) the
function vanishes on \([l_1,l_n]\)). The certificate
(\ref{eq-mk-certificate}) follows as stated after the proposition.
\(\square\)

\textbf{Proof of Proposition~\ref{prp-explicit-bernstein}, Lyapunov term
of (iii) and Bernstein step of (ii).} Write \(\tilde c = c - \mu\) and
\(Y = \hat\ell - \tilde c\). Throughout, Lemma~\ref{lem-classical-dual}
expresses \(V_\alpha(q) - \mu\) through the transform of
\(\omega_\alpha\) at \(\tilde c\), and by
Lemma~\ref{lem-transform-monotone} any pointwise majorant of the
transform majorizes the value. (iii) For \(\alpha' \leq 2\), Lyapunov's
inequality on the probability measure \(\omega_\alpha\) gives
\(\|Y_+\|_{\alpha'} \leq \|Y_+\|_2
\leq (\mathbb{E}Y^2)^{1/2} = (v + \tilde c^{\,2})^{1/2}\), so, writing
\(\kappa := \kappa_\alpha(q)\),
\(V_\alpha(q) - \mu \leq \inf_{\tilde c}\{\tilde c +
\kappa\sqrt{v + \tilde c^{\,2}}\}\). For \(\kappa > 1\) the infimum is
attained at \(\tilde c^* = -\sqrt{v/(\kappa^2-1)}\) with value
\(\sqrt{v(\kappa^2 - 1)}\); intersect with the trivial bound
\(V_\alpha - \mu \leq a\). (ii) Under a law \(\nu\) with mean \(\mu\),
variance \(v\) and support in \((-\infty, l_n]\) --- \(\omega_1\) or the
certificate \(\nu_+^{(2)}\) --- the centered variable \(\hat\ell\) has
mean \(0\), variance \(v\) and \(\hat\ell \leq a\), so Bennett's lemma
\citep[§2.7]{bennett1962, boucheron-lugosi-massart2013} gives
\(\log\int e^{t\hat\ell}d\nu \leq (v/a^2)(e^{ta} - ta - 1)
\leq v t^2/(2(1 - ta/3))\) for \(0 < t < 3/a\), the second step, to
Bernstein's majorant, being the series comparison
\(e^u - u - 1 = \sum_{k \geq 2}u^k/k! \leq
\tfrac{u^2}{2}\sum_{k\geq2}(u/3)^{k-2}\). The \(\alpha = 1\) case of
Lemma~\ref{lem-classical-dual} reads \(V_1^{\rm cl}(q;\nu) - \mu =
\inf_{t>0}\{(q + \log\int e^{t\hat\ell}d\nu)/t\}\), so the majorant
gives \(V_1(q) - \mu \leq \inf_{0<t<3/a}\{q/t + vt/(2(1-ta/3))\}\). With
\(u = ta/3 \in (0,1)\) the objective is
\(\frac{aq}{3u} + \frac{3v}{2a}\frac{u}{1-u}\), stationary where
\(\frac{1-u}{u} = \frac{3}{a}\sqrt{\frac{v}{2q}}\), and substituting
this value of \(u\) gives \(\frac{aq}{3} + \sqrt{\frac{qv}{2}} +
\sqrt{\frac{qv}{2}} = \sqrt{2vq} + \frac{aq}{3}\), which is Bernstein's
inequality \citep{boucheron-lugosi-massart2013}. \(\square\)

\section*{References}\label{references}
\addcontentsline{toc}{section}{References}

\renewcommand{\bibsection}{}

\bigskip
\noindent\textsc{Luc Rey-Bellet}\\
Department of Mathematics and Statistics, University of Massachusetts Amherst,\\
710 N.~Pleasant St., Amherst, MA 01003, USA\\
\textit{E-mail address:} \texttt{luc@cns.umass.edu}

\end{document}